# Solution of the Dark Matter Problem in the Frame of the Non-Local Physics

Boris V. Alexeev

*Moscow Lomonosov University of Fine Chemical Technologies (MITHT)*
*Prospekt Vernadskogo, 86, Moscow 119570, Russia*
Boris.Vlad.Alexeev@gmail.com

The unified generalized non-local quantum kinetic and hydrodynamic theory is applied for mathematical modeling of objects in the giant scale diapason from the galaxy and Universe scale to atom structures. The principle of universal antigravitation is considered from positions of the Newtonian theory of gravitation and non-local kinetic theory. It is found that explanation of Hubble effect in the Universe and peculiar features of the rotational speeds of galaxies need not in introduction of new essence like dark matter and dark energy. The origin of difficulties consists in total Oversimplification following from principles of local physics and reflects the general shortcomings of the local kinetic transport theory.

1.  **Introduction. Preliminary remarks.**

More than ten years ago, the accelerated cosmological expansion was discovered in direct astronomical observations at distances of a few billion light years, almost at the edge of the observable Universe. This acceleration should be explained because mutual attraction of cosmic bodies is only capable of decelerating their scattering. It means that we reach the revolutionary situation not only in physics but in the natural philosophy on the whole. Practically we are in front of the new challenge since Newton's *Mathematical Principles of Natural Philosophy* was published. As result, new idea was introduced in physics about existing of a force with the opposite sign which is called universal antigravitation. Its physical source is called as dark energy that manifests itself only because of postulated property of providing antigravitation.

It was postulated that the source of antigravitation is "dark matter" which inferred to exist from gravitational effects on visible matter. But from the other side dark matter is undetectable by emitted or scattered electromagnetic radiation. It means that new essences – dark matter, dark energy – were introduced in physics only with the aim to account for discrepancies between measurements of the mass of galaxies, clusters of galaxies and the entire universe made through dynamical and general relativistic means, measurements based on the mass of the visible "luminous" matter. It could be reasonable if we are speaking about small corrections to the system of knowledge achieved by mankind to the time we are living. But mentioned above discrepancies lead to affirmation, that dark matter constitutes 80% of the matter in the universe, while ordinary matter makes up only 20%. There is a variety in the corresponding estimations, but the situation is defined by maybe emotional, but the true exclamation which can be found between thousands Internet cues – "It is humbling, perhaps even humiliating, that we know almost nothing about 96% of what is "out there"!!

Dark matter was postulated by Swiss astrophysicist Fritz Zwicky of the California Institute of Technology in 1933. He applied the virial theorem to the Coma cluster of galaxies and obtained evidence of unseen mass. Zwicky estimated the cluster's total mass based on the motions of galaxies near its edge and compared that estimate to one based on the number of galaxies and total brightness of the cluster. He found that there was about 400 times more estimated mass than was visually observable. The gravity of the visible galaxies in the cluster would be far too small for such fast orbits, so something extra was required. This is known as the "missing mass problem". Based on these conclusions, Zwicky inferred that there must be some non-visible form of matter which would provide enough of the mass and gravity to hold the cluster together.

Observations have indicated the presence of dark matter in the universe, including the rotational speeds of galaxies, gravitational lensing of background objects by galaxy clusters such as the Bullet Cluster, and the temperature distribution of hot gas in galaxies and clusters of galaxies.

The work by Vera Rubin (see for example [1, 2]) revealed distant galaxies rotating so fast that they should fly apart. Outer stars rotated at essentially the same rate as inner ones (~254 km/s). This is in marked contrast to the solar system where planets orbit the sun with velocities that decrease as their distance from the centre increases.

By the early 1970s, flat rotation curves were routinely detected. It was not until the late 1970s, however, that the community was convinced of the need for dark matter halos around spiral galaxies. The mathematical modeling (based on Newtonian mechanics and local physics) of the rotation curves of spiral galaxies was realized for the various visible components of a galaxy (the bulge, thin disk, and thick disk). These models were unable to predict the flatness of the observed rotation curve beyond the stellar disk. The inescapable conclusion, assuming that Newton's law of gravity (and the local physics description) holds on cosmological scales, that the visible galaxy was embedded in a much larger dark matter halo that contributes roughly 50–90% of the total mass of a galaxy.

As result another models of gravitation were involved in consideration – from "improved" Newtonian laws (such as modified Newtonian dynamics and tensor-vector-scalar gravity [3]) to the Einstein's theory based on the cosmological constant [4]. This term was introduced by Einstein as a mechanism to obtain a stable solution of the gravitational field equation that would lead to a static universe, effectively using dark energy to balance gravity.

Of course it is not full list of unsolved fundamental problems. Several extremely significant problems challenge modern fundamental physics, which can be titled as **"Non-solved problems of the fundamental physics"** or more precisely – of local physical kinetics of dissipative processes: 1) Kinetic theory of entropy and the problem of the initial perturbation; 2) Strict theory of turbulence; 3) Quantum non-relativistic and relativistic hydrodynamics, theory of charges separation in the atom structure; 4) Theory of ball lightning; 5) Theory of dark matter; 6) Theory of dark energy, Hubble expansion of the Universe; 7) The destiny of anti-matter after the Big Bang. 8) A unified theory of dissipative structures – from atom structure to cosmology.

I do not intend to review the different speculations based on the principles of local physics. I see another problem. It is the problem of Oversimplification – but not "trivial" simplification of the important problem. The situation is much more serious – total Oversimplification based on principles of local physics, and obvious crisis, we see in astrophysics, simply reflects the general shortenings of the local kinetic transport theory. The antigravitation problem is solved further in the frame of non-local statistical physics and the Newtonian law of gravitation.

## 2. Elementary introduction in the basic principles of the generalized Boltzmann physical kinetics and non-local statistical physics.

I deliver here some main ideas and deductions of the generalized Boltzmann physical kinetics and non-local physics. For simplicity, the fundamental methodic aspects are considered from the qualitative standpoint of view avoiding excessively cumbersome formulas. A rigorous description can be found, for example, in the monograph [5].

In 1872 L Boltzmann [6, 7] published his kinetic equation for the one-particle distribution function (DF) $f(\mathbf{r}, \mathbf{v}, t)$. He expressed the equation in the form

$$Df/Dt = J^{st}(f), \qquad (2.1)$$

where $J^{st}$ is the collision integral, and $\frac{D}{Dt} = \frac{\partial}{\partial t} + \mathbf{v} \cdot \frac{\partial}{\partial \mathbf{r}} + \mathbf{F} \cdot \frac{\partial}{\partial \mathbf{v}}$ is the substantial (particle) derivative, $\mathbf{v}$ and $\mathbf{r}$ being the velocity and radius vector of the particle, respectively. Boltzmann equation (2.1) governs the transport processes in a one-component gas, which is sufficiently rarefied that only binary collisions between particles are of importance and valid only for *two* character scales, connected with the hydrodynamic time-scale and the time-scale between particle collisions. While we are not concerned here with the explicit form of the collision integral, note that it should satisfy conservation laws of point-like particles in binary collisions. Integrals of the distribution function (i.e. its moments) determine the macroscopic hydrodynamic characteristics of the system, in particular the number density of particles $n$ and the temperature $T$. The Boltzmann equation (BE) is not of course as simple as its symbolic form above might suggest, and it is in only a few special cases that it is amenable to a solution. One example is that of a maxwellian distribution in a locally, thermodynamically equilibrium gas in the event when no external forces are present. In this case the equality $J^{st} = 0$ and $f = f_0$ is met, giving the maxwellian distribution function $f_0$. A weak point of the classical Boltzmann kinetic theory is the way it treats the dynamic properties of interacting particles. On the one hand, as the so-called "physical" derivation of the BE suggests, Boltzmann particles are treated as material points; on the other hand, the collision integral in the BE brings into existence the cross sections for collisions between particles. A rigorous approach to the derivation of the kinetic equation for $f$ (noted in following as $KE_f$) is based on the hierarchy of the Bogolyubov-Born-Green-Kirkwood-Yvon (BBGKY) [5, 8 - 12] equations.

A $KE_f$ obtained by the multi-scale method turns into the BE if one ignores the change of the distribution function (DF) over a time of the order of the collision time (or, equivalently, over a length of the order of the particle interaction radius). It is important to note [5, 12] that accounting for *the third* of the scales mentioned above leads (*prior* to introducing any approximation destined to break the Bogolyubov chain) to additional terms, generally of the same order of magnitude, appear in the BE. If the correlation functions is used to derive $KE_f$ from the BBGKY equations, then the passage to the BE means the neglect of non-local effects.

Given the above difficulties of the Boltzmann kinetic theory, the following clearly inter related questions arise. First, what is a physically infinitesimal volume and how does its introduction (and, as the consequence, the unavoidable smoothing out of the DF) affect the kinetic equation? This question can be formulated in (from the first glance) the paradox form – what is the size of the point in the physical system? Second, how does a systematic account for the proper diameter of the particle in the derivation of the $KE_f$ affect the Boltzmann equation?

In the theory developed here I shall refer to the corresponding $KE_f$ as Generalized Boltzmann Equation (GBE). The derivation of the GBE and the applications of GBE are presented, in particular, in [5]. Accordingly, our purpose is first to explain the essence of the physical generalization of the BE.

Let a particle of finite radius be characterized, as before, by the position vector $\mathbf{r}$ and velocity $\mathbf{v}$ of its center of mass at some instant of time $t$. Let us introduce physically small volume (**PhSV**) as element of measurement of macroscopic characteristics of physical system for a point containing in this **PhSV**. We should hope that **PhSV** contains sufficient particles $N_{ph}$ for statistical description of the system. In other words, a net of physically small volumes covers the whole investigated physical system.

Every **PhSV** contains entire quantity of point-like Boltzmann particles, and *the same DF $f$ is prescribed for whole PhSV in Boltzmann physical kinetics.* Therefore, Boltzmann particles are fully "packed" in the reference volume. Let us consider two adjoining physically small volumes **PhSV₁** and **PhSV₂**. We have on principle another situation for the particles of finite size moving in physical small volumes, which are *open* thermodynamic systems.

Then, the situation is possible where, at some instant of time *t*, the particle is located on the interface between two volumes. In so doing, the lead effect is possible (say, for **PhSV₂**), when the center of mass of particle moving to the neighboring volume **PhSV₂** is still in **PhSV₁**. However, the delay effect takes place as well, when the center of mass of particle moving to the neighboring volume (say, **PhSV₂**) is already located in **PhSV₂** but a part of the particle still belongs to **PhSV₁**. *Moreover, even the point-like particles (starting after the last collision near the boundary between two mentioned volumes) can change the distribution functions in the neighboring volume. The adjusting of the particles dynamic characteristics for translational degrees of freedom takes several collisions. As result, we have in the definite sense "the Knudsen layer" between these volumes. This fact unavoidably leads to fluctuations in mass and hence in other hydrodynamic quantities. Existence of such "Knudsen layers" is not connected with the choice of space nets and fully defined by the reduced description for ensemble of particles of finite diameters in the conceptual frame of open physically small volumes, therefore – with the chosen method of measurement.*

This entire complex of effects defines non-local effects in space and time. The corresponding situation is typical for the theoretical physics – we could remind about the role of probe charge in electrostatics or probe circuit in the physics of magnetic effects.

Suppose that DF $f$ corresponds to **PhSV₁** and DF $f - \Delta f$ is connected with **PhSV₂** for Boltzmann particles. In the boundary area in the first approximation, fluctuations will be proportional to the mean free path (or, equivalently, to the mean time between the collisions). Then for **PhSV** the correction for DF should be introduced as

$$f^a = f - \tau Df/Dt \qquad (2.2)$$

in the left hand side of classical BE describing the translation of DF in phase space. As the result

$$Df^a/Dt = J^B, \qquad (2.3)$$

where $J^B$ is the Boltzmann local collision integral.

*Important to notice that it is only qualitative explanation of GBE derivation obtained earlier (see for example [5]) by different strict methods from the BBGKY – chain of kinetic equations.* The structure of the $KE_f$ is generally as follows

$$\frac{Df}{Dt} = J^B + J^{nonlocal}, \qquad (2.4)$$

where $J^{nonlocal}$ is the non-local integral term incorporating the non-local time and space effects. The generalized Boltzmann physical kinetics, in essence, involves a local approximation

$$J^{nonlocal} = \frac{D}{Dt}\left(\tau \frac{Df}{Dt}\right) \qquad (2.5)$$

for the second collision integral, here $\tau$ being proportional to the mean time *between* the particle collisions. We can draw here an analogy with the Bhatnagar - Gross - Krook (BGK) approximation for $J^B$,

$$J^B = \frac{f_0 - f}{\tau}, \qquad (2.6)$$

which popularity as a means to represent the Boltzmann collision integral is due to the huge simplifications it offers. In other words – the local Boltzmann collision integral admits approximation via the BGK algebraic expression, but more complicated non-local integral can be expressed as differential form (2.5). The ratio of the second to the first term on the right-hand side of Eq. (2.4) is given to an order of magnitude as $J^{nonlocal}/J^B \approx O(\text{Kn}^2)$ and at large Knudsen numbers (defining as ratio of mean free path of particles to the character hydrodynamic length) these terms become of the same order of magnitude. It would seem that at small Knudsen numbers answering to hydrodynamic description the contribution from the second term on the right-hand side of Eq. (2.4) is negligible.

*This is not the case, however.* When one goes over to the hydrodynamic approximation (by multiplying the kinetic equation by collision invariants and then integrating over velocities), the Boltzmann integral part vanishes, and the second term on the right-hand side of Eq. (2.4) gives a single-order contribution in the generalized Navier - Stokes description. Mathematically, we cannot neglect a term with a small parameter in front of the higher derivative. Physically, the appearing additional terms are due to viscosity and they correspond to the small-scale Kolmogorov turbulence [5, 13]. The integral term $J^{nonlocal}$ turns out to be important both at small and large Knudsen numbers in the theory of transport processes. Thus, $\tau Df/Dt$ is the distribution function fluctuation, and writing Eq. (2.3) without taking into account Eq. (2.2) makes the BE non-closed. From viewpoint of the fluctuation theory, Boltzmann employed the simplest possible closure procedure $f^a = f$.

Then, the additional GBE terms (as compared to the BE) are significant for any Kn, and the order of magnitude of the difference between the BE and GBE solutions is impossible to tell beforehand. For GBE the generalized H-theorem is proven [5, 14].

It means that Boltzmann equation does not belong even to the class of minimal physical models and corresponds so to speak to "the likelihood models". In this sense BE is wrong equation. This remark refers also to all consequences of the Boltzmann kinetic theory including "classical" hydrodynamics.

Obviously the generalized hydrodynamic equations (GHE) will explicitly involve fluctuations proportional to $\tau$. In the hydrodynamic approximation, the mean time $\tau$ between the collisions is related to the dynamic viscosity $\mu$ by

$$\tau\, p = \Pi\mu, \qquad (2.7)$$

[15, 16]. For example, the continuity equation changes its form and will contain terms proportional to viscosity. On the other hand, if the reference volume extends over the whole cavity with the hard walls, then the classical conservation laws should be obeyed, and this is exactly what the monograph [5] proves. Now several remarks of principal significance:

1. All fluctuations are found from the strict kinetic considerations and tabulated [5, 13]. The appearing additional terms in GHE are due to viscosity and they correspond to the small-scale Kolmogorov turbulence. The neglect of formally small terms is equivalent, in particular, to dropping the (small-scale) Kolmogorov turbulence from consideration and is the origin of all principal difficulties in usual turbulent theory. Fluctuations on the wall are equal to zero, from the physical point of view this fact corresponds to the laminar sub-layer. Mathematically it leads to additional boundary conditions for GHE. Major difficulties arose when the question of existence and uniqueness of solutions of the Navier - Stokes equations was addressed. O A Ladyzhenskaya has shown for three-dimensional flows that under smooth initial conditions a unique solution is only possible over a finite time interval. Ladyzhenskaya even introduced a "correction" into the Navier - Stokes equations in order that its unique solvability could be proved (see discussion in [17]). GHE do not lead to these difficulties.

2. It would appear that in continuum mechanics the idea of discreteness can be abandoned altogether and the medium under study be considered as a continuum in the literal sense of the word. Such an approach is of course possible and indeed leads to the Euler equations in hydrodynamics. However, when the viscosity and thermal conductivity effects are to be included, a totally different situation arises. As is well known, the dynamical viscosity is proportional to the mean time $\tau$ between the particle collisions, and a continuum medium in the Euler model with $\tau = 0$ implies that neither viscosity nor thermal conductivity is possible.

3. The non-local kinetic effects listed above will always be relevant to a kinetic theory using one particle description – including, in particular, applications to liquids or plasmas, where self-consistent forces with appropriately cut-off radius of their action are introduced to expand the capability of GBE [18, 19]. Fluctuation effects occur in any open thermodynamic system bounded by a control surface transparent to particles. GBE (2.3) leads to generalized hydrodynamic equations [5] as the local approximation of non local effects, for example, to the continuity equation

$$\frac{\partial \rho^a}{\partial t} + \frac{\partial}{\partial \mathbf{r}} \cdot (\rho \mathbf{v}_0)^a = 0, \qquad (2.8)$$

where $\rho^a$, $\mathbf{v}_0^a$, $(\rho \mathbf{v}_0)^a$ are calculated in view of non-locality effect in terms of gas density $\rho$, hydrodynamic velocity of flow $\mathbf{v}_0$, and density of momentum flux $\rho \mathbf{v}_0$; for locally Maxwellian distribution, $\rho^a$, $(\rho \mathbf{v}_0)^a$ are defined by the relations

$$(\rho - \rho^a)/\tau = \frac{\partial \rho}{\partial t} + \frac{\partial}{\partial \mathbf{r}} \cdot (\rho \mathbf{v}_0), \quad (\rho \mathbf{v}_0 - (\rho \mathbf{v}_0)^a)/\tau = \frac{\partial}{\partial t}(\rho \mathbf{v}_0) + \frac{\partial}{\partial \mathbf{r}} \cdot \rho \mathbf{v}_0 \mathbf{v}_0 + \vec{\mathbf{I}} \cdot \frac{\partial p}{\partial \mathbf{r}} - \rho \mathbf{a}, \qquad (2.9)$$

where $\vec{\mathbf{I}}$ is a unit tensor, and $\mathbf{a}$ is the acceleration due to the effect of mass forces.

In the general case, the parameter $\tau$ is the non-locality parameter; in quantum hydrodynamics, its magnitude is defined by the "time-energy" uncertainty relation. The violation of Bell's inequalities [20] is found for local statistical theories, and the transition to non-local description is inevitable. The following conclusion of principal significance can be done from the generalized quantum consideration [21, 22]:
1. Madelung's quantum hydrodynamics is equivalent to the Schrödinger equation (SE) and leads to description of the quantum particle evolution in the form of Euler equation and continuity equation.
2. SE is consequence of the Liouville equation as result of the local approximation of non-local equations.
3. Generalized Boltzmann physical kinetics leads to the strict approximation of non-local effects in space and time and after transmission to the local approximation leads to parameter $\tau$, which on the quantum level corresponds to the uncertainty principle "time-energy".
4. GHE lead to SE as a deep particular case of the generalized Boltzmann physical kinetics and therefore of non-local hydrodynamics.

On principal GHE needn't in using of the "time-energy" uncertainty relation for estimation of the value of the non-locality parameter $\tau$. Moreover the "time-energy" uncertainty relation does not lead to the exact relations and from position of non-local physics is only the simplest estimation of the non-local effects. Really, let us consider two neighboring physically infinitely small volumes $\mathbf{PhSV_1}$ and $\mathbf{PhSV_2}$ in a non-equilibrium system. Obviously the time $\tau$ should tends to diminish with increasing of the velocities $u$ of particles invading in the nearest neighboring physically infinitely small volume ($\mathbf{PhSV_1}$ or $\mathbf{PhSV_2}$):

$$\tau = H/u^n. \qquad (2.10)$$

But the value $\tau$ cannot depend on the velocity direction and naturally to tie $\tau$ with the particle kinetic energy, then

$$\tau = H/mu^2, \qquad (2.11)$$

where $H$ is a coefficient of proportionality, which reflects the state of physical system. In the simplest case $H$ is equal to Plank constant $\hbar$ and relation (2.11) becomes compatible with the Heisenberg relation.

Finally, we can state that introduction of open control volume by the reduced description for ensemble of particles of finite diameters leads to fluctuations (proportional to Knudsen number) of velocity moments in the volume. This fact leads to the significant reconstruction of the theory of transport processes. Obviously the mentioned non-local effects can be discussed from viewpoint of breaking of the Bell's inequalities [20] because in the non-local theory the measurement (realized in **PhSV₁**) has influence on the measurement realized in the adjoining space-time point in **PhSV₂** and verse versa.

## 3. Plasma – gravitational analogy in the generalized theory of Landau damping. Hubble expansion.

My aim consists in the application of plasma – gravitational analogy for the effect of Hubble expansion using the generalized theory of Landau damping and the generalized Boltzmann physical kinetics developed by me [5, 12 – 14, 18, 19, 21 - 24]. The collisionless damping of electron plasma waves was predicted by Landau in 1946 [25] and later was confirmed experimentally. Landau damping plays a significant role in many electronics experiments and belongs to the most well known phenomenon in statistical physics of ionized gases. The physical origin of the collisionless Landau wave damping is simple. Really, if individual electron moves in the periodic electric field, this electron can diminish its energy (electron velocity larger than phase velocity of wave) or receive additional energy from the wave (electron velocity less than phase velocity of wave). Then the total energy balance for a swarm of electrons depends on quantity of "cold" and "hot" electrons. For the Maxwellian distribution function, the quantity of "cold" electrons is more than quantity of "hot" electrons. This fact leads to, so-called, the collisionless Landau damping of the electric field perturbation. In spite of transparent physical sense, the effect of Landau damping has continued to be of great interest to theorist as well. Much of this interest is connected with counterintuitive nature of result itself coupled with the rather abstruse mathematical nature of Landau's original derivation (including so-called Landau's rule of complex integral calculation). Moreover, for these reasons there were publications containing some controversy over the reality of the phenomenon. In papers [23, 24] the difficulties originated by Landau's derivation were clarified. The mentioned consideration leads to another solution of Vlasov - Landau equation, these ones in agreement with data of experiments. The problem solved in this article consists in consideration of the generalized theory of Landau damping in gravitating systems from viewpoint of Generalized Boltzmann Physical Kinetics and non-local physics. The influence of the particle collisions is taken into account.

Plasma – gravitational analogy is well-known and frequently used effect in physical kinetics. The origin of analogy is simple and is connected with analogy between Coulomb law and Newtonian law of gravitation. From other side electrical charges can have different signs whereas there is just one kind of "gravitational charge" (i.e. masses of particles) corresponding to the force of attraction. This fact leads to the extremely important distinctions in formulation of the generalized theory of Landau damping in gravitational media. In the following, we intend to use the classical non-relativistic Newtonian law of gravitation

$$\mathbf{F}_{21} = \gamma_N \frac{m_1 m_2}{r_{12}^2} \frac{\mathbf{r}_{12}}{r_{12}}, \qquad (3.1)$$

where $\mathbf{F}_{21}$ is the force acting on the particle "1" from the particle "2", $\mathbf{r}_{12}$ is vector directed from the center-of-mass of the particle "1" to the particle "2", $\gamma_N$ is gravitational constant $\gamma_N = 6.6 \cdot 10^{-8} cm^3/(g \cdot s^2)$; the corresponding force $\mathbf{g}_{21}$ per mass unit is

$$\mathbf{g}_{21} = \mathbf{F}_{21}/m_1. \qquad (3.2)$$

The flux

$$\Phi = \int_S g_n dS \qquad (3.3)$$

for closed surface $S$ can be calculated using (3.2); one obtains

$$\int_S g_n dS = -4\pi \gamma_N \int_V \rho^a dV, \qquad (3.4)$$

where $\rho^a$ is density *inside* of volume $V$ bounded by the surface $S$. As usual, Eq. (3.4) can be rewritten as

$$\int_V (div\ \mathbf{g} + 4\pi\gamma_N\rho^a)dV = 0. \tag{3.5}$$

The definite integral (3.5) is equal to zero for arbitrary volume $V$, then

$$div\ \mathbf{g} = -4\pi\gamma_N\rho^a, \tag{3.6}$$

and after introduction the gravitational potential $\Psi$

$$\mathbf{g} = -\partial\Psi/\partial\mathbf{r} \tag{3.7}$$

we reach the known Poisson equation

$$\Delta\Psi = 4\pi\gamma_N\rho^a. \tag{3.8}$$

Generalized Boltzmann physical kinetics leads to possibility to calculate the density $\rho^a$ using the density $\rho$ (obtained with the help of the one particle DF $f$) and the fluctuation term $\rho^{fl}$. All fluctuation terms in the GBE theory were tabulated [5, 13] and for $\rho^{fl}$ we have

$$\rho^{fl} = \tau\left(\frac{\partial\rho}{\partial t} + \frac{\partial}{\partial\mathbf{r}}\cdot\rho\mathbf{v}_0\right), \tag{3.9}$$

where $\mathbf{v}_0$ is hydrodynamic velocity. After substitution of $\rho^{fl}$ in (3.8) one obtains

$$\Delta\Psi = 4\pi\gamma_N\left[\rho - \tau\left(\frac{\partial\rho}{\partial t} + \frac{\partial}{\partial\mathbf{r}}\cdot\rho\mathbf{v}_0\right)\right]. \tag{3.10}$$

From Eqs. (3.6) - (3.9) follow that classical Newtonian field equation

$$\Delta\Psi = 4\pi\gamma_N\rho \tag{3.11}$$

valid only for situation when the fluctuations terms can be omitted and then

$$\rho = \rho^a. \tag{3.12}$$

This condition can be considered as the simplest closure condition, but in the general case, the other hydrodynamic equations should be involved into consideration because Eq. (3.10) contains hydrodynamic velocity $\mathbf{v}_0$. As result, one obtains the system of moment equations, i.e. gravitation equation

$$\frac{\partial}{\partial\mathbf{r}}\cdot\mathbf{g} = -4\pi\gamma_N\left[\rho - \tau\left(\frac{\partial\rho}{\partial t} + \frac{\partial}{\partial\mathbf{r}}\cdot\rho\mathbf{v}_0\right)\right], \tag{3.13}$$

and generalized continuity, motion and energy equations which can be further applied to the theory of the rotation curves of spiral galaxies.

The following mathematical transformations will be obtained on the level of the generalized theory of Landau damping based on the generalized Boltzmann equation (GBE), and need in some preliminary additional explanations from viewpoint of so-called dark energy and dark matter.

As it was mentioned above the accelerated cosmological expansion was discovered in direct astronomical observations. For explanation of this acceleration new idea was introduced in physics about existing of a force with the opposite sign which is called universal antigravitation. In the simplest interpretation, dark energy is related usually to the Einstein cosmological constant. In review [4] the modified Newton force is written as

$$F(r) = -\frac{\gamma_N M}{r^2} + \frac{8\pi\gamma_N}{3}\rho_\mathbf{v} r, \tag{3.14}$$

where $\rho_\mathbf{v}$ is the Einstein – Gliner vacuum density introduced also in [26]. The problem can be solved without the ideology of the Einstein – Gliner vacuum. However for us is interesting the interpretation of the modified law (3.14), [4]. In the limit of large distances, the influence of central mass $M$ becomes negligibly small and the field of forces is determined only by the second term in the right side of (3.14). It follows from relation (3.14) that there is a "equilibrium" distance $r_v$, at which the sum of the gravitation and antigravitation forces is equal to zero. In other words $r_v$ is "the zero-gravitational radius". For so called Local Group of galaxies estimation of $r_v$ is about 1Mpc, [4]. There are no theoretical methods of the density $\rho_v$ calculation. Obviously, the second term in relation (3.14) should be defined as result of solution of the self-consistent gravitational problem.

Let us return now to the formulation of plasma-gravitational analogy in the frame of generalized theory of Landau damping. I intend to apply the GBE model with the aim to obtain the dispersion relation for one component gas placed in the self-consistent gravitational field and to consider effect of antigravitation in the frame of the Newton theory of gravitation.

With this aim let us admit now that there is a gravitational perturbation $\partial \Psi$ in the system of particles connected with the density perturbation $\delta \rho$. These perturbations are connected with the perturbation of DF in the system, which was before in the local equilibrium. In doing so, we will make the additional assumptions for simplification of the problem, namely:

(a) Consideration of the self-consistent gravitational field corresponds to the area of the large distance $r$ (see relation (3.14)) from the central mass $M$ where the first term is not significant and in particular the problem corresponds to the plane case. As mentioned above the second term should be defined as a self-consistent force of the Newtonian origin

$$F = -\frac{\partial \partial \Psi}{\partial x}. \qquad (3.15)$$

(b) The integral collision term is written in the Bhatnagar - Gross - Krook (BGK) form

$$J = -\frac{f - f_0}{v_r^{-1}} \qquad (3.16)$$

in the right-hand side of the GBE. Here, $f_0$ and $v_r^{-1} = \tau_r$ are respectively a certain equilibrium distribution function and the relaxation time.

(c) The evolution of particles in a self-consistent gravitational field corresponds to a non-stationary one-dimensional model, $u$ is the velocity component along the $x$ axis.

(d) The distribution function $f$ deviate little from its equilibrium counterpart $f_0$.

$$f = f_0(u) + \delta f(x, u, t). \qquad (3.17)$$

(e) A wave number $k$ and a complex frequency $\omega$ ($\omega = \omega' + i\omega''$) are appropriate to the wave mode considered;

$$\delta f = \langle \delta f \rangle e^{i(kx - \omega t)}, \qquad (3.18)$$

$$\delta \Psi = \langle \delta \varphi \rangle e^{i(kx - \omega t)}. \qquad (3.19)$$

(f) The quadratic GBE terms determining the deviation from the equilibrium DF are neglected.

Under these assumptions listed above, the GBE is written as follows (see also (2.2) – (2.6)):

$$\frac{\partial f}{\partial t} + u\frac{\partial f}{\partial x} + F\frac{\partial f}{\partial u} - \tau \left\{ \frac{\partial^2 f}{\partial t^2} + 2u\frac{\partial^2 f}{\partial t \partial x} + u^2 \frac{\partial^2 f}{\partial x^2} + 2F\frac{\partial^2 f}{\partial t \partial u} + \right.$$
$$\left. + \frac{\partial F}{\partial t}\frac{\partial f}{\partial u} + F\frac{\partial f}{\partial x} + u\frac{\partial F}{\partial x}\frac{\partial f}{\partial u} + F^2 \frac{\partial^2 f}{\partial u^2} + 2uF\frac{\partial^2 f}{\partial u \partial x} \right\} = -v_r \delta f \qquad (3.20)$$

where the following relations take place for the corresponding terms in Eq. (3.20)

$$\frac{\partial f}{\partial t} = -i\omega \delta f, \quad u\frac{\partial f}{\partial x} = iku\delta f, \quad F\frac{\partial f}{\partial u} = -\frac{\partial \delta \Psi}{\partial x}\frac{\partial f_0}{\partial u}, \quad \frac{\partial^2 f}{\partial t^2} = -\omega^2 \delta f, \quad 2u\frac{\partial^2 f}{\partial t \partial x} = 2\omega uk\delta f,$$

$$u^2 \frac{\partial^2 f}{\partial x^2} = -u^2 k^2 \delta f, \quad 2F\frac{\partial^2 f}{\partial t \partial u} = 0, \quad \frac{\partial F}{\partial t}\frac{\partial f}{\partial u} = -\frac{\partial}{\partial t}\frac{\partial \delta \Psi}{\partial x}\frac{\partial f}{\partial u} = -\omega k\delta \Psi \frac{\partial f_0}{\partial u}, \quad F\frac{\partial f}{\partial x} = 0, \quad (3.21)$$

$$u\frac{\partial f}{\partial u}\frac{\partial F}{\partial x} = -u\frac{\partial f}{\partial u}\frac{\partial}{\partial x}\frac{\partial \delta \Psi}{\partial x} = k^2 u\delta \Psi \frac{\partial f_0}{\partial u}, \quad F^2 \frac{\partial^2 f}{\partial u^2} = 0, \quad \frac{\partial^2 f}{\partial u \partial x} 2uF = 0.$$

We are concerned with developing (within the GBE framework) the dispersion relation for gravitational field, and substitution of (3.21) into Eq. (3.20) yields

$$\{i(ku-\omega) + v_r + \tau(ku-\omega)^2\}\langle \delta f \rangle - \langle \delta \Psi \rangle \frac{\partial f_0}{\partial u} k\{i + \tau(ku-\omega)\} = 0. \quad (3.22)$$

For the physical system under consideration, the influence of the collision term $v_r\langle \delta f \rangle$ is rather small. For this case the Poisson equation in the form (3.11) can be used and from (3.22), (3.23)

$$k^2 \langle \delta \Psi \rangle = -4\pi \gamma_N \langle \delta n \rangle, \quad (3.23)$$

follows

$$\langle \delta f \rangle = \frac{4\pi \gamma_N m}{k} \frac{[i - \tau(\omega - ku)]\frac{\partial f_0}{\partial u}}{i(\omega - ku) - \tau(\omega - ku)^2 - v_r} \langle \delta n \rangle. \quad (3.24)$$

After integration over all $u$ we arrive at the dispersion relation

$$1 = \frac{4\pi \gamma_N m}{k} \int_{-\infty}^{+\infty} \frac{\frac{\partial f_0}{\partial u}[i - \tau(\omega - ku)]}{i(\omega - ku) - \tau(\omega - ku)^2 - v_r} du. \quad (3.25)$$

Let us suppose that the velocity depending part of DF $f_0$ corresponds to the Maxwell DF. Then after differentiating in Eq. (3.25) under the sign of integral and some transformations we obtain the integral dispersion equation

$$1 - \frac{1}{r_A^2 k^2}\left[1 - \sqrt{\frac{m}{2\pi k_B T}} \int_{-\infty}^{+\infty} \frac{\{[i - \tau(\omega - ku)]\omega - v_r\}e^{-mu^2/2k_B T}}{i(\omega - ku) - \tau(\omega - ku)^2 - v_r} du\right] = 0, \quad (3.26)$$

where

$$r_A = \sqrt{\frac{k_B T}{4\pi \gamma_N m^2 n}}. \quad (3.27)$$

Poisson equation (3.11) has the structure like the Poisson equation for the electrical potential, as result the relation for $r_A$ is analogous to the Debye - Hueckel radius $r_D = \sqrt{k_B T/(4\pi e^2 n)}$.

Introducing now the dimensionless variables

$$\hat{u} = u\sqrt{\frac{m}{2k_B T}}, \quad \hat{\omega} = \omega \frac{1}{k}\sqrt{\frac{m}{2k_B T}}, \quad \hat{v}_r = v_r \frac{1}{k}\sqrt{\frac{m}{2k_B T}}, \quad \hat{\tau} = \tau k\sqrt{\frac{2k_B T}{m}} \quad (3.28)$$

we can rewrite Eq (3.26) in the form

$$1 - \frac{1}{r_A^2 k^2}\left[1 - \frac{1}{\sqrt{\pi}} \int_{-\infty}^{+\infty} \frac{\{[i - \hat{\tau}(\hat{\omega} - \hat{u})]\hat{\omega} - \hat{v}_r\}e^{-\hat{u}^2}}{i(\hat{\omega} - \hat{u}) - \hat{\tau}(\hat{\omega} - \hat{u})^2 - \hat{v}_r} d\hat{u}\right] = 0. \quad (3.29)$$

Now consider a situation in which the denominator of the complex integrand in Eq. (3.29) becomes zero. The quadratic equation

$$\hat{\tau} y^2 - iy + \hat{v}_r = 0, \quad y = \hat{\omega} - \hat{u} \quad (3.30)$$

has the roots

$$y_1 = \frac{i}{2\hat{\tau}}\left(1+\sqrt{1+4\hat{\tau}\hat{v}_r}\right), \quad y_2 = \frac{i}{2\hat{\tau}}\left(1-\sqrt{1+4\hat{\tau}\hat{v}_r}\right). \tag{3.31}$$

Hence, Eq. (3.29) can be rewritten as

$$1 - \frac{1}{r_A^2 k^2}\left[1 + \frac{1}{\tau\sqrt{\pi}}\int_{-\infty}^{+\infty}\frac{\{[i+\hat{\tau}(\hat{u}-\hat{\omega})]\hat{\omega}-\hat{v}_r\}e^{-\hat{u}^2}}{(\hat{u}-\hat{u}_1)(\hat{u}-\hat{u}_2)}d\hat{u}\right] = 0 \tag{3.32}$$

where

$$\hat{u}_1 = \hat{\omega} - y_1, \quad \hat{u}_2 = \hat{\omega} - y_2. \tag{3.33}$$

Let us transform equation (3.32) to the following one:

$$1 - \frac{1}{r_A^2 k^2}\left\{1 + \frac{1}{\sqrt{\pi}}\left[\left(\frac{i\hat{v}_r + 0.5\hat{\omega}}{\sqrt{1+4\hat{\tau}\hat{v}_r}} - 0.5\hat{\omega}\right)\int_{-\infty}^{+\infty}\frac{e^{-\hat{u}^2}}{(\hat{u}_1-\hat{u})}d\hat{u} - \left(\frac{i\hat{v}_r + 0.5\hat{\omega}}{\sqrt{1+4\hat{\tau}\hat{v}_r}} + 0.5\hat{\omega}\right)\int_{-\infty}^{+\infty}\frac{e^{-\hat{u}^2}}{(\hat{u}_2-\hat{u})}d\hat{u}\right]\right\} = 0 \tag{3.34}$$

Equation (3.34) contains improper Cauchy type integrals. From the theory of complex variables is known Cauchy's integral formula: if the function $f(z)$ is analytic inside and on a simple closed curve $C$, and $z_0$ is any point inside C, then

$$f(z_0) = -\frac{1}{2\pi i}\oint_C \frac{f(z)}{z_0 - z}dz \tag{3.35}$$

where $C$ is traversed in the positive (counterclockwise) sense.

Let $C$ be the boundary of a simple closed curve placed in lower half plane (for example a semicircle of radius $R$) with the corresponding element of real axis, $z_0$ is an interior point. As usual after adding to this semicircle a cross-cut connecting semicircle $C$ with the interior circle (surrounding $z_0$) of the infinite small radius for analytic $f(z)$ the following formula obtains

$$\oint_C \frac{f(z)}{z_0 - z}dz = -\int_{-R}^{R}\frac{f(\tilde{x})}{z_0 - \tilde{x}}d\tilde{x} + \int_{C_R}\frac{f(z)}{z_0 - z}dz + 2\pi i f(z_0), \tag{3.36}$$

because the integrals along cross-cut cancel each other, ($z = \tilde{x} + i\tilde{y}$).

Analogous for upper half plane

$$\oint_C \frac{f(z)}{z_0 - z}dz = \int_{-R}^{R}\frac{f(\tilde{x})}{z_0 - \tilde{x}}d\tilde{x} + \int_{C_R}\frac{f(z)}{z_0 - z}dz + 2\pi i f(z_0). \tag{3.37}$$

The formulae (3.36), (3.37) could be used for calculation (including the case $R \to \infty$) of the integrals along the real axis with the help of the residual theory *for arbitrary* $z_0$ if analytical function $f(z)$ satisfies the special conditions of decreasing by $R \to \infty$.

Let us consider now integral $\int_{C_R}\frac{e^{-z^2}}{z_0 - z}dz$. Generally speaking, for function $f(z) = e^{-z^2}$ Cauchy's conditions are not satisfied. Really for a point $z = \tilde{x} + i\tilde{y}$ this function is $f(z) = e^{\tilde{y}^2 - \tilde{x}^2}[\cos(2\tilde{x}\tilde{y}) - i\sin(2\tilde{x}\tilde{y})]$ and $f(z)$ is growing by $|\tilde{y}| > |\tilde{x}|$ for this part of $C_R$.

But from physical point of view in **the linear problem** of interaction of individual particles **only** with waves of potential self-consistent gravitational field the natural assumption can be introduced that solution depends **only** of concrete $z_0 = \hat{\omega}' + i\hat{\omega}''$, but does not depend of another possible modes of oscillations in physical system.

It can be realized only if the calculations do not depend of choosing of contour $C_R$. This fact leads to the additional conditions, for lower half plane

$$\int_{-\infty}^{\infty} \frac{f(\tilde{x})}{z_0 - \tilde{x}} d\tilde{x} = 2\pi i f(z_0), \qquad (3.38)$$

and for upper half plane

$$\int_{-\infty}^{\infty} \frac{f(\tilde{x})}{z_0 - \tilde{x}} d\tilde{x} = -2\pi i f(z_0). \qquad (3.39)$$

As it is shown in [23, 24] the Landau approximation for the improper integral contains (in the implicit form) restrictions (valid only for close vicinity of $\tilde{x}$-axis) for the choice of contour $C$; these restrictions lead to the continuous spectrum. The question arises, is it possible to find solutions of the equation (3.34) by the restrictions (3.38), (3.39)? In the following will be shown that the conditions (3.38), (3.39) together with (3.34) lead to the discrete spectrum of $z_0 = \hat{\omega}' + i\hat{\omega}''$ and from physical point of view conditions (3.38), (3.39) can be considered as conditions of quantization.

The relations (3.38), (3.39) are the additional conditions which physical sense consists in the extraction of independent oscillations – oscillations which existence does not depend on presence of other oscillations in considering physical system.

Then Eq. (3.34) produces the dispersion relation, which admits a damped gravitational wave solution $(\hat{\omega}'' < 0)$ for small influence of the collision integral (see also [23, 24]):

$$\mp e^{\hat{u}_2^2} \frac{1 - r_A^2 k^2}{2\sqrt{\pi}} = \frac{\hat{v}_r}{\sqrt{1 + 4\hat{\tau}\hat{v}_r}} - \frac{i\hat{\omega}}{2}\left(1 + \frac{1}{\sqrt{1 + 4\hat{\tau}\hat{v}_r}}\right). \qquad (3.40)$$

where

$$\hat{u}_2^2 = \hat{\omega}'^2 - \hat{\omega}''^2 - \hat{\omega}''\frac{\sqrt{1 + 4\hat{\tau}\hat{v}_r} - 1}{\hat{\tau}} - \frac{1 + 2\hat{\tau}\hat{v}_r - \sqrt{1 + 4\hat{\tau}\hat{v}_r}}{2\hat{\tau}^2} + i\left(2\hat{\omega}'' + \frac{\sqrt{1 + 4\hat{\tau}\hat{v}_r} - 1}{\hat{\tau}}\right)\hat{\omega}'. \qquad (3.41)$$

The time of the collision relaxation $\tau_{rel} = v_r^{-1}$ for gravitational physical system can be estimated in terms of the mean time $\tau$ between close collisions and the Coulomb logarithm:

$$\tau v_r = \Lambda, \ \hat{\tau}\hat{v}_r = \Lambda. \qquad (3.42)$$

We separate the real and imaginary parts in Eq. (3.40). One obtains for the real part

$$\mp \frac{1 - r_a^2 k^2}{2\sqrt{\pi}} \exp\left\{\hat{\omega}'^2 - \hat{\omega}''^2 - \hat{\omega}''\hat{v}_r \frac{\sqrt{1 + 4\Lambda} - 1}{\Lambda} - \hat{v}_r^2 \frac{1 + 2\Lambda - \sqrt{1 + 4\Lambda}}{2\Lambda^2}\right\} =$$

$$= \left[\frac{\hat{v}_r}{\sqrt{1 + 4\Lambda}} + 0.5\hat{\omega}'' + \frac{0.5\hat{\omega}''}{\sqrt{1 + 4\Lambda}}\right]\cos\left[\hat{\omega}'\left(2\hat{\omega}'' + \hat{v}_r \frac{\sqrt{1 + 4\Lambda} - 1}{\Lambda}\right)\right] - \qquad (3.43)$$

$$- 0.5\hat{\omega}'\left[1 + \frac{1}{\sqrt{1 + 4\Lambda}}\right]\sin\left[\hat{\omega}'\left(2\hat{\omega}'' + \hat{v}_r \frac{\sqrt{1 + 4\Lambda} - 1}{\Lambda}\right)\right].$$

Similarly, for the imaginary part we find

$$0.5\hat{\omega}'\left[1 + \frac{1}{\sqrt{1 + 4\Lambda}}\right]\cos\left[\hat{\omega}'\left(2\hat{\omega}'' + \hat{v}_r \frac{\sqrt{1 + 4\Lambda} - 1}{\Lambda}\right)\right] +$$

$$+ \left[\frac{\hat{v}_r}{\sqrt{1 + 4\Lambda}} + 0.5\hat{\omega}'' + \frac{0.5\hat{\omega}''}{\sqrt{1 + 4\Lambda}}\right]\sin\left[\hat{\omega}'\left(2\hat{\omega}'' + \hat{v}_r \frac{\sqrt{1 + 4\Lambda} - 1}{\Lambda}\right)\right] = 0. \qquad (3.44)$$

Coulomb logarithm $\Lambda$ is large for such objects like galaxies, the typical value $\Lambda \sim 200$ and the system of equations (3.43), (3.44) for the large Coulomb logarithm $\Lambda$ simplifies to

$$\mp \frac{1-r_a^2 k^2}{\sqrt{\pi}} e^{\tilde{\omega}'^2 - \tilde{\omega}''^2} = \tilde{\omega}'' \cos(2\tilde{\omega}'\tilde{\omega}'') - \tilde{\omega}' \sin(2\tilde{\omega}'\tilde{\omega}''), \qquad (3.45)$$

$$\tilde{\omega}' \cos(2\tilde{\omega}'\tilde{\omega}'') + \tilde{\omega}'' \sin(2\tilde{\omega}'\tilde{\omega}'') = 0. \qquad (3.46)$$

Let us introduce the notation

$$\alpha = 2\tilde{\omega}'\tilde{\omega}'', \quad \beta = 1 - r_a^2 k^2, \qquad (3.47)$$

we obtain the universal equation

$$-e^{\sigma \cot \sigma} \sin \sigma = \frac{\pi}{2\beta^2} \sigma, \qquad (3.48)$$

where $\sigma = -2\alpha = -4\tilde{\omega}'\tilde{\omega}''$. This equation does not depend on the sign in front of parameter $\beta$ in (3.45). The exact solution of equation (3.48) can be found with the help of the $W$-function of Lambert

$$\sigma_n = \text{Im}\left[W_n\left(\frac{2\beta^2}{\pi}\right)\right], \qquad (3.49)$$

frequencies $\tilde{\omega}'_n, \tilde{\omega}''_n$ are (see also (3.18), (3.28))

$$\omega'_n = k\sqrt{-\frac{k_B T}{2m} \sigma_n \tan \frac{\sigma_n}{2}}, \quad \omega''_n = -k\sqrt{-\frac{k_B T}{2m} \sigma_n \cot \frac{\sigma_n}{2}} \qquad (3.50)$$

In asymptotic for large entire positive $n$ (singular point $r_A k = 1$ is considered further in this section)

$$\sigma_n = \left(n + \frac{1}{2}\right)\pi, \quad \tilde{\omega}'_n = \frac{\sqrt{\sigma_n}}{2} = \frac{1}{2}\sqrt{\pi\left(n+\frac{1}{2}\right)}, \quad \tilde{\omega}''_n = -\frac{\sqrt{\sigma_n}}{2} = -\frac{1}{2}\sqrt{\pi\left(n+\frac{1}{2}\right)}. \qquad (3.51)$$

The exact solution for the $n$ – th discrete solution from the spectrum of oscillations follows from (3.49), (3.50):

$$\tilde{\omega}_n = \frac{1}{2}\sqrt{-\text{Im}\left[W_n\left(\frac{2(1-r_A^2 k^2)^2}{\pi}\right)\right] \tan\left[\frac{1}{2}\text{Im}\left[W_n\left(\frac{2(1-r_A^2 k^2)^2}{\pi}\right)\right]\right]} - $$

$$-\frac{i}{2}\sqrt{-\text{Im}\left[W_n\left(\frac{2(1-r_A^2 k^2)^2}{\pi}\right)\right] \cot\left[\frac{1}{2}\text{Im}\left[W_n\left(\frac{2(1-r_A^2 k^2)^2}{\pi}\right)\right]\right]}. \qquad (3.52)$$

The square of the oscillation frequency of the longitudinal gravitational waves $\tilde{\omega}'^2_n$ is proportional to the wave energy. Hence, the energy of waves is quantized, and as $n$ grows one obtains the asymptotic expression analogous to quantum levels of quantum oscillator in one dimension

$$\tilde{\omega}'^2_n = \frac{\pi}{4}\left(n + \frac{1}{2}\right), \qquad (3.53)$$

the squares of possible dimensionless frequencies become equally spaced:

$$\tilde{\omega}'^2_{n+1} - \tilde{\omega}'^2_n = \frac{\pi}{4}. \qquad (3.54)$$

or

$$\omega'^2_{n+1} - \omega'^2_n = \frac{\pi}{2} k^2 \frac{k_B T}{m}. \tag{3.55}$$

This difference can be connected with energy of Newtonian graviton. Figures 1 and 2 reflect the result of calculations for 200 discrete levels for the case of the large Coulomb logarithm $\Lambda$. For high levels this spectrum contains many very close equidistant curves with partly practically straight lines, which human eyes can perceive as background. Moreover plotter from the technical point of view has no possibility to reflect the small curvature of lines approximating this curvature as a long step. My suggestion is to turn this shortcoming into merit in the explication of topology of high quantum levels in quantum systems.

Really, extremely interesting that this (from the first glance) grave shortcoming of plotters lead to the automatic construction of approximation for derivatives $d(r_D k)/d\hat\omega'$ and $d(r_D k)/d\hat\omega''$.

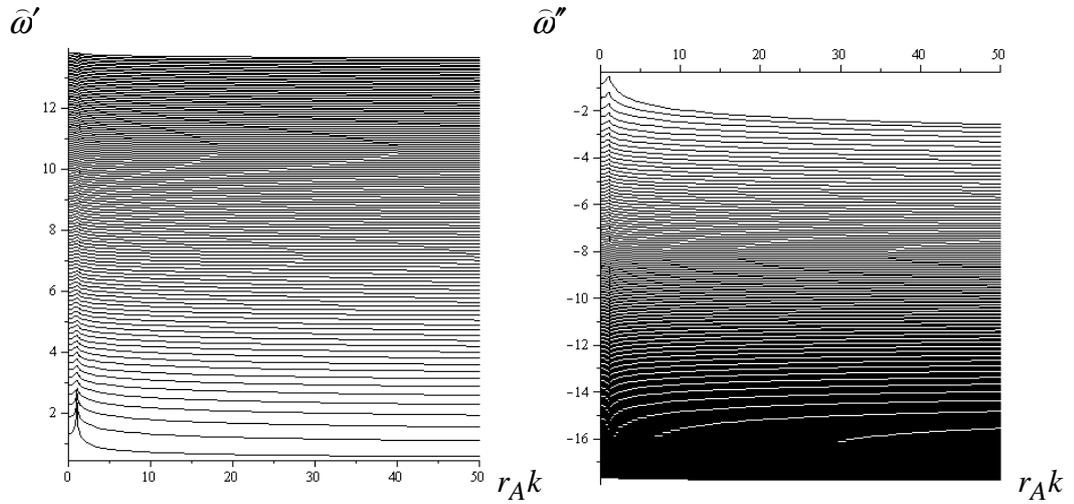

Fig. 1. The dimensionless frequency $\hat\omega'$ versus parameter $r_A k$, (left).
Fig. 2. The dimensionless frequency $\hat\omega''$ versus parameter $r_A k$, (right).

You can see this very complicated topology of curves in Figs. 1, 2 including the discrete spectrum of the bell-like curves in the mentioned figures. This singularity is connected with the existence of the generalized derivatives $d(r_A k)/d\hat\omega'$, $d(r_A k)/d\hat\omega''$ for discontinuous functions. This effect has no attitude to the mathematical programming and looks in the definite sense like effect of "shroud of Christ" – self-organization of visible information in the human conscience. Enlarging of scaling shows that the complicated curves topology exists also in the black domain. Then Figs. 1, 2 can be used for understanding of the future development of events in physical system after the initial linear stage.

For example Fig. 1 shows the discrete set of frequencies which vicinity corresponds to passing over from abnormal to normal dispersion (for example, by $\hat\omega' \sim 7$) for discrete systems of $r_A k$. Of course the non-linear stage needs the special investigation with using of another methods including the method of direct mathematical modeling  It seems that the curves of high levels have different topology in comparison with the low levels. Nevertheless, it is far from reality, the high-level frequencies have the same character features as low frequencies.

Look at Figs. 3, 4 and you see for frequencies $\hat\omega'_{200}$ and for $\hat\omega''_{200}$ the same character features as for lower frequencies.

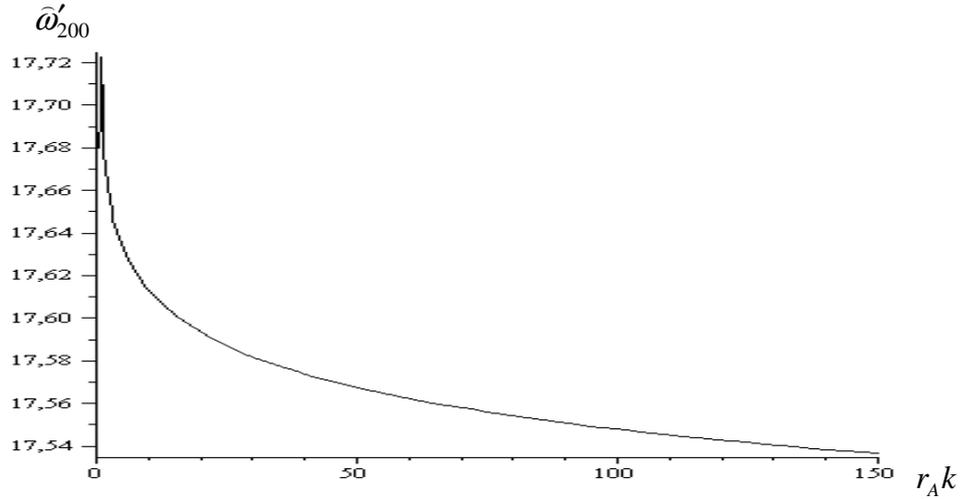

Fig. 3. The dimensionless frequency $\hat{\omega}'_{200}$ versus parameter $r_A k$.

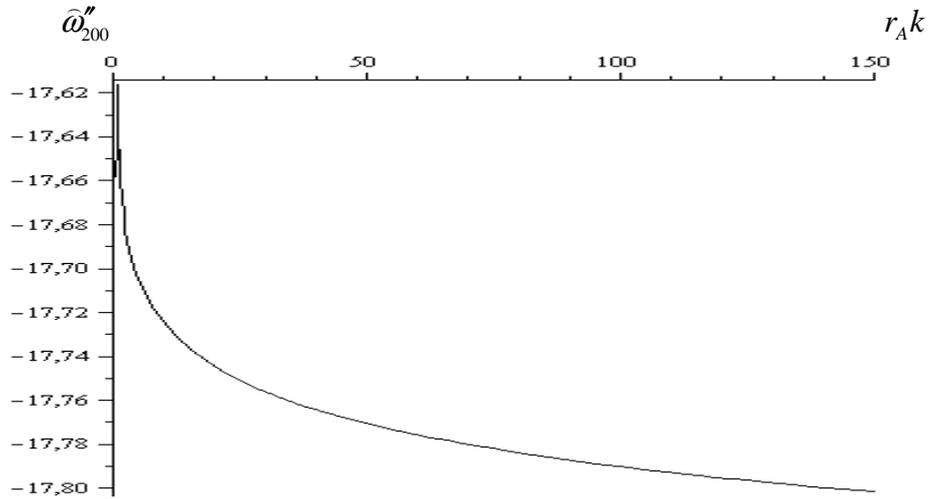

Fig. 4. The dimensionless frequency $\hat{\omega}''_{200}$ versus parameter $r_A k$.

It is of interest to investigate the singular point where

$$r_A k = 1. \tag{3.56}$$

Note the solution $\sigma \to \pi + 0$ and therefore $\hat{\omega}' \to \infty$ and $\hat{\omega}'' \to 0$. But phase velocity of wave $u_\phi = \omega' r_a$ and phase velocity of gravitational wave turns into infinity (in the frame of non-relativistic theory) and damping is equal to zero.

In vicinity of $r_A k = 1$ one obtains "gravitational window" with increasing of frequency $\hat{\omega}'_n$ and decreasing of damping; the corresponding wave lengths $\lambda_A$ is

$$\lambda_A = 2\pi r_A. \tag{3.57}$$

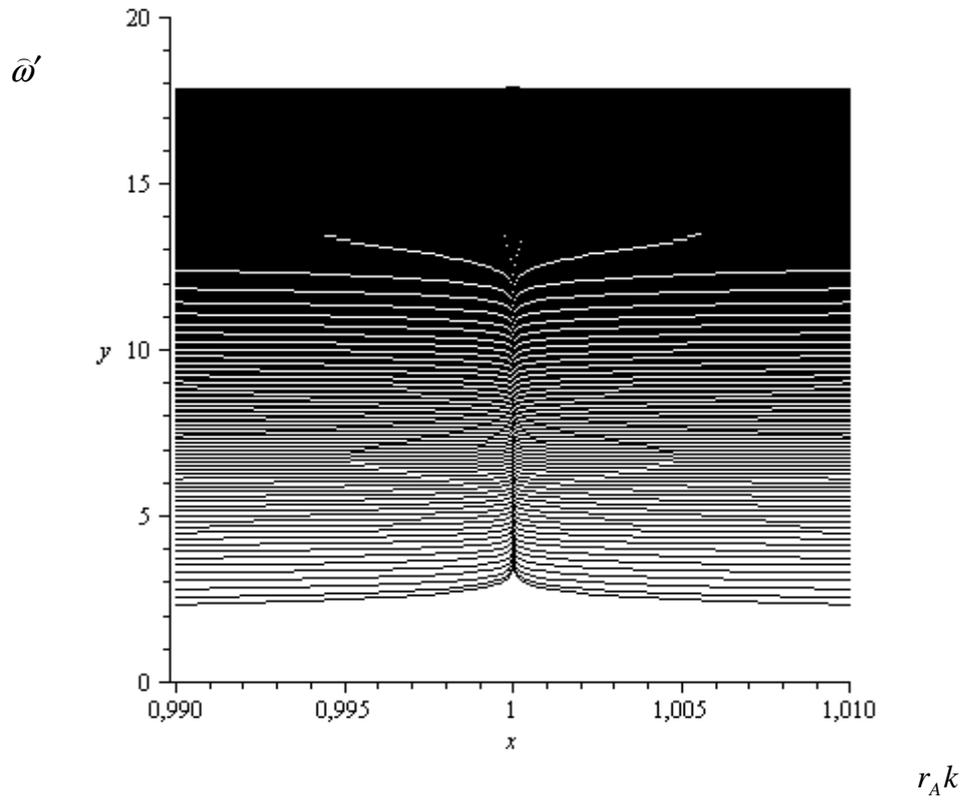

Fig. 5. Topology of the dispersion curves $\widetilde{\omega}'$ in the vicinity of the gravitational window.

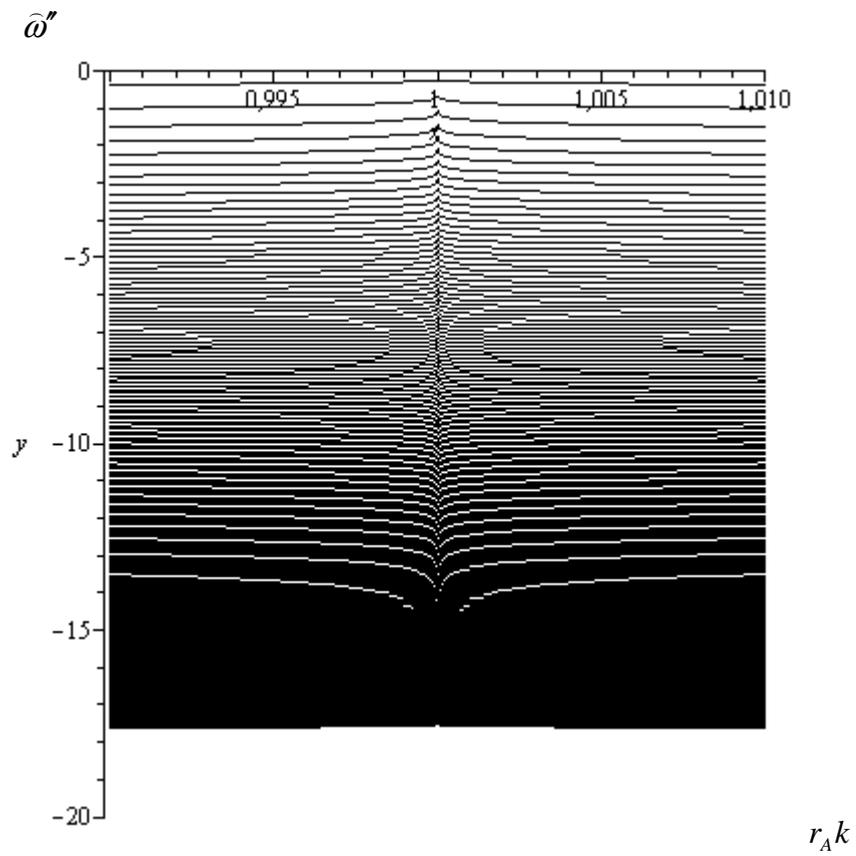

Fig. 6. Topology of the dispersion curves $\widetilde{\omega}''$ in the vicinity of the gravitational window.

Figs. 5, 6 reflect the topology of the dispersion curves in the vicinity of the gravitational window. Now we can create the physical picture leading to the Hubble flow:

*The main origin of Hubble effect (including the matter expansion with acceleration) is self – catching of expanding matter by the self – consistent Newtonian gravitational field in conditions of weak influence of central massive bodies.*

The relation (3.57) is the condition of this self – catching as result of explosion with the appearance of waves for which the wave lengths is of about $\lambda_A$. Gravitational self – catching takes place for Big Bang having given birth to the global expansion of Universe, but also for Little Bang [4] in so-called Local Group of galaxies. The gravitationally bound system of the Local Group can exist only within region $r < r_v$. In this case $r_v$ need not to be connected with the modified Newton force and can be considered as character value where gravitation of the central mass is not significant. Outside the Group at distances $r > r_v$ the Hubble flow of Galaxies starts. This "no reentering radius" was found as result of direct observations of the Local Group: $r_v$ is of order 1 *Mpc*.

All concrete calculations for the evolution of the Local Group and the local flow will be done in the frame of non-linear theory (see Section 5).

Some important remarks should be done:

1. Effects of gravitational self-catching should be typical for Universe. The existence of "Hubble boxes" discussed in review [4] as typical blocks of the nearby Universe.
2. As it follows from Figs. 1, 2, 5, 6 effect of gravitational self-catching exists *in finite region* close to $r_A k = 1$, the phase velocity is defined by discrete spectrum $u_{\phi,n} = \widehat{\omega}'_n \sqrt{2k_B T / m}$.
3. Gravitational self-catching can be significant in the Earth conditions.

The last remark needs to be explained. Gravitational self-catching can be essential in the physical systems which character lengths correlates with $r_A$ *in conditions of weak influence of central massive bodies.* The corresponding conditions are realizing in tsunami waves. For water by the earth conditions $T = 300$ K, $\rho = 1 \text{ g/cm}^3$; $\gamma_N = 6.6 \cdot 10^{-8} \text{cm}^3/(\text{g} \cdot \text{s}^2)$, the lengths $r_A = 407.43$ km and $\lambda_A = 2558.66$ km. For close collisions $r_c \sim 10^{-7}$ cm and Coulomb logarithm $\Lambda_a = \ln \frac{r_A}{r_c} \sim 10^2$.

The delivered theory can be applied in the Earth conditions if the influence of central mass can be excluded from consideration. This situation realizes in the tsunami motion because the direction to the Earth center supposes perpendicular to the direction of additional self-consistent gravitational force acting in the tsunami wave. In essence, the catching of water mass is realizing by longitudinal self-consistent gravitational wave. I don't intend here to go into details, but the origin of effects of the small attenuation can be qualitatively explained from position of kinetic theory. Let us calculate the mean velocity $\bar{u}_+$ of particles moving in a chosen direction. If this direction is considering as the positive ones, then $u > 0$ and for the Maxwellian function $f_0$

$$\bar{u}_+ = \sqrt{\frac{m}{2\pi k_B T}} \int_0^{+\infty} e^{-mu^2/2k_B T} u\, du = \sqrt{\frac{k_B T}{2\pi m}}. \qquad (3.58)$$

Kinetic energy, connected with this motion is

$$m\bar{u}_+^2 / 2 = k_B T /(4\pi). \qquad (3.59)$$

From relations (3.27), (3.59) follow

$$\bar{u}_+ = r_A \sqrt{2\rho \gamma_N} \qquad (3.60)$$

Therefore, if the selected direction is opposite to the direction of the wave motion, energy of gravitational field $E_a = \gamma_N m \rho r_a^2$ (per particle) should be applied for exclusion of such kind of particles. For water in considered estimation one obtains $E_a = \gamma_N m \rho r_A^2 = 3.293 \cdot 10^{-15}$ erg, $\bar{u}_+ = r_A \sqrt{2\rho\gamma_N} = 533$ km/hour - the typical value of tsunami in ocean. Otherwise, the wave expansion leads to the energy dissipation of the directional motion in the form of the chaotic heat motion. But in the case if the forces of gravitation attraction counteract (or keep to a minimum) these losses, the wave is moving without attenuation.

**4. Disk galaxy rotation curves and the problem of dark matter.**

About forty years after Zwicky's initial observations, in the late 1960s and early 1970s, Vera Rubin, astronomer at the Department of Terrestrial Magnetism at the Carnegie Institution of Washington presented findings based on a new sensitive spectrograph that could measure the velocity curve of edge-on spiral galaxies to a greater degree of accuracy than had ever before been achieved. Together with Kent Ford, Rubin announced at a 1975 meeting of the American Astronomical Society the astonishing discovery that most stars in spiral galaxies orbit at roughly the same speed reflected schematically on Fig. 7. The following extensive radio observations determined the detailed rotation curve of spiral disk galaxies to be flat (as the curve B), much beyond as seen in the optical band. Obviously the trivial balance between the gravitational and centrifugal forces leads to relation between orbital speed $V$ and galactocentric distance $r$ as $V^2 = \gamma_N M / r$ beyond the physical extent of the galaxy of mass $M$ (the curve A). The obvious contradiction with the velocity curve B having a 'flat' appearance out to a large radius, was explained by introduction of a new physical essence – dark matter because for spherically symmetric case the hypothetical density distribution $\rho(r) \sim 1/r^2$ leads to $V = const$. The result of this activity is known – undetectable dark matter which does not emit radiation, inferred solely from its gravitational effects. But it means that upwards of 50% of the mass of galaxies was contained in the dark galactic halo.

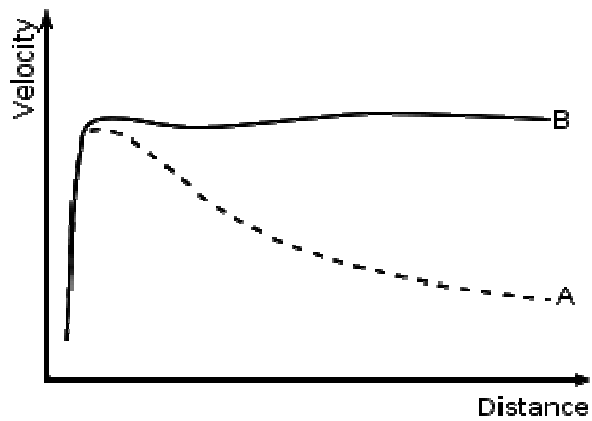

Fig. 7. Rotation curve of a typical spiral galaxy: predicted (**A**) and observed (**B**).

In the following I intend to show that the character features reflected on Fig. 7 can be explained in the frame of Newtonian gravitation law and the non-local kinetic description created by me. With this aim let us consider the formation of the soliton's type of solution of the generalized hydrodynamic equations for gravitational media like galaxy in the self consistent gravitational field. Strict consideration leads to the following system of the generalized hydrodynamic equations (GHE) [13 - 17] written in the generalized Euler form taking into account the force of gravitation:

(Poisson equation)
$$\Delta\Psi = 4\pi\gamma_N \left[ \rho - \tau\left(\frac{\partial\rho}{\partial t} + \frac{\partial}{\partial \mathbf{r}}\cdot(\rho\mathbf{v_0})\right)\right], \tag{4.1}$$

(continuity equation)
$$\frac{\partial}{\partial t}\left\{\rho - \tau\left[\frac{\partial\rho}{\partial t} + \frac{\partial}{\partial \mathbf{r}}\cdot(\rho\mathbf{v_0})\right]\right\} +$$
$$+ \frac{\partial}{\partial \mathbf{r}}\cdot\left\{\rho\mathbf{v_0} - \tau\left[\frac{\partial}{\partial t}(\rho\mathbf{v_0}) + \frac{\partial}{\partial \mathbf{r}}\cdot\rho\mathbf{v_0}\mathbf{v_0} + \vec{\mathbf{I}}\cdot\frac{\partial p}{\partial \mathbf{r}} - \rho\mathbf{g}\right]\right\} = 0, \tag{4.2}$$

(momentum equation)
$$\frac{\partial}{\partial t}\left\{\rho\mathbf{v_0} - \tau\left[\frac{\partial}{\partial t}(\rho\mathbf{v_0}) + \frac{\partial}{\partial \mathbf{r}}\cdot\rho\mathbf{v_0}\mathbf{v_0} + \frac{\partial p}{\partial \mathbf{r}} - \rho\mathbf{g}\right]\right\} -$$
$$- \mathbf{g}\left[\rho - \tau\left(\frac{\partial\rho}{\partial t} + \frac{\partial}{\partial \mathbf{r}}\cdot(\rho\mathbf{v_0})\right)\right] + \frac{\partial}{\partial \mathbf{r}}\cdot\left\{\rho\mathbf{v_0}\mathbf{v_0} + \right.$$
$$+ p\vec{\mathbf{I}} - \tau\left[\frac{\partial}{\partial t}(\rho\mathbf{v_0}\mathbf{v_0} + p\vec{\mathbf{I}}) + \frac{\partial}{\partial \mathbf{r}}\cdot\rho(\mathbf{v_0}\mathbf{v_0})\mathbf{v_0} + \right.$$
$$\left.\left. + 2\vec{\mathbf{I}}\left[\frac{\partial}{\partial \mathbf{r}}\cdot(p\mathbf{v_0})\right] + \frac{\partial}{\partial \mathbf{r}}\cdot(\vec{\mathbf{I}}p\mathbf{v_0}) - \mathbf{g}\rho\mathbf{v_0} - \mathbf{v_0}\mathbf{g}\rho\right]\right\} = 0, \tag{4.3}$$

(energy equation)
$$\frac{\partial}{\partial t}\left\{\frac{\rho v_0^2}{2} + \frac{3}{2}p - \tau\left[\frac{\partial}{\partial t}\left(\frac{\rho v_0^2}{2} + \frac{3}{2}p\right) + \right.\right.$$
$$\left.\left. + \frac{\partial}{\partial \mathbf{r}}\cdot\left(\frac{1}{2}\rho v_0^2\mathbf{v_0} + \frac{5}{2}p\mathbf{v_0}\right) - \mathbf{g}\cdot\rho\mathbf{v_0}\right]\right\} +$$
$$+ \frac{\partial}{\partial \mathbf{r}}\cdot\left\{\frac{1}{2}\rho v_0^2\mathbf{v_0} + \frac{5}{2}p\mathbf{v_0} - \tau\left[\frac{\partial}{\partial t}\left(\frac{1}{2}\rho v_0^2\mathbf{v_0} + \right.\right.\right.$$
$$\left. + \frac{5}{2}p\mathbf{v_0}\right) + \frac{\partial}{\partial \mathbf{r}}\cdot\left(\frac{1}{2}\rho v_0^2\mathbf{v_0}\mathbf{v_0} + \frac{7}{2}p\mathbf{v_0}\mathbf{v_0} + \frac{1}{2}pv_0^2\vec{\mathbf{I}} + \right.$$
$$\left.\left.\left. + \frac{5}{2}\frac{p^2}{\rho}\vec{\mathbf{I}}\right) - \rho\mathbf{g}\cdot\mathbf{v_0}\mathbf{v_0} - p\mathbf{g}\cdot\vec{\mathbf{I}} - \frac{1}{2}\rho v_0^2\mathbf{g} - \frac{3}{2}\mathbf{g}p\right]\right\} -$$
$$- \left\{\rho\mathbf{g}\cdot\mathbf{v_0} - \tau\left[\mathbf{g}\cdot\left(\frac{\partial}{\partial t}(\rho\mathbf{v_0}) + \frac{\partial}{\partial \mathbf{r}}\cdot\rho\mathbf{v_0}\mathbf{v_0} + \right.\right.\right.$$
$$\left.\left.\left. + \frac{\partial}{\partial \mathbf{r}}\cdot p\vec{\mathbf{I}} - \rho\mathbf{g}\right)\right]\right\} = 0, \tag{4.4}$$

where $\mathbf{g} = -\frac{\partial\Psi}{\partial \mathbf{r}}$ is acceleration in gravitational field, $\Delta$ is Laplacian, $\gamma_N$ is Newtonian gravitation constant.

Our aim consists in calculation of the self-consistent hydrodynamic moments of possible formation like gravitational soliton. In the first approximation for spiral galaxies the problem can be considered in the frame of the non-stationary 1D formulation. Then the system of GHE consist from the generalized Poisson equation (reflecting the effects of the density and the density flux perturbations), continuity equation, motion and energy equations. This system of four equations for non-stationary 1D case is written in the form (see also (3.13)):

(Poisson equation)
$$\frac{\partial^2 \Psi}{\partial x^2} = 4\pi\gamma_N \left[ \rho - \tau\left(\frac{\partial \rho}{\partial t} + \frac{\partial}{\partial x}(\rho u)\right)\right], \tag{4.5}$$

(continuity equation)
$$\frac{\partial}{\partial t}\left\{\rho - \tau\left[\frac{\partial \rho}{\partial t} + \frac{\partial}{\partial x}(\rho v_0)\right]\right\} + \frac{\partial}{\partial x}\left\{\rho v_0 - \tau\left[\frac{\partial}{\partial t}(\rho v_0) + \right.\right.$$
$$\left.\left. + \frac{\partial}{\partial x}(\rho v_0^2) + \frac{\partial p}{\partial x} + \rho\frac{\partial \Psi}{\partial x}\right]\right\} = 0, \tag{4.6}$$

(motion equation)
$$\frac{\partial}{\partial t}\left\{\rho u - \tau\left[\frac{\partial}{\partial t}(\rho u) + \frac{\partial}{\partial x}(\rho u^2) + \frac{\partial p}{\partial x} + \rho\frac{\partial \Psi}{\partial x}\right]\right\} + \frac{\partial \Psi}{\partial x}\left[\rho - \tau\left(\frac{\partial \rho}{\partial t} + \frac{\partial}{\partial x}(\rho u)\right)\right] +$$
$$+ \frac{\partial}{\partial x}\left\{\rho u^2 + p - \tau\left[\frac{\partial}{\partial t}(\rho u^2 + p) + \frac{\partial}{\partial x}(\rho u^3 + 3pu) + 2\rho u\frac{\partial \Psi}{\partial x}\right]\right\} = 0, \tag{4.7}$$

(energy equation)
$$\frac{\partial}{\partial t}\left\{\rho u^2 + 3p - \tau\left[\frac{\partial}{\partial t}(\rho u^2 + 3p) + \frac{\partial}{\partial x}(\rho u^3 + 5pu) + 2\rho u\frac{\partial \Psi}{\partial x}\right]\right\} +$$
$$+ \frac{\partial}{\partial x}\left\{\rho u^3 + 5pu - \tau\left[\frac{\partial}{\partial t}(\rho u^3 + 5pu) + \frac{\partial}{\partial x}\left(\rho u^4 + \right.\right.\right.$$
$$\left.\left.\left. + 8pu^2 + 5\frac{p^2}{\rho}\right) + \frac{\partial \Psi}{\partial x}(3\rho u^2 + 5p)\right]\right\} +$$
$$+ 2\frac{\partial \Psi}{\partial x}\left\{\rho u - \tau\left[\frac{\partial}{\partial t}(\rho u) + \frac{\partial}{\partial x}(\rho u^2 + p) + \rho\frac{\partial \Psi}{\partial x}\right]\right\} = 0, \tag{4.8}$$

where $u$ is translational velocity of the one species object, $\Psi$ - self consistent gravitational potential, $\rho$ is density and $p$ is pressure, $\tau$ is non-locality parameter.

Let us introduce the coordinate system moving along the positive direction of $x$- axis in ID space with velocity $C = u_0$ equal to phase velocity of considering object
$$\xi = x - Ct. \tag{4.9}$$
Taking into account the De Broglie relation we should wait that the group velocity $u_g$ is equal $2u_0$. In moving coordinate system all dependent hydrodynamic values are function of $(\xi, t)$. We investigate the possibility of the object formation of the soliton type. For this solution there is no explicit dependence on time for coordinate system moving with the phase velocity $u_0$. Write down the system of equations (4.5) - (4.8) in the dimensionless form, where dimensionless symbols are marked by tildes. For the scales $\rho_0$, $u_0, x_0 = u_0 t_0, \Psi_0 = u_0^2, \gamma_{N0} = \frac{u_0^2}{\rho_0 x_0^2}$ $p_0 = \rho_0 u_0^2$
and conditions $\tilde{C} = C/u_0 = 1$, the equations take the form:

(generalized Poisson equation)
$$\frac{\partial^2 \tilde{\Psi}}{\partial \tilde{\xi}^2} = 4\pi \tilde{\gamma}_N \left[ \tilde{\rho} - \tilde{\tau} \left( -\frac{\partial \tilde{\rho}}{\partial \tilde{\xi}} + \frac{\partial}{\partial \tilde{\xi}} (\tilde{\rho}\tilde{u}) \right) \right], \tag{4.10}$$

(continuity equation)
$$\frac{\partial \tilde{\rho}}{\partial \tilde{\xi}} - \frac{\partial \tilde{\rho}\tilde{u}}{\partial \tilde{\xi}} + \frac{\partial}{\partial \tilde{\xi}} \left\{ \tilde{\tau} \left[ \frac{\partial}{\partial \tilde{\xi}} \left[ \tilde{p} + \tilde{\rho}\tilde{u}^2 + \tilde{\rho} - 2\tilde{\rho}\tilde{u} \right] + \tilde{\rho} \frac{\partial \tilde{\Psi}}{\partial \tilde{\xi}} \right] \right\} = 0, \tag{4.11}$$

(motion equation)
$$\frac{\partial}{\partial \tilde{\xi}} \left( \tilde{\rho}\tilde{u}^2 + \tilde{p} - \tilde{\rho}\tilde{u} \right) + \frac{\partial}{\partial \tilde{\xi}} \left\{ \tilde{\tau} \left[ \frac{\partial}{\partial \tilde{\xi}} \left( 2\tilde{\rho}\tilde{u}^2 - \tilde{\rho}\tilde{u} + 2\tilde{p} - \tilde{\rho}\tilde{u}^3 - 3\tilde{p}\tilde{u} \right) + \tilde{\rho} \frac{\partial \tilde{\Psi}}{\partial \tilde{\xi}} \right] \right\} +$$
$$+ \frac{\partial \tilde{\Psi}}{\partial \tilde{\xi}} \left\{ \tilde{\rho} - \tilde{\tau} \left[ -\frac{\partial \tilde{\rho}}{\partial \tilde{\xi}} + \frac{\partial}{\partial \tilde{\xi}} (\tilde{\rho}\tilde{u}) \right] \right\} - 2 \frac{\partial}{\partial \tilde{\xi}} \left\{ \tilde{\tau} \tilde{\rho} \tilde{u} \frac{\partial \tilde{\Psi}}{\partial \tilde{\xi}} \right\} = 0, \tag{4.12}$$

(energy equation)
$$\frac{\partial}{\partial \tilde{\xi}} \left( \tilde{\rho}\tilde{u}^2 + 3\tilde{p} - \tilde{\rho}\tilde{u}^3 - 5\tilde{p}\tilde{u} \right) -$$
$$- \frac{\partial}{\partial \tilde{\xi}} \left\{ \tilde{\tau} \frac{\partial}{\partial \tilde{\xi}} \left( 2\tilde{\rho}\tilde{u}^3 + 10\tilde{p}\tilde{u} - \tilde{\rho}\tilde{u}^2 - 3\tilde{p} - \tilde{\rho}\tilde{u}^4 - 8\tilde{p}\tilde{u}^2 - 5\frac{\tilde{p}^2}{\tilde{\rho}} \right) \right\} +$$
$$+ \frac{\partial}{\partial \tilde{\xi}} \left\{ \tilde{\tau} (3\tilde{\rho}\tilde{u}^2 + 5\tilde{p}) \frac{\partial \tilde{\Psi}}{\partial \tilde{\xi}} \right\} - 2\tilde{\rho}\tilde{u} \frac{\partial \tilde{\Psi}}{\partial \tilde{\xi}} - 2 \frac{\partial}{\partial \tilde{\xi}} \left\{ \tilde{\tau}\tilde{\rho}\tilde{u} \frac{\partial \tilde{\Psi}}{\partial \tilde{\xi}} \right\} +$$
$$+ 2\tilde{\tau} \frac{\partial \tilde{\Psi}}{\partial \tilde{\xi}} \left[ -\frac{\partial}{\partial \tilde{\xi}} (\tilde{\rho}\tilde{u}) + \frac{\partial}{\partial \tilde{\xi}} (\tilde{\rho}\tilde{u}^2 + \tilde{p}) + \tilde{\rho} \frac{\partial \tilde{\Psi}}{\partial \tilde{\xi}} \right] = 0, \tag{4.13}$$

Some comments to the system of four ordinary non-linear equations (4.10) – (4.13):
1. Every equation from the system is of the second order and needs two conditions. The problem belongs to the class of Cauchy problems.
2. In comparison for example, with the Schrödinger theory connected with behavior of the wave function, no special conditions are applied for dependent variables including the domain of the solution existing. This domain is defined automatically in the process of the numerical solution of the concrete variant of calculations.
3. From the introduced scales

$$\rho_0, \ u_0, \ x_0 = u_0 t_0, \ \Psi_0 = u_0^2, \ \gamma_{N0} = \frac{u_0^2}{\rho_0 x_0^2}, \ p_0 = \rho_0 u_0^2, \tag{4.14}$$

only three parameters are independent, namely, $\rho_0, \ u_0, x_0$.

4. Approximation for the dimensionless non-local parameter $\tilde{\tau}$ should be introduced (see (2.10, (2.11)). In the definite sense it is not the problem of the hydrodynamic level of the physical system description (like the calculation of the kinetic coefficients in the classical hydrodynamics). Interesting to notice that quantum GHE were applied with success for calculation of atom structure [21, 22], which is considered as two species charged $e, i$ mixture. The corresponding approximations for the non-local parameters $\tau_i$, $\tau_e$ and $\tau_{ei}$ are proposed in [21, 22]. In the theory of the atom structure [21], [22] after taking into account the Balmer's relation, (2.11) transforms into

$$\tau_e = n\hbar \big/ m_e u^2, \tag{4.15}$$

where $n = 1, 2, ...$ is principal quantum number. As result the length scale relation was written as $x_0 = \dfrac{H}{m_e u_0} = \dfrac{n\hbar}{m_e u_0}$. But the value $v^{qu} = \hbar / m_e$ has the dimension [$cm^2/s$] and can be named as quantum viscosity, $v^{qu} = 1.1577 \ cm^2/s$. Then

$$\tau_e = n v^{qu} \big/ u^2. \tag{4.16}$$

Introduce now the quantum Reynolds number

$$\text{Re}^{qu} = \frac{u_0 x_0}{v^{qu}}. \tag{4.17}$$

As result from (4.16), (4.17) follows the condition of quantization for $\text{Re}^{qu}$. Namely

$$\text{Re}^{qu} = n, \quad n = 1, 2, ... \tag{4.18}$$

5. Taking into account the previous considerations I introduce the following approximation for the dimensionless non-local parameter

$$\tilde{\tau} = \frac{1}{\tilde{u}^2}, \tag{4.19}$$

or

$$\tau = \frac{1}{u^2} u_0 x_0 = \frac{v_0^k}{u^2}, \tag{4.20}$$

where the scale for the kinematical viscosity is introduced $v_0^k = u_0 x_0$, (compare with (4.15)).

Then we have the physically transparent result – non-local parameter is proportional to the kinematical viscosity and in inverse proportion to the square of velocity.

Now we are ready to display the results of the mathematical modeling realized with the help of Maple (the versions Maple 9 or more can be used). The system of generalized hydrodynamic equations (4.10) – (4.13) have the great possibilities of mathematical modeling as result of changing of eight Cauchy conditions describing the character features of initial perturbations which lead to the soliton formation.

The following Maple notations on figures are used: r- density $\tilde{\rho}$, u- velocity $\tilde{u}$ (solid line), p - pressure $\tilde{p}$ (dashed line) and v - self consistent potential $\tilde{\Psi}$. Explanations placed under all following figures, Maple program contains Maple's notations – for example the expression $D(u)(0) = 0$ means in the usual notations $\dfrac{\partial \tilde{u}}{\partial \tilde{\xi}}(0) = 0$, independent variable $t$ responds to $\tilde{\xi}$.

We begin with investigation of the problem of principle significance – is it possible after a perturbation (defined by Cauchy conditions) to obtain the gravitational object of the soliton's kind as result of the self-organization of the matter? With this aim let us consider the initial perturbations (SYSTEM 1)

```
u(0)=1,p(0)=1,r(0)=1,D(u)(0)=0,D(p)(0)=0,D(r)(0)=0,D(v)(0)=0,
v(0)=1
```

The following figures reflect the result of solution of Eqs. (4.10) – (4.13) with the choice of scales leading to $\tilde{\gamma}_N = 1$. Figs. 8 – 11 correspond to the approximation of the non-local parameter $\tilde{\tau}$ in the form (4.19).

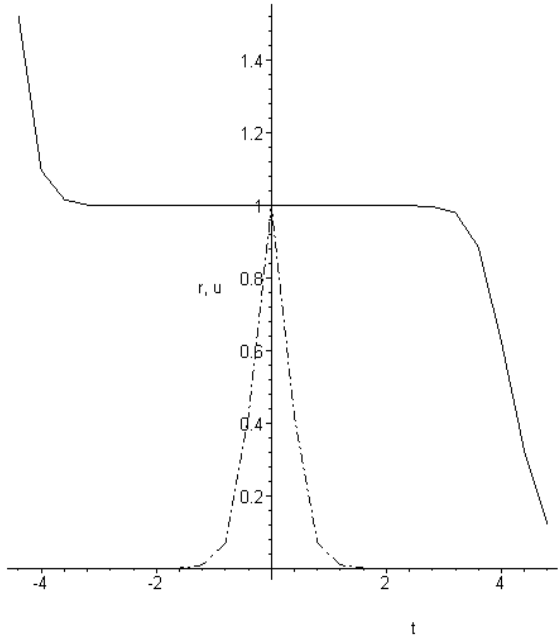 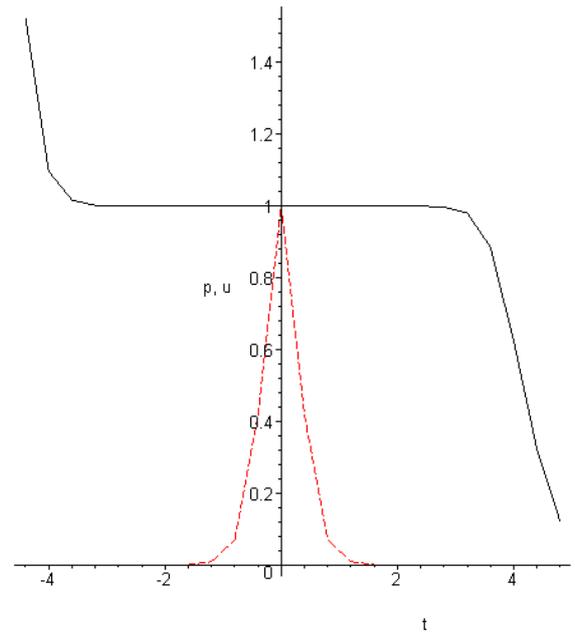

Fig. 8. r - density $\tilde{\rho}$ (dash dotted line), u - velocity $\tilde{u}$ in gravitational soliton.

Fig. 9. p - pressure $\tilde{p}$, u- velocity $\tilde{u}$ in gravitational soliton.

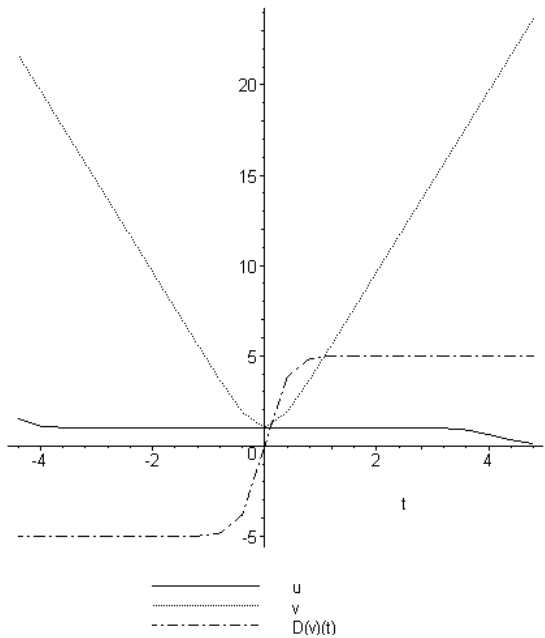 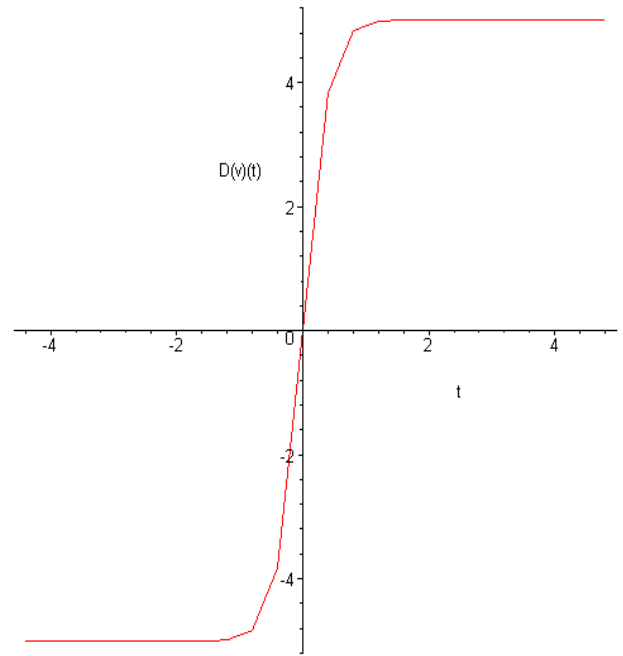

Fig. 10. u- velocity $\tilde{u}$, v - self consistent potential $\tilde{\Psi}$, $D(v)(t) = \dfrac{\partial \tilde{\Psi}}{\partial \tilde{\xi}}$ in quantum soliton, (left).

Fig. 11. $D(v)(t) = \dfrac{\partial \tilde{\Psi}}{\partial \tilde{\xi}}$ in quantum soliton, (right).

Fig. 8 displays the gravitational object placed in bounded region of 1D space, all parts of this object are moving with the same velocity. Important to underline that no special boundary conditions were used for this and all following cases. Then this soliton is product of the self-organization of gravitational matter. Figs. 9, 10 contain the answer for formulated above question about stability of the object. The derivative (see Fig. 11)

$$\frac{\partial \tilde{\Psi}}{\partial \tilde{\xi}} = \frac{\partial \Psi}{\partial \xi}\frac{x_0}{u_0^2} = -g(\xi)\Big/\left(u_0^2/x_0\right) = -\tilde{g}(\xi)$$ is proportional to the self-consistent gravitational force acting on the soliton and in its vicinity. Therefore the stability of the object is result of the self-consistent influence of the gravitational potential and pressure.

Extremely important that the self-consistent gravitational force has the character of the flat area which exists in the vicinity of the object. This solution exists only in the restricted area of space; the corresponding character length is defined automatically as result of the numerical solution of the problem.

The non-local parameter $\tilde{\tau}$, in the definite sense plays the role analogous to kinetic coefficients in the usual Boltzmann kinetic theory. The influence on the results of calculations is not too significant. The same situation exists in the generalized hydrodynamics.

Really, let us use the another approximation for $\tilde{\tau}$ in the simplest possible form, namely

$$\tilde{\tau} = 1. \qquad (4.21)$$

The following figures 12 – 15 reflect the results of solution of Eqs. (4.10) – (4.13) with the choice of scales leading to $\tilde{\gamma}_N = 1$, but with the approximation of the non-local parameter $\tilde{\tau}$ in the form (4.21).

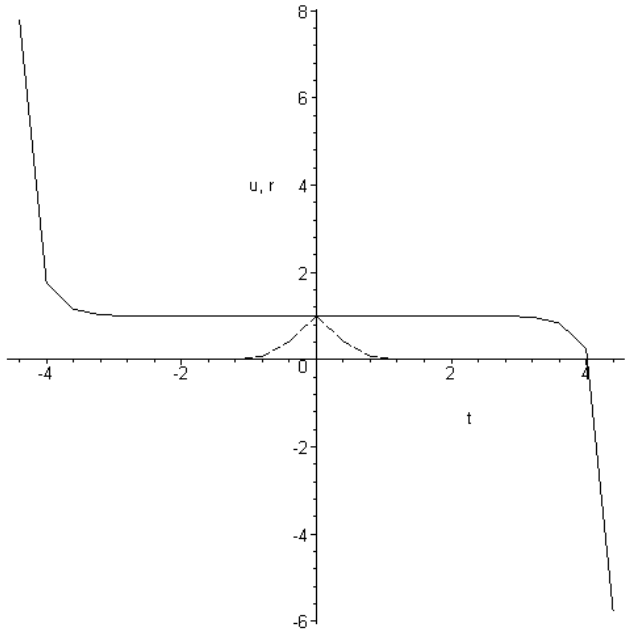 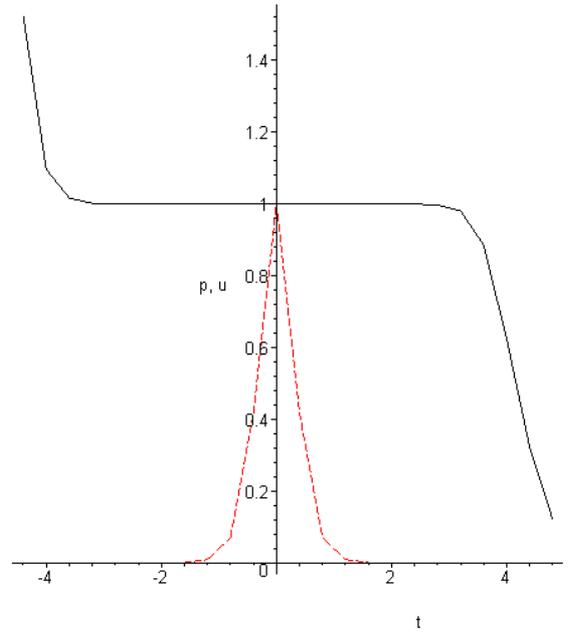

Fig. 12. r - density $\tilde{\rho}$ (dashed line), u - velocity $\tilde{u}$ in gravitational soliton.

Fig. 13. p - pressure $\tilde{p}$ (dashed line), u - velocity $\tilde{u}$ in gravitational soliton.

As it is follows from Figs. 8 – 15 in the vicinity of the central massive galaxy kernel – gravitational soliton, exists the domain with the constant gravitational force acting on the unit of mass. As result rotation curve of a typical spiral galaxy follows the curve (B) instead of (A) on Fig. 7. These peculiar features of the halo movement can be explained without new concepts like "dark matter".

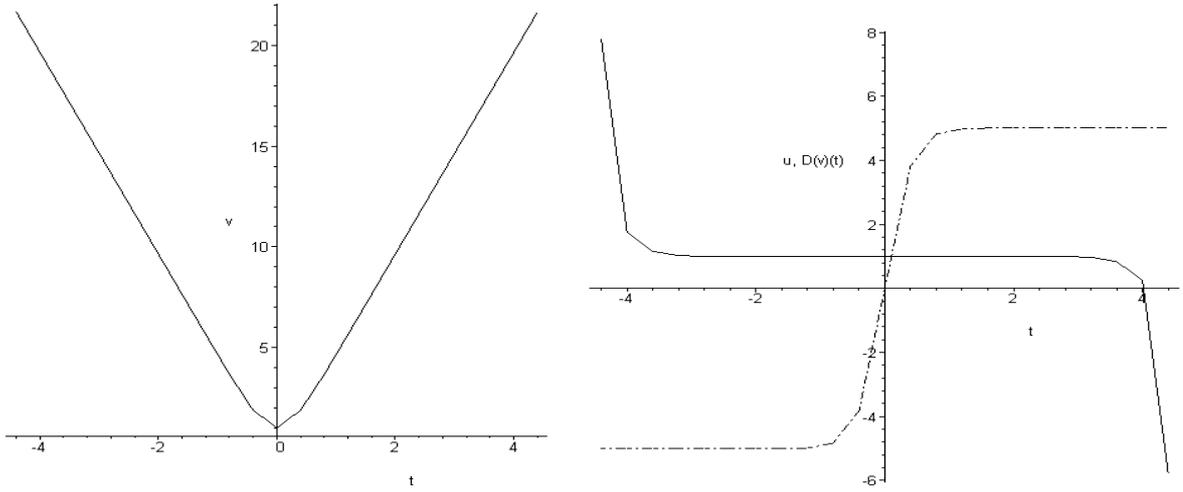

Fig. 14. v – self consistent potential $\tilde{\Psi}$ in gravitational soliton, (left).

Fig. 15. u- velocity $\tilde{u}$ (solid line), $D(v)(t) = \dfrac{\partial \tilde{\Psi}}{\partial \tilde{\xi}}$ in quantum soliton, (right).

Spiral galaxies have rather complicated geometrical forms and 3D calculations can be used. Appendix contains the full system of 3D non-local hydrodynamic equations in moving (along x axis) Cartesian coordinate system. But reasonable to suppose that influence of halo on galaxy kernel is not too significant and use for calculations the spherical coordinate system. Appendix contains also the corresponding expression for derivatives in the spherical coordinate system we need. The 1D calculations in the Cartesian coordinate system correspond to calculations in the spherical coordinate system by the large radii of curvature, but have also the independent significance in the another character scales. Namely for explanations of the meteorological front motion (without taking into account the Earth rotation). In this theory cyclone or anticyclone correspond to moving solitons. By the way Figs. 11, 15 reflect correctly the wind orientation in front and behind the soliton.

The following figures reflect the result of soliton calculations for the case of spherical symmetry for galaxy kernel. The velocity $\tilde{u}$ corresponds to the direction of the soliton movement for the spherical coordinate system on the following figures. Self-consistent gravitational force $F$ acting on the unit of mass permits to define the orbital velocity $w$ of objects in halo, $w = \sqrt{Fr}$, or

$$\tilde{w} = \sqrt{\tilde{r} \, \partial \tilde{\Psi} / \partial \tilde{r}}, \qquad (4.22)$$

where $r$ is the distance from the center of galaxy. All calculations are realized for the conditions (SYSTEM I) but for different parameter

$$G = \tilde{\gamma}_N = \gamma_N / \gamma_{N0} = \gamma_N \rho_0 x_0^2 / u_0^2. \qquad (4.23)$$

Parameter $G$ plays the role of the similarity criteria in traditional hydrodynamics.

Important conclusions:
1. The following Figures 16 – 26 demonstrate the evolution of the rotation curves from the Kepler regime (curve **A**, small G) to observed (**B**, large G) for typical spiral galaxies.
2. The stars with planets (like Sun) correspond to gravitational soliton with small G and therefore originate the Kepler rotation regime.
3. Regime **B** cannot be obtained in the frame of local statistical physics on principal and authors of many papers introduce different approximations for additional "dark matter density" (as usual in Poisson equation) trying to find coincidence with data of observations.
4. From the wrong position of local theories Poisson equation (4.1) contains "dark matter density", continuity equation (4.2) contains the "flux of dark matter density", motion equation (4.3) includes "dark energy", energy equation (4.3) has "the flux of dark energy" and so on to the "senior dark velocity moments". This entire situation is similar to the turbulent theories based on local statistical physics and empirical corrections for velocity moments.

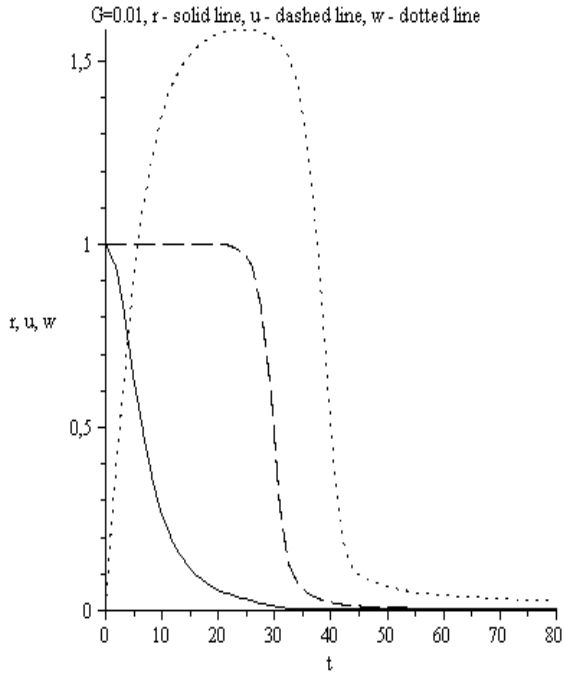
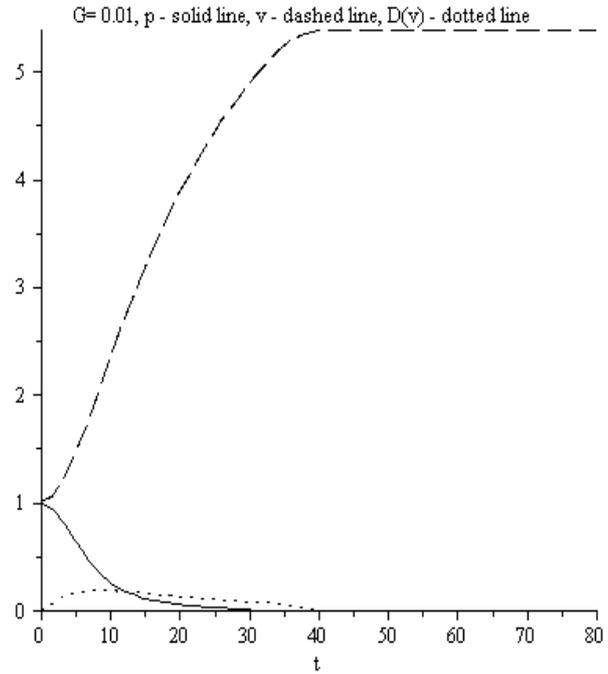

Fig. 16. r - density $\tilde{\rho}$, u - velocity $\tilde{u}$ in gravitational soliton, w - orbital velocity $\tilde{w}$. $G = 0.01$.

Fig. 17. p - pressure $\tilde{p}$, v - self consistent potential $\tilde{\Psi}$, $D(v)(t) = \partial \tilde{\Psi} / \partial \tilde{\xi}$ in gravitational soliton.

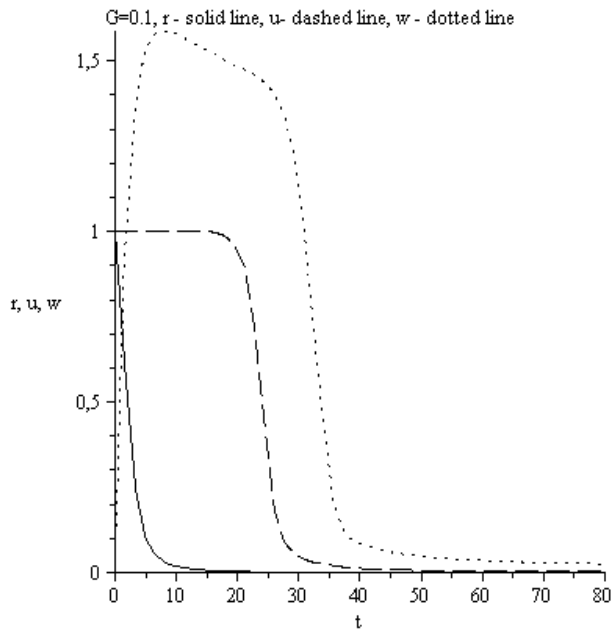
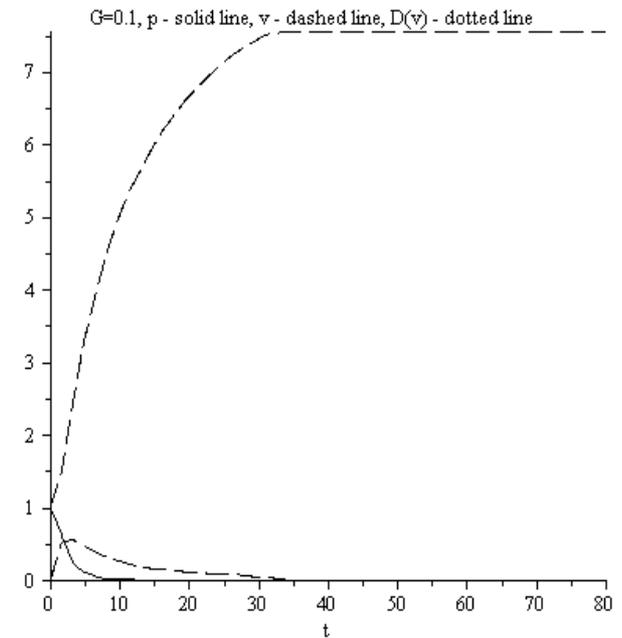

Fig. 18. r - density $\tilde{\rho}$, u - velocity $\tilde{u}$ in gravitational soliton, w - orbital velocity $\tilde{w}$. $G = 0.1$.

Fig. 19. p - pressure $\tilde{p}$, v - self consistent potential $\tilde{\Psi}$, $D(v)(t) = \dfrac{\partial \tilde{\Psi}}{\partial \tilde{\xi}}$ in gravitational soliton.

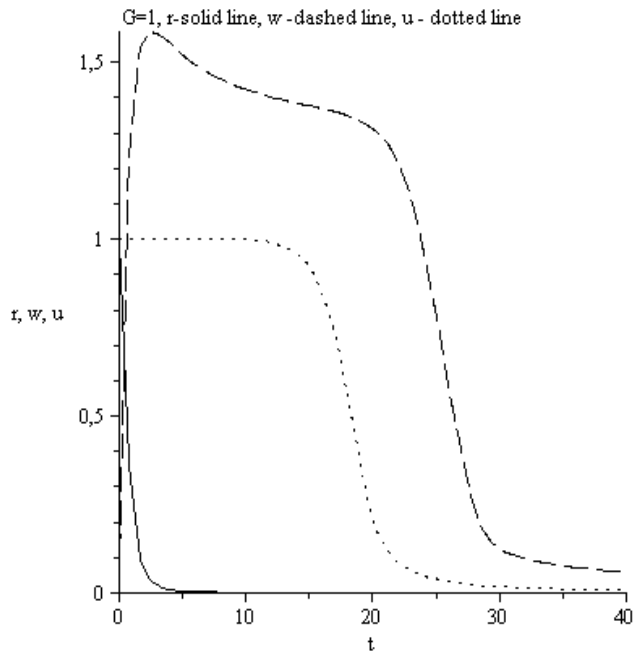 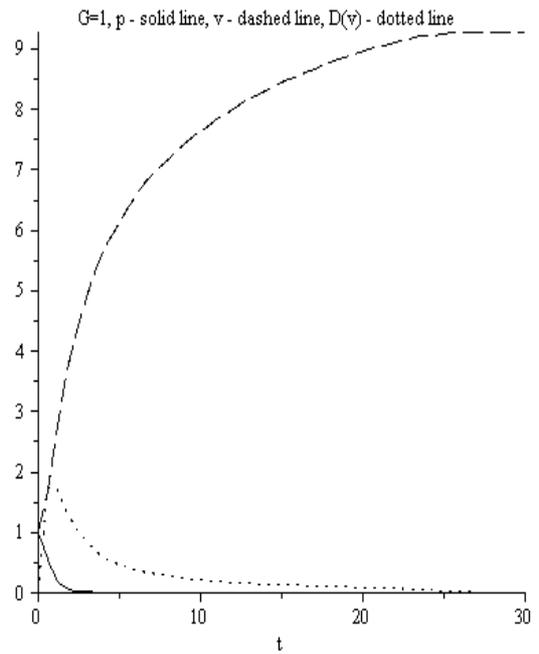

Fig. 20. r - density $\tilde{\rho}$, u - velocity $\tilde{u}$ in gravitational soliton, $w$ - orbital velocity $\tilde{w}$. $G=1$.

Fig. 21. p - pressure $\tilde{p}$, v - self consistent potential $\tilde{\Psi}$, $D(\mathrm{v})(t)=\dfrac{\partial \tilde{\Psi}}{\partial \tilde{\xi}}$ in gravitational soliton.

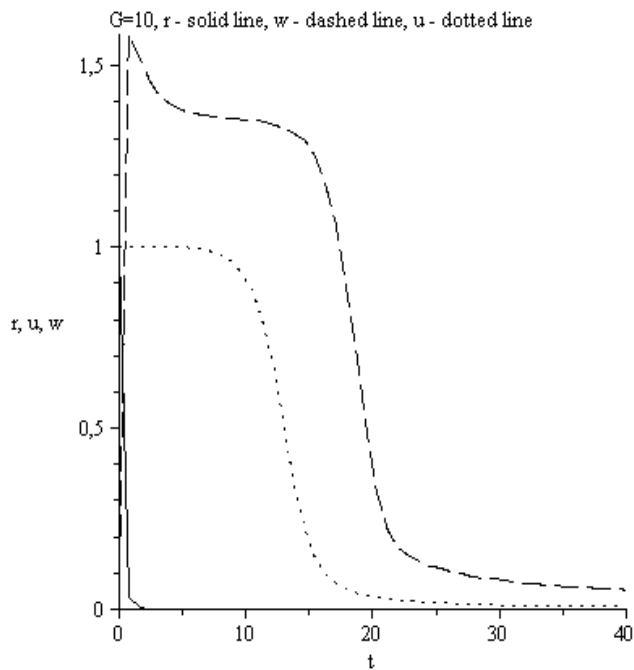 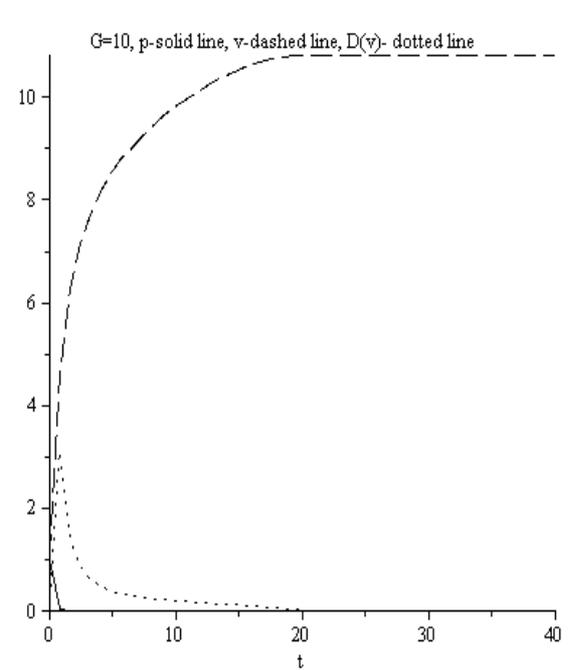

Fig. 22. r - density $\tilde{\rho}$, u - velocity $\tilde{u}$ in gravitational soliton, $w$ - orbital velocity $\tilde{w}$. $G=10$.

Fig. 23. p - pressure $\tilde{p}$, v - self consistent potential $\tilde{\Psi}$, $D(\mathrm{v})(t)=\dfrac{\partial \tilde{\Psi}}{\partial \tilde{\xi}}$ in gravitational soliton.

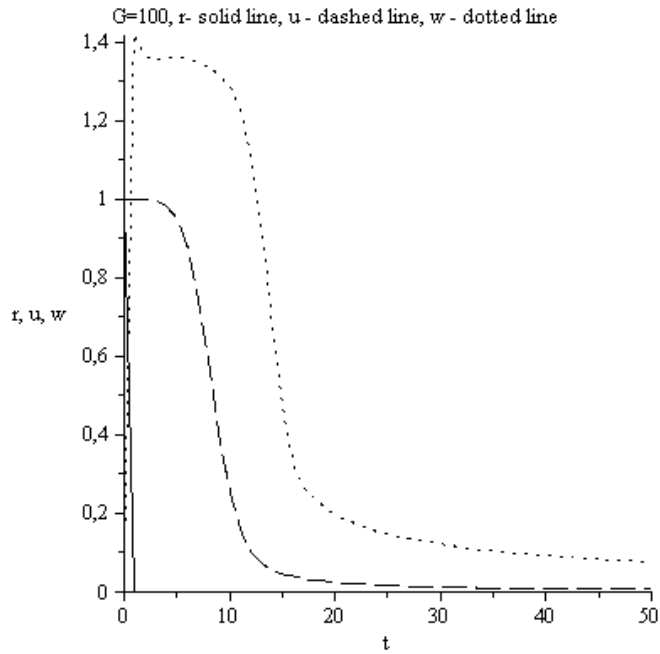 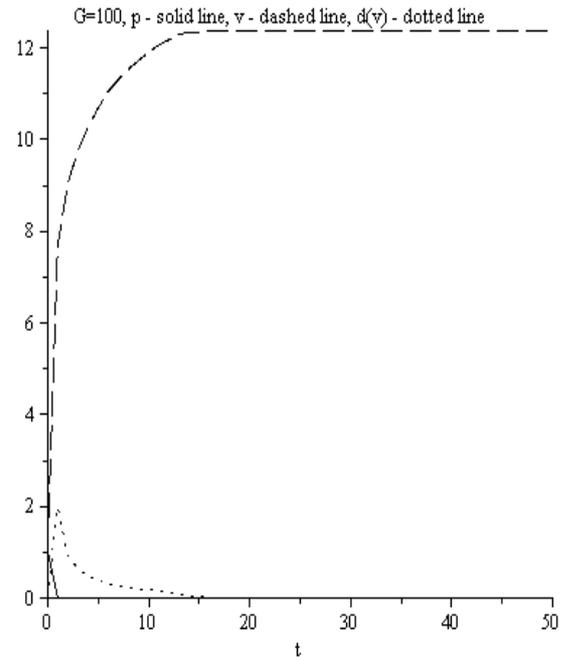

Fig. 24. r - density $\tilde{\rho}$, u - velocity $\tilde{u}$ in gravitational soliton, w - orbital velocity $\tilde{w}$. $G = 100$.

Fig. 25. p - pressure $\tilde{p}$, v - self consistent potential $\tilde{\Psi}$, $D(v)(t) = \dfrac{\partial \tilde{\Psi}}{\partial \tilde{\xi}}$ in gravitational soliton.

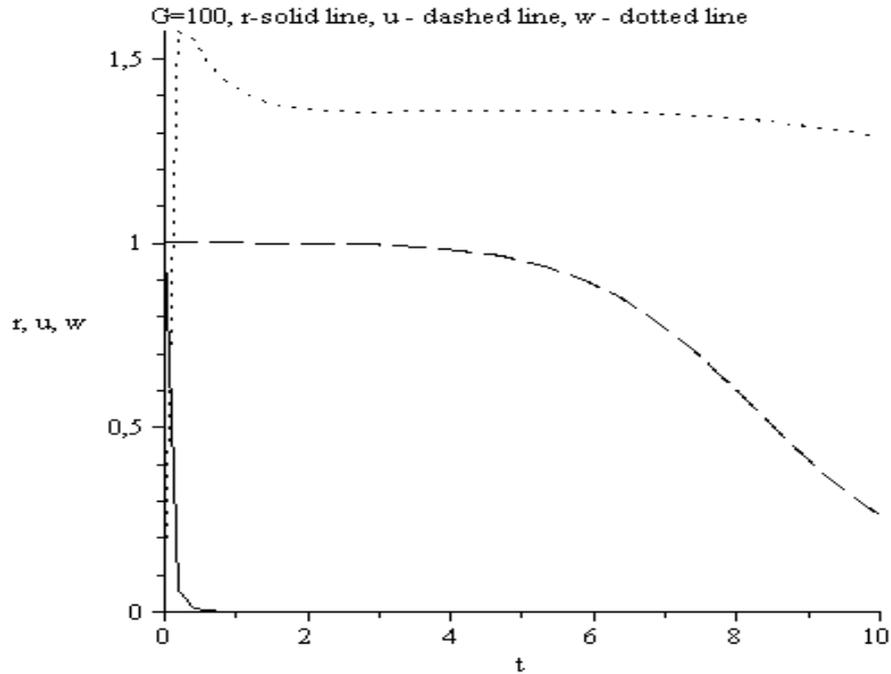

Fig. 26. r - density $\tilde{\rho}$, u - velocity $\tilde{u}$ in gravitational soliton, w - orbital velocity $\tilde{w}$.

As we see peculiar features of the halo movement can be explained without new concepts like "dark matter".

## 5. Investigation of the gravitational self-catching effect.

In the Section 3 the physical picture (following only from the non-local statistical theory and leading to the Hubble flow) is formulated. Namely

**The main origin of Hubble effect (including the matter expansion with acceleration) is self – catching of the expanding matter by the self – consistent gravitational field in conditions of weak influence of the central massive bodies.**

The formulated result is obtained in the frame of the linear theory. Is it possible to obtain the corresponding result on the level of the general non-linear description?

This investigation was realized with the success and has the direct attitude to the known investigations of S. Perlmutter, A. Riess (USA) and B. Schmidt (Australia), [27 – 29]. The trio studied what are called Type 1a supernovae, determining that more distant objects seem to move faster. Their observations suggest that not only is the Universe expanding, its expansion is relentlessly speeding up (in the following PRS-regime, Nobel Prize of 2011 year).

Effects of gravitational self-catching should be typical for Universe. The existence of "Hubble boxes" is discussed in review [4] as typical blocks of the nearby Universe.

Gravitational self – catching takes place for Big Bang having given birth to the global expansion of Universe, but also for Little Bang in so called Local Group (using the Hubble's terminology) of galaxies. Then the evolution of the Local Group (the typical Hubble box) is really fruitful field for testing of different theoretical constructions (see Fig. 27 taken from review [4]).

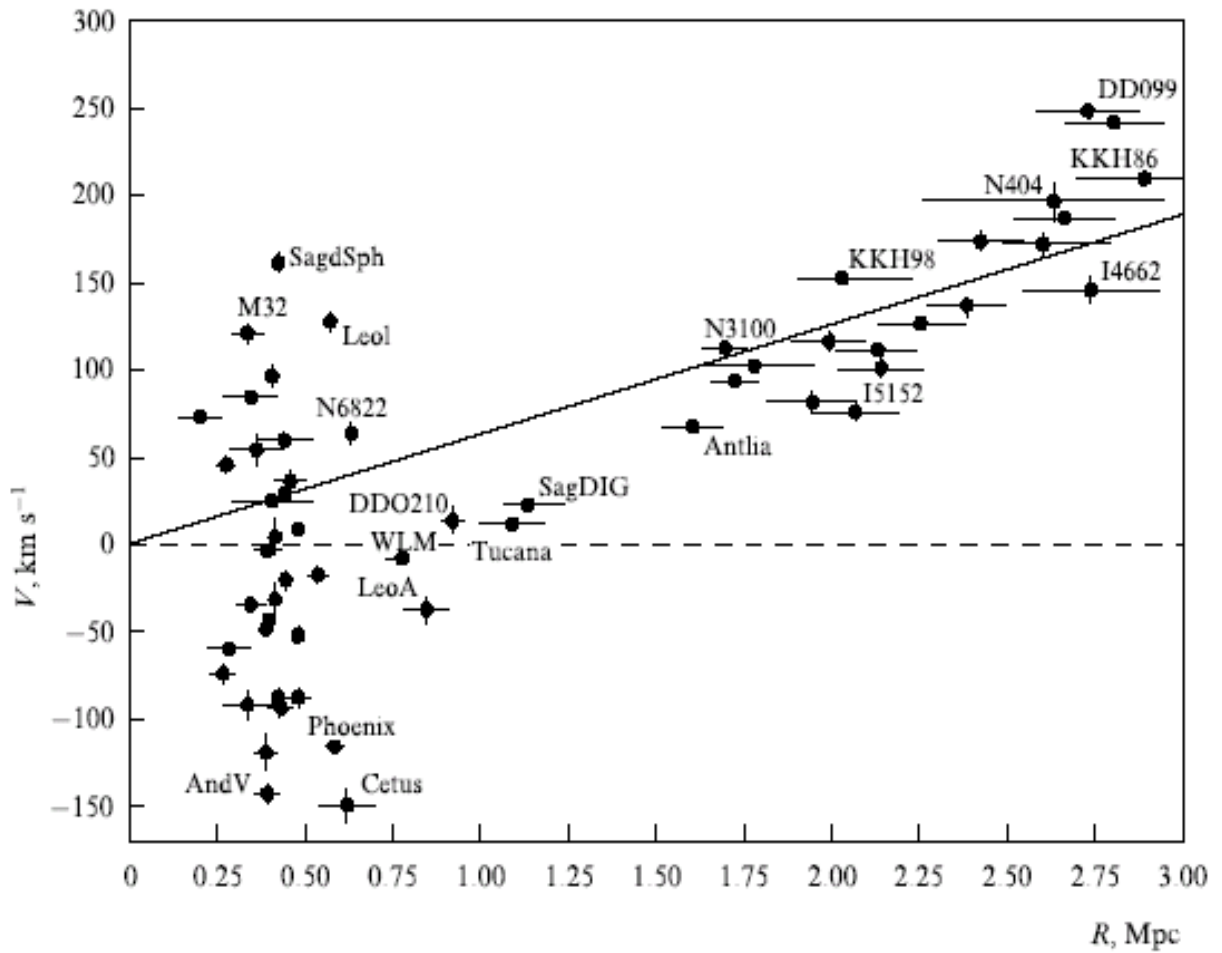

Fig. 27. Velocity-distance diagram for galaxies at distances of up to 3 *Mpc*.

The data were obtained by Karachentsev and his collaborators in 2002 – 2007 in observation with the Hubble Space Telescope. Each point corresponds to a galaxy with measured values of distance and line-of-site velocity in the reference frame related to the center of the Local Group. The diagram shows two distinct structures, the Local Group and the local flow of galaxies.

The galaxies of the Local Group occupy a volume with the radius up to ~ 1.1 – 1.2 *Mpc*, but there no galaxies in the volume which radius is less than 0.25 *Mpc*. These galaxies move both away from the center (positive velocities) and toward the center (negative velocities). These galaxies form a gravitationally bound quasi-stationary system. Their average radial velocity is equal to zero.

The galaxies of the local flow are located outside the group and all of them are moving from the center (positive velocities) beginning their motion near $R \approx 1$ *Mpc* with the velocity $v \sim 50$ *km/s*. By the way the measured by Karachentsev the average Hubble parameter for the Local Group is $72 \pm 6$ $km \cdot s^{-1} Mpc^{-1}$. Let us choose these values as scales:

$$x_0 = 1 \ Mpc, \ u_0 = 50 \ km/s. \tag{5.1}$$

Recession velocities increase as the distance increases in accordance with the Hubble law. The straight line corresponds to the dependence from observations

$$v = H(r)r \tag{5.2}$$

for the region outside of the Local Group. In the dimensional form (see scales (4.14)) we have

$$\tilde{v} = \tilde{H}(\tilde{r})\tilde{r} \tag{5.3}$$

where

$$\tilde{H} = \frac{x_0}{u_0} H(\tilde{r}). \tag{5.4}$$

For the following calculations we should choose the corresponding scales, especially for estimation of $G$ for evolution of the Local Group (see also (4.23))

$$G = \tilde{\gamma}_N = \gamma_N / \gamma_{N0} = \gamma_N \rho_0 x_0^2 / u_0^2$$

For the density scale estimation the avearge density of the local flow could be used. But the corresponding data are not accessible and I use the average density of the Local Group which can be taken from references [30, 31] with $\rho_0 = 4.85 \cdot 10^{-29} \ g/cm^3$. Then from (4.23) we have $G \cong 1$.

Let us go now to the mathematical modeling. The non-local system of hydrodynamic equations describing the explosion with the spherical symmetry [32, 33] is written as

$$g_r = -\frac{\partial \psi}{\partial r}, \tag{5.5}$$

Poisson equation

$$\frac{1}{r^2}\frac{\partial}{\partial r}\left(r^2 \frac{\partial \psi}{\partial r}\right) = 4\pi\gamma_N \left[\rho - \tau\left(\frac{\partial \rho}{\partial t} + \frac{1}{r^2}\frac{\partial (r^2 \rho v_r)}{\partial r}\right)\right], \tag{5.6}$$

Continuity equation

$$\frac{\partial}{\partial t}\left\{\rho - \tau\left[\frac{\partial \rho}{\partial t} + \frac{1}{r^2}\frac{\partial (r^2 \rho v_r)}{\partial r}\right]\right\} + \frac{1}{r^2}\frac{\partial}{\partial r}\left\{r^2\left\{\rho v_r - \tau\left[\frac{\partial}{\partial t}(\rho v_r) + \frac{1}{r^2}\frac{\partial (r^2 \rho v_r^2)}{\partial r} - \rho g_r\right]\right\}\right\} - \frac{1}{r^2}\frac{\partial}{\partial r}\left(\tau r^2 \frac{\partial p}{\partial r}\right) = 0, \tag{5.7}$$

Motion equation

$$\frac{\partial}{\partial t}\left\{\rho v_r - \tau\left[\frac{\partial}{\partial t}(\rho v_r) + \frac{1}{r^2}\frac{\partial(r^2\rho v_r^2)}{\partial r} + \frac{\partial p}{\partial r} - \rho g_r\right]\right\} - g_r\left[\rho - \tau\left(\frac{\partial \rho}{\partial t} + \frac{1}{r^2}\frac{\partial(r^2\rho v_r)}{\partial r}\right)\right] +$$

$$\frac{1}{r^2}\frac{\partial}{\partial r}\left\{r^2\left\{\rho v_r^2 - \tau\left[\frac{\partial}{\partial t}(\rho v_r^2) + \frac{1}{r^2}\frac{\partial(r^2\rho v_r^3)}{\partial r} - 2g_r\rho v_r\right]\right\}\right\} +$$

$$\frac{\partial p}{\partial r} - \frac{\partial}{\partial r}\left(\tau\frac{\partial p}{\partial t}\right) - 2\frac{\partial}{\partial r}\left(\frac{\tau}{r^2}\frac{\partial(r^2 pv_r)}{\partial r}\right) - \frac{1}{r^2}\frac{\partial}{\partial r}\left(\tau r^2\frac{\partial(pv_r)}{\partial r}\right) = 0, \qquad (5.8)$$

Energy equation

$$\frac{\partial}{\partial t}\left\{\frac{1}{2}\rho v_r^2 + \frac{3}{2}p - \tau\left[\frac{\partial}{\partial t}\left(\frac{1}{2}\rho v_r^2 + \frac{3}{2}p\right) + \frac{1}{r^2}\frac{\partial}{\partial r}\left(r^2 v_r\left(\frac{1}{2}\rho v_r^2 + \frac{5}{2}p\right)\right) - \rho g_r v_r\right]\right\} +$$

$$\frac{1}{r^2}\frac{\partial}{\partial r}\left\{r^2\left\{\left(\frac{1}{2}\rho v_r^2 + \frac{5}{2}p\right)v_r - \tau\left[\frac{\partial}{\partial t}\left(\left(\frac{1}{2}\rho v_r^2 + \frac{5}{2}p\right)v_r\right) + \frac{1}{r^2}\frac{\partial}{\partial r}\left(r^2\left(\frac{1}{2}\rho v_r^2 + \frac{7}{2}p\right)v_r^2\right) - \right.\right.\right.$$

$$\left.\left.\left. \rho g_r v_r^2 - \left(\frac{1}{2}\rho v_r^2 + \frac{3}{2}p\right)g_r\right]\right\}\right\} - \left\{\rho g_r v_r - \tau\left[g_r\left(\frac{\partial}{\partial t}(\rho v_r) + \frac{1}{r^2}\frac{\partial}{\partial r}(r^2\rho v_r^2) + \frac{\partial p}{\partial r} - \rho g_r\right)\right]\right\} -$$

$$\frac{1}{r^2}\frac{\partial}{\partial r}\left(\tau r^2\frac{\partial}{\partial r}\left(\frac{1}{2}pv_r^2 + \frac{5}{2}\frac{p^2}{\rho}\right)\right) + \frac{1}{r^2}\frac{\partial}{\partial r}(r^2\tau p g_r) = 0. \qquad (5.9)$$

The system of Eqs. (5.5) – (5.9) belongs to the class of the 1D non-stationary equations and can be solved by known numerical methods. But for the aims of the transparent vast mathematical modeling of self – catching of the expanding matter by the self – consistent gravitational field I introduce the following assumption. Let us allot the quasi-stationary Hubble regime when only the implicit dependence on time exists for the unknown values (see also (5.3)).

It means that for intermediate regime the substitution

$$\frac{\partial}{\partial t} = \frac{\partial}{\partial r}\frac{\partial r}{\partial t} = v_r\frac{\partial}{\partial r} \qquad (5.10)$$

can be introduced. As result we have the following system of the 1D dimensionless equations

$$\frac{1}{\tilde{r}^2}\frac{\partial}{\partial \tilde{r}}\left(\tilde{r}^2\frac{\partial \tilde{\psi}}{\partial \tilde{r}}\right) = 4\pi G\left[\tilde{\rho} - \tilde{\tau}\left(\tilde{v}_r\frac{\partial \tilde{\rho}}{\partial \tilde{r}} + \frac{1}{\tilde{r}^2}\frac{\partial(\tilde{r}^2\tilde{\rho}\tilde{v}_r)}{\partial \tilde{r}}\right)\right], \qquad (5.11)$$

$$\tilde{v}_r\frac{\partial}{\partial \tilde{r}}\left\{\tilde{\rho} - \tilde{\tau}\left[\tilde{v}_r\frac{\partial \tilde{\rho}}{\partial \tilde{r}} + \frac{1}{\tilde{r}^2}\frac{\partial(\tilde{r}^2\tilde{\rho}\tilde{v}_r)}{\partial \tilde{r}}\right]\right\} + \frac{1}{\tilde{r}^2}\frac{\partial}{\partial \tilde{r}}\left\{\tilde{r}^2\left\{\tilde{\rho}\tilde{v}_r - \tilde{\tau}\left[\tilde{v}_r\frac{\partial}{\partial \tilde{r}}(\tilde{\rho}\tilde{v}_r) + \right.\right.\right.$$

$$\left.\left.\left. + \frac{1}{\tilde{r}^2}\frac{\partial(\tilde{r}^2\tilde{\rho}\tilde{v}_r^2)}{\partial \tilde{r}} + \tilde{\rho}\frac{\partial \tilde{\psi}}{\partial \tilde{r}}\right]\right\}\right\} - \frac{1}{\tilde{r}^2}\frac{\partial}{\partial \tilde{r}}\left(\tilde{\tau}\tilde{r}^2\frac{\partial \tilde{p}}{\partial \tilde{r}}\right) = 0, \qquad (5.12)$$

$$\tilde{v}_r \frac{\partial}{\partial \tilde{r}} \left\{ \tilde{\rho}\tilde{v}_r - \tilde{\tau}\left[ \tilde{v}_r \frac{\partial}{\partial \tilde{r}}(\tilde{\rho}\tilde{v}_r) + \frac{1}{\tilde{r}^2}\frac{\partial(\tilde{r}^2 \tilde{\rho}\tilde{v}_r^2)}{\partial \tilde{r}} + \frac{\partial \tilde{p}}{\partial \tilde{r}} + \tilde{\rho}\frac{\partial \tilde{\psi}}{\partial \tilde{r}} \right] \right\} +$$

$$+ \frac{\partial \tilde{\psi}}{\partial \tilde{r}} \left[ \tilde{\rho} - \tilde{\tau}\left( \tilde{v}_r \frac{\partial \tilde{\rho}}{\partial \tilde{r}} + \frac{1}{\tilde{r}^2}\frac{\partial(\tilde{r}^2 \tilde{\rho}\tilde{v}_r)}{\partial \tilde{r}} \right) \right] +$$

$$+ \frac{1}{\tilde{r}^2}\frac{\partial}{\partial \tilde{r}} \left\{ \tilde{r}^2 \left\{ \tilde{\rho}\tilde{v}_r^2 - \tilde{\tau}\left[ \tilde{v}_r \frac{\partial}{\partial r}(\tilde{\rho}\tilde{v}_r^2) + \frac{1}{\tilde{r}^2}\frac{\partial(\tilde{r}^2 \tilde{\rho}\tilde{v}_r^3)}{\partial \tilde{r}} + 2\frac{\partial \tilde{\psi}}{\partial \tilde{r}}\tilde{\rho}\tilde{v}_r \right] \right\} \right\} +$$

$$+ \frac{\partial \tilde{p}}{\partial \tilde{r}} - \frac{\partial}{\partial \tilde{r}}\left( \tilde{v}_r \tilde{\tau} \frac{\partial \tilde{p}}{\partial \tilde{r}} \right) - 2\frac{\partial}{\partial \tilde{r}}\left( \frac{\tilde{\tau}}{\tilde{r}^2}\frac{\partial(\tilde{r}^2 \tilde{p}\tilde{v}_r)}{\partial \tilde{r}} \right) - \frac{1}{\tilde{r}^2}\frac{\partial}{\partial \tilde{r}}\left( \tilde{\tau}\tilde{r}^2 \frac{\partial(\tilde{p}\tilde{v}_r)}{\partial \tilde{r}} \right) = 0, \qquad (5.13)$$

$$\tilde{v}_r \frac{\partial}{\partial \tilde{r}} \left\{ \tilde{\rho}\tilde{v}_r^2 + 3\tilde{p} - \tilde{\tau}\left[ \tilde{v}_r \frac{\partial}{\partial \tilde{r}}(\tilde{\rho}\tilde{v}_r^2 + 3\tilde{p}) + \frac{1}{\tilde{r}^2}\frac{\partial}{\partial \tilde{r}}(\tilde{r}^2 \tilde{v}_r(\tilde{\rho}\tilde{v}_r^2 + 5\tilde{p})) + 2\tilde{\rho}\frac{\partial \tilde{\psi}}{\partial \tilde{r}}\tilde{v}_r \right] \right\} +$$

$$+ \frac{1}{\tilde{r}^2}\frac{\partial}{\partial \tilde{r}} \left\{ \tilde{r}^2 \left\{ (\tilde{\rho}\tilde{v}_r^2 + 5\tilde{p})\tilde{v}_r - \tilde{\tau}\left[ \tilde{v}_r \frac{\partial}{\partial \tilde{r}}((\tilde{\rho}\tilde{v}_r^2 + 5\tilde{p})\tilde{v}_r) + \frac{1}{\tilde{r}^2}\frac{\partial}{\partial \tilde{r}}(\tilde{r}^2(\tilde{\rho}\tilde{v}_r^2 + 7\tilde{p})\tilde{v}_r^2) + \right. \right. \right.$$

$$\left. \left. \left. + 2\tilde{\rho}\frac{\partial \tilde{\psi}}{\partial \tilde{r}}\tilde{v}_r^2 + (\tilde{\rho}\tilde{v}_r^2 + 3\tilde{p})\frac{\partial \tilde{\psi}}{\partial \tilde{r}} \right] \right\} \right\} + 2\left\{ \tilde{\rho}\frac{\partial \tilde{\psi}}{\partial \tilde{r}}\tilde{v}_r - \right.$$

$$\left. - \tilde{\tau}\frac{\partial \tilde{\psi}}{\partial \tilde{r}}\left[ \tilde{v}_r \frac{\partial}{\partial \tilde{r}}(\tilde{\rho}\tilde{v}_r) + \frac{1}{\tilde{r}^2}\frac{\partial}{\partial \tilde{r}}(\tilde{r}^2 \tilde{\rho}\tilde{v}_r^2) + \frac{\partial \tilde{p}}{\partial \tilde{r}} + \tilde{\rho}\frac{\partial \tilde{\psi}}{\partial \tilde{r}} \right] \right\} - \frac{1}{\tilde{r}^2}\frac{\partial}{\partial \tilde{r}}\left( \tilde{\tau}\tilde{r}^2 \frac{\partial}{\partial \tilde{r}}\left( \tilde{p}\tilde{v}_r^2 + 5\frac{\tilde{p}^2}{\tilde{\rho}} \right) \right) -$$

$$- 2\frac{1}{\tilde{r}^2}\frac{\partial}{\partial \tilde{r}}\left( \tilde{r}^2 \tilde{\tau}\tilde{p} \frac{\partial \tilde{\psi}}{\partial \tilde{r}} \right) = 0. \qquad (5.14)$$

Now we are ready to display the results of the mathematical modeling realized with the help of Maple. The system of generalized hydrodynamic equations (5.11) – (5.14) have the great possibilities of mathematical modeling as result of changing of eight Cauchy conditions describing the character features of the local flow evolution.

The following Maple notations on figures are used: r - density $\tilde{\rho}$, u - velocity $\tilde{v}_r$, p - pressure $\tilde{p}$ and v - self consistent potential $\tilde{\psi}$, h - $\tilde{H}$, independent variable t is $\tilde{r}$. Explanations placed under all following figures, Maple program contains Maple's notations – for example the expression $D(u)(0) = 0$ means in the usual notations $(\partial \tilde{u}/\partial \tilde{r})(0) = 0$.

As mentioned before the non-local parameter $\tilde{\tau}$, in the definite sense plays the role analogous to kinetic coefficients in the usual Boltzmann kinetic theory. The influence on the results of calculations is not too significant. The same situation exists in the generalized hydrodynamics.

As before I introduce the following approximations for the dimensionless non-local parameter (see (4.19))

$$\tilde{\tau} = 1/\tilde{u}^2,$$

and another approximation for $\tilde{\tau}$ in the simplest possible form, namely (see (4.21))

$$\tilde{\tau} = 1.$$

One obtains for the approximation (4.19) and SYSTEM 2

```
v(1)=1,u(1)=1,r(1)=1,p(1)=1,
D(v)(1)=0,D(u)(1)=0,D(r)(1)=0,D(p)(1)=0
```

Figures 28 – 32 correspond to G=1, the non-local $\tilde{\tau}$ approximation is (4.19). From these calculations follow:

1. As was waiting the quasi-stationary regime exists only in the restricted (on the left and on the right sides) area. Out of these boundaries the explicit time dependent regime should be considered. But it is not the Hubble regime.
2. In the Hubble regime one obtains the negative area (magenta curve). It corresponds to the self-consistent force acting along the expansion of the local flow.
3. The dependence of $\tilde{H}(\tilde{r})$ is not linear (see Fig. 29), more over the curvature contains maximum.
4. The regime with the acceleration is shown separately on the Fig. 30. This area of acceleration placed between two areas of the deceleration.

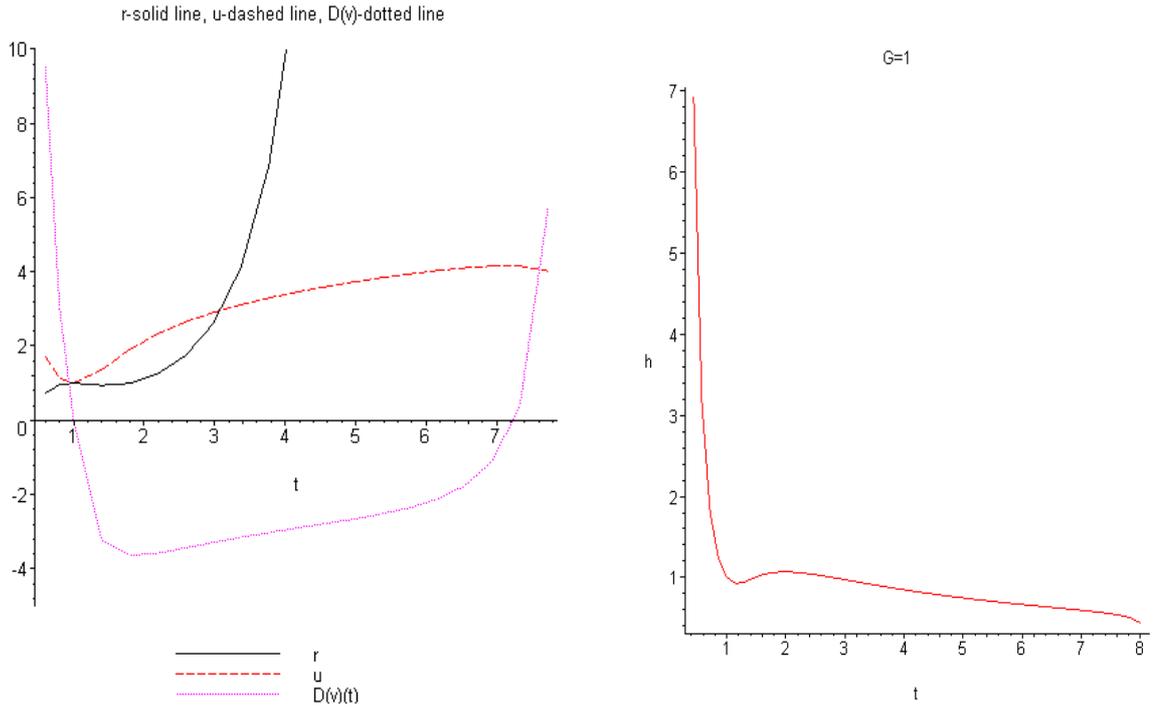

Fig. 28. r - density $\tilde{\rho}$, u - $\tilde{u}$, $D(v)(t) = \partial \tilde{\psi} / \partial \tilde{r}$, (left).

Fig. 29. Dependence of the dimensionless Hubble parameter on the radial distance, (right).

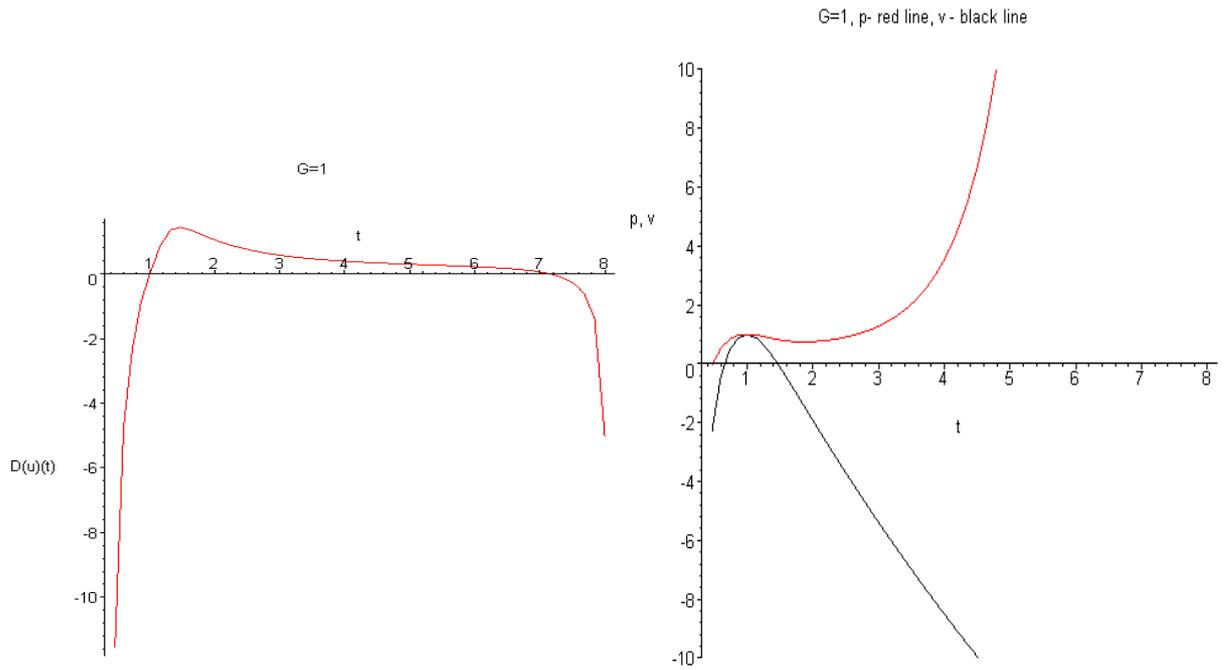

Fig. 30. Dependence of the acceleration – deceleration $D(u)(t) = \partial \tilde{u}/\partial \tilde{r}$ on $\tilde{r}$, (left)

Fig 31. p - pressure $\tilde{p}$, v – self consistent potential $\tilde{\psi}$, (right)

Let us show now the results of calculations for another $\tilde{\tau}$ approximation in the simplest possible form, namely (see (4.21) $\tilde{\tau} = 1$. One obtains for the approximation (4.21), SYSTEM 2 and G=1, see Figs. 32 – 35.

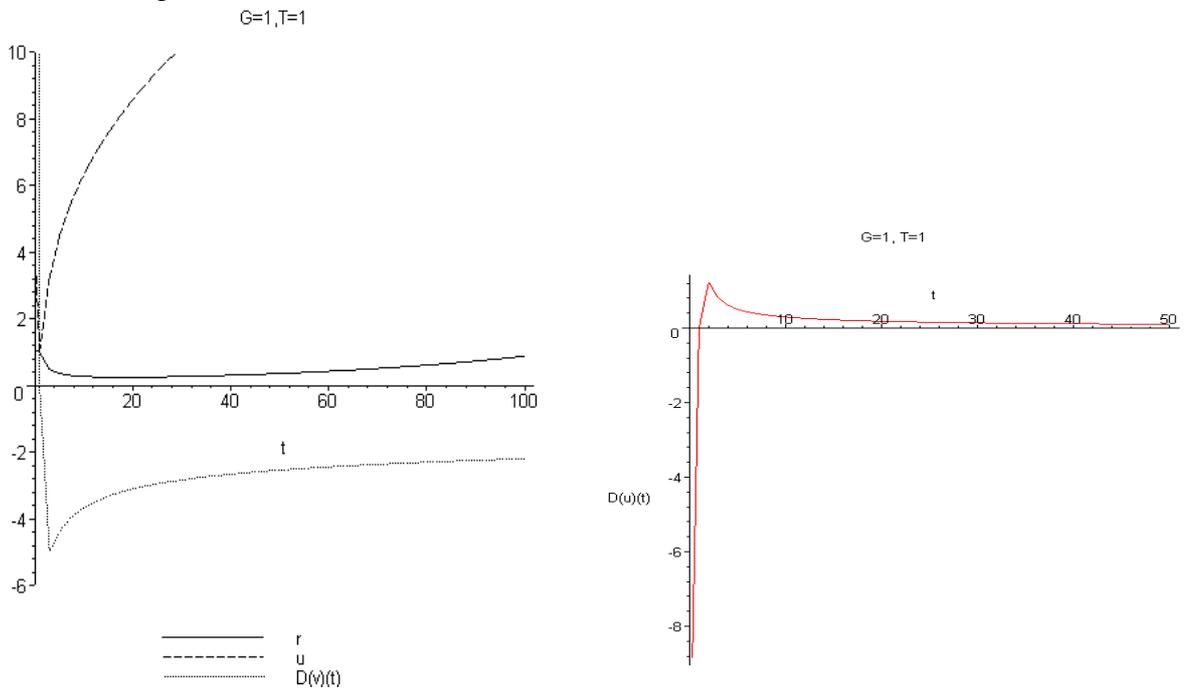

Fig. 32. r - density $\tilde{\rho}$, u - $\tilde{u}$, $D(v)(t) = \partial \tilde{\psi}/\partial \tilde{r}$, (left).

Fig. 33. Dependence of the acceleration – deceleration $D(u)(t) = \partial \tilde{u}/\partial \tilde{r}$ on $\tilde{r}$, (right).

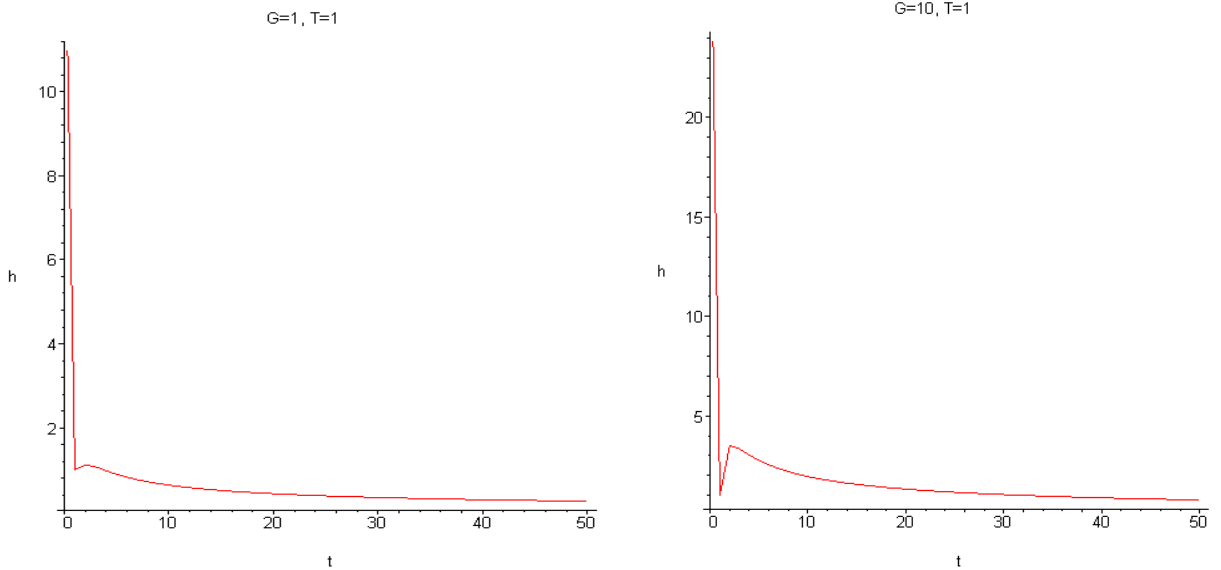

Fig. 34. Dependence of the dimensionless Hubble parameter on the radial distance for G=1, (left)
Fig. 35. Dependence of the dimensionless Hubble parameter on the radial distance for G=10, (right)

Continue now the previous conclusions:
5. Approximation $\tilde{\tau} = const$ conserves all principal features of the previous dependences, but the area of the Hubble regime becomes much large.
6. The PRS regime on Fig. 35 is reflected (see also Figs. 32, 33).
7. Approximation $\tilde{\tau} = const$ allows considering the numerical transmission to the "classical" gas dynamics of explosions. By the $\tilde{\tau} \to 0$ there NO HUBBLE AND PRS REGIMES ON PRINCIPAL.

Let us consider the influence of diminishing of the G parameter using the approximation in the form (4.19), see Figs. 36 – 40.

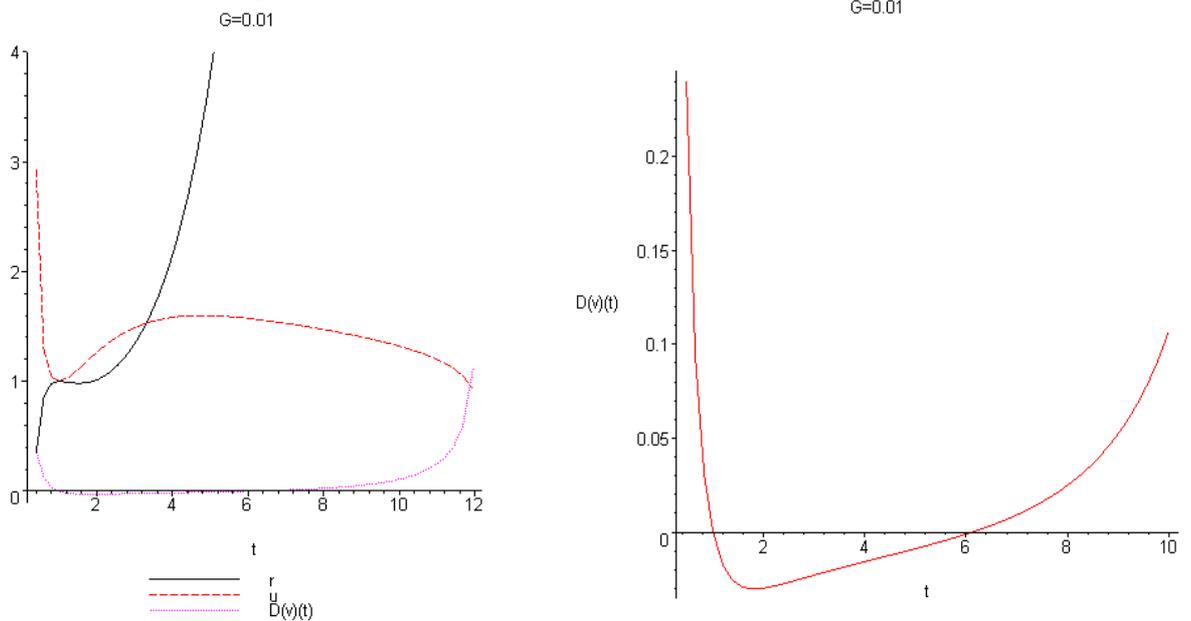

Fig. 36. r - density $\tilde{\rho}$, u - $\tilde{u}$, $D(v)(t) = \partial \tilde{\psi}/\partial \tilde{r}$, (left).
Fig. 37. $D(v)(t) = \partial \tilde{\psi}/\partial \tilde{r}$, (right).

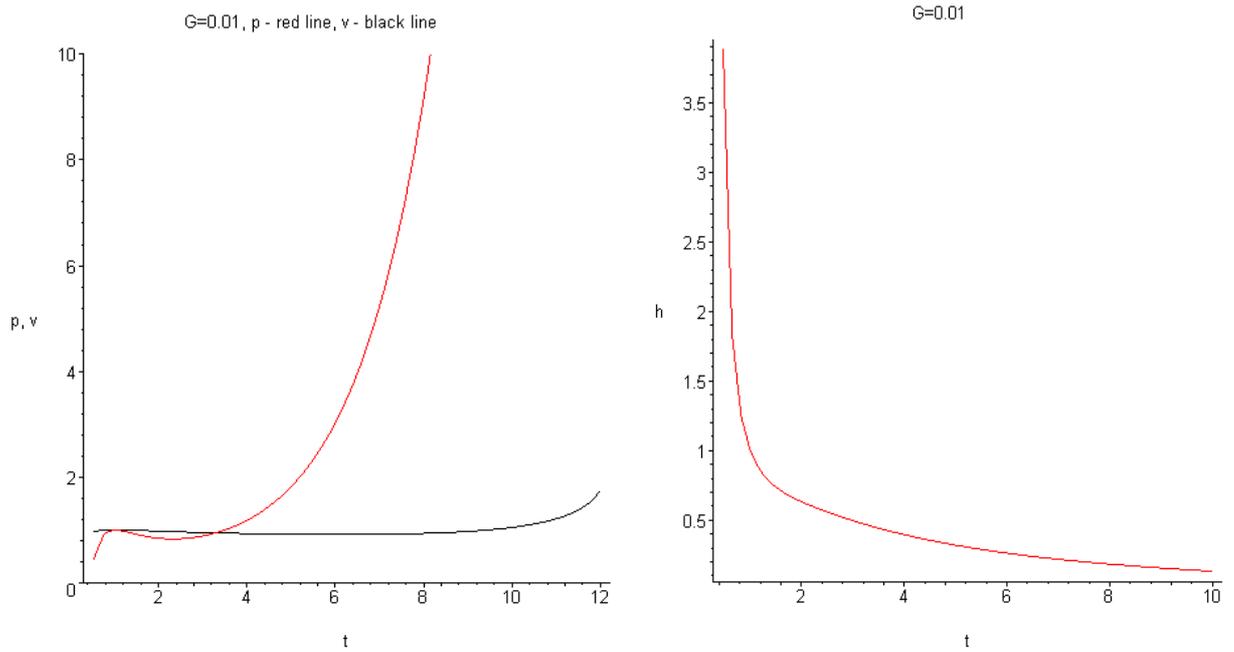

Fig 38. p - pressure $\tilde{p}$, v – self consistent potential $\tilde{\psi}$, (left).

Fig. 39. Dependence of the dimensionless Hubble parameter on the radial distance, (right).

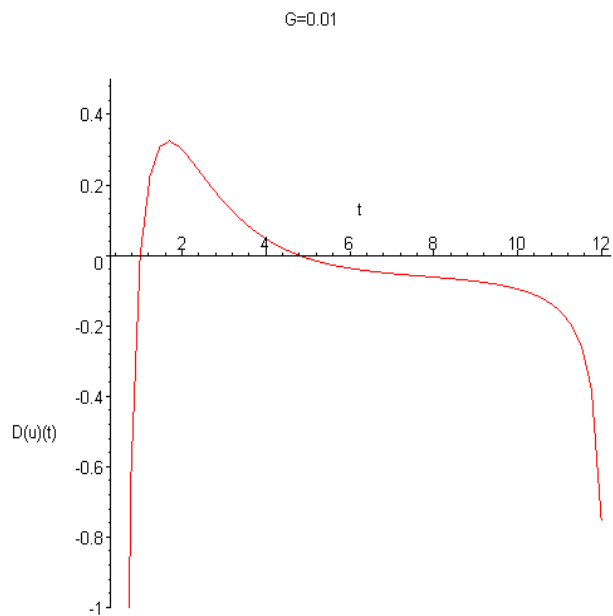

Fig. 40. Dependence of the acceleration – deceleration $D(\mathrm{u})(t) = \partial \tilde{u} / \partial \tilde{r}$ on $\tilde{r}$.

Continue the previous conclusions:

8. Diminishing of G leads to diminishing of the area of the Hubble regime with the positive acceleration of the matter catching by the self-consistent gravitational field.

9. Dependence of $\tilde{H}(\tilde{r})$ does not contain maximum by the small G.

10. Proposal to Observers – to find the Hubble boxes where is realizing the mentioned regimes of the Hubble enlargement (see Fig. 29, regime of the first kind, or Fig. 39, regime of the second kind). Looking at Fig. 27 reasonable to conclude that after R=1.75 *Mpc* the Hubble strait line should be transformed in the steeper curve following the first kind regime.

## 6. Nonstationary generalized hydrodynamic equations in the self consistent electrical field. Quantization in the generalized quantum hydrodynamics.

In the following I intend to demonstrate the validity of the unified description in the absolutely another scale diapason related with atom structure. With this aim let us write down the system of the non-local quantum hydrodynamic equations with taking into account the self consistent electromagnetic forces.

Strict consideration leads to the following system of the generalized hydrodynamic equations (GHE) [5, 13] written in the generalized Euler form:

Continuity equation for species $\alpha$:

$$\frac{\partial}{\partial t}\left\{\rho_\alpha - \tau_\alpha^{(0)}\left[\frac{\partial \rho_\alpha}{\partial t} + \frac{\partial}{\partial \mathbf{r}} \cdot (\rho_\alpha \mathbf{v}_0)\right]\right\} + $$
$$+ \frac{\partial}{\partial \mathbf{r}} \cdot \left\{\rho_\alpha \mathbf{v}_0 - \tau_\alpha^{(0)}\left[\frac{\partial}{\partial t}(\rho_\alpha \mathbf{v}_0) + \frac{\partial}{\partial \mathbf{r}} \cdot (\rho_\alpha \mathbf{v}_0 \mathbf{v}_0) + \vec{\mathbf{I}} \cdot \frac{\partial p_\alpha}{\partial \mathbf{r}} - \rho_\alpha \mathbf{F}_\alpha^{(1)} - \frac{q_\alpha}{m_\alpha}\rho_\alpha \mathbf{v}_0 \times \mathbf{B}\right]\right\} = R_\alpha. \quad (6.1)$$

Continuity equation for mixture:

$$\frac{\partial}{\partial t}\left\{\rho - \sum_\alpha \tau_\alpha^{(0)}\left[\frac{\partial \rho_\alpha}{\partial t} + \frac{\partial}{\partial \mathbf{r}} \cdot (\rho_\alpha \mathbf{v}_0)\right]\right\} + $$
$$+ \frac{\partial}{\partial \mathbf{r}} \cdot \left\{\rho \mathbf{v}_0 - \sum_\alpha \tau_\alpha^{(0)}\left[\frac{\partial}{\partial t}(\rho_\alpha \mathbf{v}_0) + \frac{\partial}{\partial \mathbf{r}} \cdot (\rho_\alpha \mathbf{v}_0 \mathbf{v}_0) + \vec{\mathbf{I}} \cdot \frac{\partial p_\alpha}{\partial \mathbf{r}} - \rho_\alpha \mathbf{F}_\alpha^{(1)} - \frac{q_\alpha}{m_\alpha}\rho_\alpha \mathbf{v}_0 \times \mathbf{B}\right]\right\} = 0, \quad (6.2)$$

Momentum equation for component

$$\frac{\partial}{\partial t}\left\{\rho_\alpha \mathbf{v}_0 - \tau_\alpha^{(0)}\left[\frac{\partial}{\partial t}(\rho_\alpha \mathbf{v}_0) + \frac{\partial}{\partial \mathbf{r}} \cdot \rho_\alpha \mathbf{v}_0 \mathbf{v}_0 + \frac{\partial p_\alpha}{\partial \mathbf{r}} - \rho_\alpha \mathbf{F}_\alpha^{(1)} - \frac{q_\alpha}{m_\alpha}\rho_\alpha \mathbf{v}_0 \times \mathbf{B}\right]\right\} - $$
$$- \mathbf{F}_\alpha^{(1)}\left[\rho_\alpha - \tau_\alpha^{(0)}\left(\frac{\partial \rho_\alpha}{\partial t} + \frac{\partial}{\partial \mathbf{r}}(\rho_\alpha \mathbf{v}_0)\right)\right] - \frac{q_\alpha}{m_\alpha}\left\{\rho_\alpha \mathbf{v}_0 - \tau_\alpha^{(0)}\left[\frac{\partial}{\partial t}(\rho_\alpha \mathbf{v}_0) + \frac{\partial}{\partial \mathbf{r}} \cdot \rho_\alpha \mathbf{v}_0 \mathbf{v}_0 + \frac{\partial p_\alpha}{\partial \mathbf{r}} - \right.\right.$$
$$\left.\left. - \rho_\alpha \mathbf{F}_\alpha^{(1)} - \frac{q_\alpha}{m_\alpha}\rho_\alpha \mathbf{v}_0 \times \mathbf{B}\right]\right\} \times \mathbf{B} + \frac{\partial}{\partial \mathbf{r}} \cdot \left\{\rho_\alpha \mathbf{v}_0 \mathbf{v}_0 + p_\alpha \vec{\mathbf{I}} - \tau_\alpha^{(0)}\left[\frac{\partial}{\partial t}(\rho_\alpha \mathbf{v}_0 \mathbf{v}_0 + p_\alpha \vec{\mathbf{I}}) + \right.\right. \quad (6.3)$$
$$+ \frac{\partial}{\partial \mathbf{r}} \cdot \rho_\alpha (\mathbf{v}_0 \mathbf{v}_0)\mathbf{v}_0 + 2\vec{\mathbf{I}}\left(\frac{\partial}{\partial \mathbf{r}} \cdot (p_\alpha \mathbf{v}_0)\right) + \frac{\partial}{\partial \mathbf{r}} \cdot (\vec{\mathbf{I}} p_\alpha \mathbf{v}_0) - \mathbf{F}_\alpha^{(1)}\rho_\alpha \mathbf{v}_0 - \rho_\alpha \mathbf{v}_0 \mathbf{F}_\alpha^{(1)} - $$
$$\left.\left. - \frac{q_\alpha}{m_\alpha}\rho_\alpha[\mathbf{v}_0 \times \mathbf{B}]\mathbf{v}_0 - \frac{q_\alpha}{m_\alpha}\rho_\alpha \mathbf{v}_0[\mathbf{v}_0 \times \mathbf{B}]\right]\right\} = \int m_\alpha \mathbf{v}_\alpha J_\alpha^{st,el} d\mathbf{v}_\alpha + \int m_\alpha \mathbf{v}_\alpha J_\alpha^{st,inel} d\mathbf{v}_\alpha.$$

Energy equation for component:
$$\frac{\partial}{\partial t}\left\{\frac{\rho_\alpha v_0^2}{2} + \frac{3}{2} p_\alpha + \varepsilon_\alpha n_\alpha - \tau_\alpha^{(0)}\left[\frac{\partial}{\partial t}\left(\frac{\rho_\alpha v_0^2}{2} + \frac{3}{2} p_\alpha + \varepsilon_\alpha n_\alpha\right) + \right.\right.$$
$$\left.\left. + \frac{\partial}{\partial \mathbf{r}} \cdot \left(\frac{1}{2}\rho_\alpha v_0^2 \mathbf{v}_0 + \frac{5}{2} p_\alpha \mathbf{v}_0 + \varepsilon_\alpha n_\alpha \mathbf{v}_0\right) - \mathbf{F}_\alpha^{(1)} \cdot \rho_\alpha \mathbf{v}_0 \right]\right\} +$$
$$+ \frac{\partial}{\partial \mathbf{r}} \cdot \left\{\frac{1}{2}\rho_\alpha v_0^2 \mathbf{v}_0 + \frac{5}{2} p_\alpha \mathbf{v}_0 + \varepsilon_\alpha n_\alpha \mathbf{v}_0 - \tau_\alpha^{(0)}\left[\frac{\partial}{\partial t}\left(\frac{1}{2}\rho_\alpha v_0^2 \mathbf{v}_0 + \right.\right.\right.$$
$$\left. + \frac{5}{2} p_\alpha \mathbf{v}_0 + \varepsilon_\alpha n_\alpha \mathbf{v}_0\right) + \frac{\partial}{\partial \mathbf{r}} \cdot \left(\frac{1}{2}\rho_\alpha v_0^2 \mathbf{v}_0 \mathbf{v}_0 + \frac{7}{2} p_\alpha \mathbf{v}_0 \mathbf{v}_0 + \frac{1}{2} p_\alpha v_0^2 \vec{\mathbf{I}} + \right.$$
$$\left. + \frac{5}{2}\frac{p_\alpha^2}{\rho_\alpha}\vec{\mathbf{I}} + \varepsilon_\alpha n_\alpha \mathbf{v}_0 \mathbf{v}_0 + \varepsilon_\alpha \frac{p_\alpha}{m_\alpha}\vec{\mathbf{I}}\right) - \rho_\alpha \mathbf{F}_\alpha^{(1)} \cdot \mathbf{v}_0 \mathbf{v}_0 - p_\alpha \mathbf{F}_\alpha^{(1)} \cdot \vec{\mathbf{I}} - \quad (6.4)$$
$$- \frac{1}{2}\rho_\alpha v_0^2 \mathbf{F}_\alpha^{(1)} - \frac{3}{2}\mathbf{F}_\alpha^{(1)} p_\alpha - \frac{\rho_\alpha v_0^2}{2}\frac{q_\alpha}{m_\alpha}[\mathbf{v}_0 \times \mathbf{B}] - \frac{5}{2} p_\alpha \frac{q_\alpha}{m_\alpha}[\mathbf{v}_0 \times \mathbf{B}] -$$
$$\left.\left. - \varepsilon_\alpha n_\alpha \frac{q_\alpha}{m_\alpha}[\mathbf{v}_0 \times \mathbf{B}] - \varepsilon_\alpha n_\alpha \mathbf{F}_\alpha^{(1)}\right]\right\} - \left\{\rho_\alpha \mathbf{F}_\alpha^{(1)} \cdot \mathbf{v}_0 - \tau_\alpha^{(0)}\left[\mathbf{F}_\alpha^{(1)} \cdot \right.\right.$$
$$\left.\left. \cdot \left(\frac{\partial}{\partial t}(\rho_\alpha \mathbf{v}_0) + \frac{\partial}{\partial \mathbf{r}} \cdot \rho_\alpha \mathbf{v}_0 \mathbf{v}_0 + \frac{\partial}{\partial \mathbf{r}} \cdot p_\alpha \vec{\mathbf{I}} - \rho_\alpha \mathbf{F}_\alpha^{(1)} - q_\alpha n_\alpha[\mathbf{v}_0 \times \mathbf{B}]\right)\right]\right\} =$$
$$= \int\left(\frac{m_\alpha v_\alpha^2}{2} + \varepsilon_\alpha\right) J_\alpha^{st,el} d\mathbf{v}_\alpha + \int\left(\frac{m_\alpha v_\alpha^2}{2} + \varepsilon_\alpha\right) J_\alpha^{st,inel} d\mathbf{v}_\alpha.$$

Energy equation for mixture:
$$\frac{\partial}{\partial t}\left\{\frac{\rho v_0^2}{2} + \frac{3}{2} p + \sum_\alpha \varepsilon_\alpha n_\alpha - \sum_\alpha \tau_\alpha^{(0)}\left[\frac{\partial}{\partial t}\left(\frac{\rho_\alpha v_0^2}{2} + \frac{3}{2} p_\alpha + \varepsilon_\alpha n_\alpha\right) + \right.\right.$$
$$\left.\left. + \frac{\partial}{\partial \mathbf{r}} \cdot \left(\frac{1}{2}\rho_\alpha v_0^2 \mathbf{v}_0 + \frac{5}{2} p_\alpha \mathbf{v}_0 + \varepsilon_\alpha n_\alpha \mathbf{v}_0\right) - \mathbf{F}_\alpha^{(1)} \cdot \rho_\alpha \mathbf{v}_0\right]\right\} +$$
$$+ \frac{\partial}{\partial \mathbf{r}} \cdot \left\{\frac{1}{2}\rho v_0^2 \mathbf{v}_0 + \frac{5}{2} p \mathbf{v}_0 + \mathbf{v}_0 \sum_\alpha \varepsilon_\alpha n_\alpha - \sum_\alpha \tau_\alpha^{(0)}\left[\frac{\partial}{\partial t}\left(\frac{1}{2}\rho_\alpha v_0^2 \mathbf{v}_0 + \right.\right.\right.$$
$$\left. + \frac{5}{2} p_\alpha \mathbf{v}_0 + \varepsilon_\alpha n_\alpha \mathbf{v}_0\right) + \frac{\partial}{\partial \mathbf{r}} \cdot \left(\frac{1}{2}\rho_\alpha v_0^2 \mathbf{v}_0 \mathbf{v}_0 + \frac{7}{2} p_\alpha \mathbf{v}_0 \mathbf{v}_0 + \frac{1}{2} p_\alpha v_0^2 \vec{\mathbf{I}} + \right.$$
$$\left. + \frac{5}{2}\frac{p_\alpha^2}{\rho_\alpha}\vec{\mathbf{I}} + \varepsilon_\alpha n_\alpha \mathbf{v}_0 \mathbf{v}_0 + \varepsilon_\alpha \frac{p_\alpha}{m_\alpha}\vec{\mathbf{I}}\right) - \rho_\alpha \mathbf{F}_\alpha^{(1)} \cdot \mathbf{v}_0 \mathbf{v}_0 - p_\alpha \mathbf{F}_\alpha^{(1)} \cdot \vec{\mathbf{I}} - \quad (6.5)$$
$$- \frac{1}{2}\rho_\alpha v_0^2 \mathbf{F}_\alpha^{(1)} - \frac{3}{2}\mathbf{F}_\alpha^{(1)} p_\alpha - \frac{\rho_\alpha v_0^2}{2}\frac{q_\alpha}{m_\alpha}[\mathbf{v}_0 \times \mathbf{B}] - \frac{5}{2} p_\alpha \frac{q_\alpha}{m_\alpha}[\mathbf{v}_0 \times \mathbf{B}] -$$
$$\left.\left. - \varepsilon_\alpha n_\alpha \frac{q_\alpha}{m_\alpha}[\mathbf{v}_0 \times \mathbf{B}] - \varepsilon_\alpha n_\alpha \mathbf{F}_\alpha^{(1)}\right]\right\} - \left\{\mathbf{v}_0 \cdot \sum_\alpha \rho_\alpha \mathbf{F}_\alpha^{(1)} - \right.$$
$$\left. - \sum_\alpha \tau_\alpha^{(0)}\left[\mathbf{F}_\alpha^{(1)} \cdot \left(\frac{\partial}{\partial t}(\rho_\alpha \mathbf{v}_0) + \frac{\partial}{\partial \mathbf{r}} \cdot \rho_\alpha \mathbf{v}_0 \mathbf{v}_0 + \frac{\partial}{\partial \mathbf{r}} \cdot p_\alpha \vec{\mathbf{I}} - \rho_\alpha \mathbf{F}_\alpha^{(1)} - q_\alpha n_\alpha[\mathbf{v}_0 \times \mathbf{B}]\right)\right]\right\} = 0.$$

Here $\mathbf{F}_\alpha^{(1)}$ are the forces of the non-magnetic origin, $\mathbf{B}$ - magnetic induction, $\vec{\mathbf{I}}$ - unit tensor, $q_\alpha$ - charge of the $\alpha$-component particle, $p_\alpha$ - static pressure for $\alpha$-component, $\varepsilon_\alpha$ - internal energy for the particles of $\alpha$-component, $\mathbf{v}_0$ - hydrodynamic velocity for mixture.

In the following we intend to obtain the soliton's type of solution of the generalized hydrodynamic equations for plasma in the self consistent electrical field. All elements of possible formation like quantum soliton should move with the same translational velocity. Then the system of GHE (see also (6.1) – (6.5)) consists from the generalized Poisson equation reflecting the effects of the charge and the charge flux perturbations, two continuity equations for positive and negative species (in particular, for ion and electron components), one motion equation for mixture and two energy equations for ion and electron components. This system of six equations for non-stationary 1D case can be written in the form [19, 21-24]:

(Poisson equation)
$$\frac{\partial^2 \varphi}{\partial x^2} = -4\pi e \left\{ \left[ n_i - \tau_i \left( \frac{\partial n_i}{\partial t} + \frac{\partial}{\partial x}(n_i u) \right) \right] - \left[ n_e - \tau_e \left( \frac{\partial n_e}{\partial t} + \frac{\partial}{\partial x}(n_e u) \right) \right] \right\}, \qquad (6.6)$$

(continuity equation for positive ion component)
$$\frac{\partial}{\partial t}\left\{ \rho_i - \tau_i \left[ \frac{\partial \rho_i}{\partial t} + \frac{\partial}{\partial x}(\rho_i u) \right] \right\} + \frac{\partial}{\partial x}\left\{ \rho_i u - \tau_i \left[ \frac{\partial}{\partial t}(\rho_i u) + \frac{\partial}{\partial x}(\rho_i u^2) + \frac{\partial p_i}{\partial x} - \rho_i F_i \right] \right\}, \qquad (6.7)$$

(continuity equation for electron component)
$$\frac{\partial}{\partial t}\left\{ \rho_e - \tau_e \left[ \frac{\partial \rho_e}{\partial t} + \frac{\partial}{\partial x}(\rho_e u) \right] \right\} + \frac{\partial}{\partial x}\left\{ \rho_e u - \tau_e \left[ \frac{\partial}{\partial t}(\rho_e u) + \frac{\partial}{\partial x}(\rho_e u^2) + \frac{\partial p_e}{\partial x} - \rho_e F_e \right] \right\}, \qquad (6.8)$$

(momentum equation)
$$\frac{\partial}{\partial t}\left\{ \rho u - \tau_i \left[ \frac{\partial}{\partial t}(\rho_i u) + \frac{\partial}{\partial x}(p_i + \rho_i u^2) - \rho_i F_i \right] - \tau_e \left[ \frac{\partial}{\partial t}(\rho_e u) + \frac{\partial}{\partial x}(p_e + \rho_e u^2) - \rho_e F_e \right] \right\} -$$
$$- \rho_i F_i - \rho_e F_e + F_i \tau_i \left( \frac{\partial \rho_i}{\partial t} + \frac{\partial}{\partial x}(\rho_i u) \right) + F_e \tau_e \left( \frac{\partial \rho_e}{\partial t} + \frac{\partial}{\partial x}(\rho_e u) \right) + \qquad (6.9)$$
$$+ \frac{\partial}{\partial x}\left\{ \begin{array}{l} \rho u^2 + p - \tau_i \left[ \frac{\partial}{\partial t}(\rho_i u^2 + p_i) + \frac{\partial}{\partial x}(\rho_i u^3 + 3p_i u) - 2\rho_i u F_i \right] - \\ - \tau_e \left[ \frac{\partial}{\partial t}(\rho_e u^2 + p_e) + \frac{\partial}{\partial x}(\rho_e u^3 + 3p_e u) \right] - 2\rho_e u F_e \end{array} \right\} = 0,$$

(energy equation for positive ion component)
$$\frac{\partial}{\partial t}\left\{ \rho_i u^2 + 3p_i - \tau_i \left[ \frac{\partial}{\partial t}(\rho_i u^2 + 3p_i) + \frac{\partial}{\partial x}(\rho_i u^3 + 5p_i u) - 2\rho_i F_i u \right] \right\} +$$
$$+ \frac{\partial}{\partial x}\left\{ \rho_i u^3 + 5p_i u - \tau_i \left[ \frac{\partial}{\partial t}(\rho_i u^3 + 5p_i u) + \frac{\partial}{\partial x}\left( \rho_i u^4 + 8p_i u^2 + 5\frac{p_i^2}{\rho_i} \right) - F_i(3\rho_i u^2 + 5p_i) \right] \right\} \quad (6.10)$$
$$- 2u\rho_i F_i + 2\tau_i F_i \left[ \frac{\partial}{\partial t}(\rho_i u) + \frac{\partial}{\partial x}(\rho_i u^2 + p_i) - \rho_i F_i \right] = -\frac{p_i - p_e}{\tau_{ei}},$$

(energy equation for electron component)

$$\frac{\partial}{\partial t}\left\{\rho_e u^2 + 3p_e - \tau_e\left[\frac{\partial}{\partial t}(\rho_e u^2 + 3p_e) + \frac{\partial}{\partial x}(\rho_e u^3 + 5p_e u) - 2\rho_e F_e u\right]\right\} +$$

$$+ \frac{\partial}{\partial x}\left\{\rho_e u^3 + 5p_e u - \tau_e\left[\frac{\partial}{\partial t}(\rho_e u^3 + 5p_e u) + \frac{\partial}{\partial x}\left(\rho_e u^4 + 8p_e u^2 + 5\frac{p_e^2}{\rho_e}\right) - F_e(3\rho_e u^2 + 5p_e)\right]\right\} \quad (6.11)$$

$$- 2u\rho_e F_e + 2\tau_e F_e\left[\frac{\partial}{\partial t}(\rho_e u) + \frac{\partial}{\partial x}(\rho_e u^2 + p_e) - \rho_e F_e\right] = -\frac{p_e - p_i}{\tau_{ei}},$$

where $u$ is translational velocity of the quantum object, $\varphi$ - scalar potential, $n_i$ and $n_e$ are the number density of the charged species, $F_i$ and $F_e$ are the forces acting on the unit mass of ion and electron. Approximations for the non-local parameters $\tau_i$, $\tau_e$ and $\tau_{ei}$ need the special consideration. In the following for the $\tau_i$ and $\tau_i$ approximation the relation (2.11) is used in the forms

$$\tau_i = H/m_i u^2, \quad \tau_e = H/m_e u^2. \qquad (6.12)$$

For non-local parameter of electron-ion interaction $\tau_{ei}$ is applicable the relation

$$\frac{1}{\tau_{ei}} = \frac{1}{\tau_e} + \frac{1}{\tau_i}. \qquad (6.13)$$

In this case parameter $\tau_{ei}$ serves as relaxation time in the process of the particle interaction of different kinds. Transformation (6.13) for the case $H = \hbar$ leads to the obvious compatibility with the Heisenberg principle

$$\frac{1}{\tau_{ei}} = \frac{\tau_e + \tau_i}{\tau_e \tau_i} = \frac{u^2}{\hbar}(m_e + m_i). \qquad (6.14)$$

Equality (6.14) is the consequence of the "time-energy" uncertainty relation for the combined particle with mass $m_i + m_e$. On principal the time values $\tau_i$ and $\tau_e$ should be considered as sums of mean times between collisions ($\tau_i^{tr}, \tau_e^{tr}$) and discussed above non-local quantum values ($\tau_i^{qu}, \tau_e^{qu}$), namely for example

$$\tau_i = \tau_i^{tr} + \tau_i^{qu}. \qquad (6.15)$$

For molecular hydrogen in standard conditions mean time between collisions is equal to $6.6 \cdot 10^{-11}$ s. For quantum objects moving with velocities typical for plasmoids or lightning balls $\tau^{qu}$ is much more than $\tau^{tr}$ and the usual static pressure $p$ transforms in the pressure which can be named as the rest non-local pressure. In the definite sense this kind of pressure can be considered as analogue of the Bose condensate pressure.

**7. Quantum solitons in the self consistent electric field.**

Let us introduce the coordinate system moving along the positive direction of $x$- axis in ID space with velocity $C = u_0$ equal to phase velocity of considering quantum object $\xi = x - Ct$, (see also (4.9)). As before taking into account the De Broglie relation we should wait that the group velocity $u_g$ is equal to $2u_0$. In moving coordinate system all dependent hydrodynamic values are function of $(\xi, t)$. We investigate the possibility of the quantum object formation of the soliton type. For this solution there is no explicit dependence on time for coordinate system moving with the phase velocity $u_0$. Write down the system of equations (6.6) - (6.11) for the two component mixture of charged particles without taking into account the component's internal energy in the dimensionless form, where dimensionless symbols are marked by tildes.

The following formulae are valid for acting forces

$$F_i = -\frac{e}{m_i}\frac{\partial \varphi}{\partial x}, \quad F_e = \frac{e}{m_e}\frac{\partial \varphi}{\partial x}. \tag{7.1}$$

Comments to the following system of six dimensionless ordinary non-linear equations (7.2) – (7.7):

1. The following scales are introduced in (7.2) – (7.7)

$$u_0, \quad x_0 = \frac{\hbar}{m_e}\frac{1}{u_0}, \quad \varphi_0 = \frac{m_e}{e}u_0^2, \quad \rho_0 = \frac{m_e^4}{4\pi\hbar^2 e^2}u_0^4, \quad p_0 = \rho_0 u_0^2 = \frac{m_e^4}{4\pi\hbar^2 e^2}u_0^6.$$

From the introduced scales only two parameters are independent – the phase velocity $u_0$ of the quantum object, and the external parameter $H$, which is proportional to Plank constant $\hbar$ and in general case should be inserted in the scale relation as $x_0 = H/(m_e u_0) = n\hbar/(m_e u_0)$. It leads to exchange in all scales $\hbar \leftrightarrow H$. The value $v^{qu} = \hbar/m_e$ is quantum viscosity. Of course on principal the electron mass can be replaced in scales by mass of other particles with the negative charge. From this point of view the obtained solutions have the universal character defined only by Cauchy conditions.

2. Every equation from the system is of the second order and needs two conditions. The problem belongs to the class of Cauchy problems.

3. In comparison with the Schrödinger theory connected with the behavior of the wave function, no special conditions are applied for the dependent variables including the domain of the solution existing. This domain is defined automatically in the process of the numerical solution of the concrete variant of calculations.

$$\frac{\partial^2 \tilde{\varphi}}{\partial \tilde{\xi}^2} = -\left\{\frac{m_e}{m_i}\left[\tilde{\rho}_i - \frac{1}{\tilde{u}^2}\frac{m_e}{m_i}\left(-\frac{\partial \tilde{\rho}_i}{\partial \tilde{\xi}} + \frac{\partial}{\partial \tilde{\xi}}(\tilde{\rho}_i \tilde{u})\right)\right] - \left[\tilde{\rho}_e - \frac{1}{\tilde{u}^2}\left(-\frac{\partial \tilde{\rho}_e}{\partial \tilde{\xi}} + \frac{\partial}{\partial \tilde{\xi}}(\tilde{\rho}_e \tilde{u})\right)\right]\right\}, \tag{7.2}$$

$$\frac{\partial \tilde{\rho}_i}{\partial \tilde{\xi}} - \frac{\partial \tilde{\rho}_i \tilde{u}}{\partial \tilde{\xi}} + \frac{m_e}{m_i}\frac{\partial}{\partial \tilde{\xi}}\left\{\frac{1}{\tilde{u}^2}\left[\frac{\partial}{\partial \tilde{\xi}}\left(\tilde{p}_i + \tilde{\rho}_i + \tilde{\rho}_i \tilde{u}^2 - 2\tilde{\rho}_i \tilde{u}_i\right) + \frac{m_e}{m_i}\tilde{\rho}_i\frac{\partial \tilde{\varphi}}{\partial \tilde{\xi}}\right]\right\} = 0, \tag{7.3}$$

$$\frac{\partial \tilde{\rho}_e}{\partial \tilde{\xi}} - \frac{\partial \tilde{\rho}_e \tilde{u}}{\partial \tilde{\xi}} + \frac{\partial}{\partial \tilde{\xi}}\left\{\frac{1}{\tilde{u}^2}\left[\frac{\partial}{\partial \tilde{\xi}}\left(\tilde{p}_e + \tilde{\rho}_e + \tilde{\rho}_e \tilde{u}^2 - 2\tilde{\rho}_e \tilde{u}_e\right) - \tilde{\rho}_e\frac{\partial \tilde{\varphi}}{\partial \tilde{\xi}}\right]\right\} = 0, \tag{7.4}$$

$$\frac{\partial}{\partial \tilde{\xi}}\left\{(\tilde{\rho}_i + \tilde{\rho}_e)\tilde{u}^2 + (\tilde{p}_i + \tilde{p}_e) - (\tilde{\rho}_i + \tilde{\rho}_e)\tilde{u}\right\} +$$

$$+ \frac{\partial}{\partial \tilde{\xi}}\left\{\begin{array}{l}\frac{1}{\tilde{u}^2}\frac{m_e}{m_i}\left[\frac{\partial}{\partial \tilde{\xi}}\left(2\tilde{p}_i + 2\tilde{\rho}_i \tilde{u}^2 - \tilde{\rho}_i \tilde{u} - \tilde{\rho}_i \tilde{u}^3 - 3\tilde{p}_i \tilde{u}\right) + \tilde{\rho}_i \frac{m_e}{m_i}\frac{\partial \tilde{\varphi}}{\partial \tilde{\xi}}\right] + \\ + \frac{1}{\tilde{u}^2}\left[\frac{\partial}{\partial \tilde{\xi}}\left(2\tilde{p}_e + 2\tilde{\rho}_e \tilde{u}^2 - \tilde{\rho}_e \tilde{u} - \tilde{\rho}_e \tilde{u}^3 - 3\tilde{p}_e \tilde{u}\right) - \tilde{\rho}_e \frac{\partial \tilde{\varphi}}{\partial \tilde{\xi}}\right]\end{array}\right\} +$$

$$+ \tilde{\rho}_i \frac{m_e}{m_i}\frac{\partial \tilde{\varphi}}{\partial \tilde{\xi}} - \tilde{\rho}_e \frac{\partial \tilde{\varphi}}{\partial \tilde{\xi}} - \frac{\partial \tilde{\varphi}}{\partial \tilde{\xi}}\frac{1}{\tilde{u}^2}\left(\frac{m_e}{m_i}\right)^2\left(-\frac{\partial \tilde{\rho}_i}{\partial \tilde{\xi}} + \frac{\partial}{\partial \tilde{\xi}}(\tilde{\rho}_i \tilde{u})\right) +$$

$$+ \frac{\partial \tilde{\varphi}}{\partial \tilde{\xi}}\frac{1}{\tilde{u}^2}\left(-\frac{\partial \tilde{\rho}_e}{\partial \tilde{\xi}} + \frac{\partial}{\partial \tilde{\xi}}(\tilde{\rho}_e \tilde{u})\right) - 2\frac{\partial}{\partial \tilde{\xi}}\left\{\frac{1}{\tilde{u}}\frac{\partial \tilde{\varphi}}{\partial \tilde{\xi}}\left[\left(\frac{m_e}{m_i}\right)^2 \tilde{\rho}_i - \tilde{\rho}_e\right]\right\} = 0, \tag{7.5}$$

$$\frac{\partial}{\partial \tilde{\xi}}\left\{\tilde{\rho}_i\tilde{u}^3 + 5\tilde{p}_i\tilde{u} - \tilde{\rho}_i\tilde{u}^2 - 3\tilde{p}_i\right\} +$$

$$+\frac{\partial}{\partial \tilde{\xi}}\left\{\frac{1}{\tilde{u}^2}\frac{m_e}{m_i}\left[\begin{array}{l}\frac{\partial}{\partial \tilde{\xi}}\left(2\tilde{\rho}_i\tilde{u}^3 + 10\tilde{p}_i\tilde{u} - \tilde{\rho}_i\tilde{u}^4 - 8\tilde{p}_i\tilde{u}^2 - 5\frac{\tilde{p}_i^2}{\tilde{\rho}_i} - \tilde{\rho}_i\tilde{u}^2 - 3\tilde{p}_i\right) + \\ +\frac{m_e}{m_i}\frac{\partial \tilde{\varphi}}{\partial \tilde{\xi}}\left(2\tilde{\rho}_i\tilde{u} - 3\tilde{\rho}_i\tilde{u}^2 - 5\tilde{p}_i\right)\end{array}\right]\right\}$$

$$+2\frac{m_e}{m_i}\tilde{\rho}_i\tilde{u}\frac{\partial \tilde{\varphi}}{\partial \tilde{\xi}} -$$

$$-2\frac{\partial \tilde{\varphi}}{\partial \tilde{\xi}}\frac{1}{\tilde{u}^2}\left(\frac{m_e}{m_i}\right)^2\left[\frac{\partial}{\partial \tilde{\xi}}\left(\tilde{\rho}_i\tilde{u}^2 + \tilde{p}_i - \tilde{\rho}_i\tilde{u}\right) + \tilde{\rho}_i\frac{m_e}{m_i}\frac{\partial \tilde{\varphi}}{\partial \tilde{\xi}}\right] = -(\tilde{p}_i - \tilde{p}_e)\tilde{u}^2\left(1 + \frac{m_i}{m_e}\right).$$

(7.6)

$$\frac{\partial}{\partial \tilde{\xi}}\left(\tilde{\rho}_e\tilde{u}^3 + 5\tilde{p}_e\tilde{u} - \tilde{\rho}_e\tilde{u}^2 - 3\tilde{p}_e\right) - 2\tilde{\rho}_e\tilde{u}\frac{\partial \tilde{\varphi}}{\partial \tilde{\xi}} +$$

$$+\frac{\partial}{\partial \tilde{\xi}}\left\{\frac{1}{\tilde{u}^2}\left[\begin{array}{l}\frac{\partial}{\partial \tilde{\xi}}\left(2\tilde{\rho}_e\tilde{u}^3 + 10\tilde{p}_e\tilde{u} - \tilde{\rho}_e\tilde{u}^4 - 8\tilde{p}_e\tilde{u}^2 - 5\frac{\tilde{p}_e^2}{\tilde{\rho}_e} - \tilde{\rho}_e\tilde{u}^2 - 3\tilde{p}_e\right) + \\ +\frac{\partial \tilde{\varphi}}{\partial \tilde{\xi}}\left(3\tilde{\rho}_e\tilde{u}^2 + 5\tilde{p}_e - 2\tilde{\rho}_e\tilde{u}\right)\end{array}\right]\right\} +$$

(7.7)

$$+2\frac{\partial \tilde{\varphi}}{\partial \tilde{\xi}}\frac{1}{\tilde{u}^2}\left[\frac{\partial}{\partial \tilde{\xi}}\left(\tilde{\rho}_e\tilde{u}^2 + \tilde{p}_e - \tilde{\rho}_e\tilde{u}\right) - \tilde{\rho}_e\frac{\partial \tilde{\varphi}}{\partial \tilde{\xi}}\right] = -(\tilde{p}_e - \tilde{p}_i)\left(1 + \frac{m_i}{m_e}\right)\tilde{u}^2.$$

## 8. Some results of mathematical modeling.

The system of generalized quantum hydrodynamic equations (7.2) – (7.7) have the great possibilities of mathematical modeling as result of changing of twelve Cauchy conditions describing the character features of initial perturbations which lead to the soliton formation. The following figures reflect some results of calculations realized according to the system of equations (7.2) - (7.7) with the help of Maple 9. The following notations on figures are used: r-density $\tilde{\rho}_i$, s- density $\tilde{\rho}_e$, u- velocity $\tilde{u}$, p - pressure $\tilde{p}_i$, q - pressure $\tilde{p}_e$ and v - self consistent potential $\tilde{\varphi}$. Explanations placed under all following figures, Maple program contains Maple's notations – for example the expression $D(u)(0) = 0$ means in usual notations $(\partial \tilde{u}/\partial \tilde{\xi})(0) = 0$, independent variable $t$ responds to $\tilde{\xi}$. We begin with investigation of the problem of principle significance – is it possible after a perturbation (defined by Cauchy conditions) to obtain the quantum object of the soliton's kind as result of the self-organization of ionized matter? In the case of the positive answer, what is the origin of existence of this stable object? By the way the mentioned questions belong to the typical problem in the theory of the ball lightning. With this aim let us consider the initial perturbations (SYSTEM 3):

```
v(0)=1,r(0)=1,s(0)=1,u(0)=1,p(0)=1,q(0)=.95,
D(v)(0)=0,D(r)(0)=0,D(s)(0)=0,D(u)(0)=0,D(p)(0)=0,D(q)(0)=0,
```

in the mixture of positive and negative ions of equal masses, the pressure $\tilde{p}_i(0)$ of positive particles is larger than $\tilde{p}_e(0)$ of the negative ones.

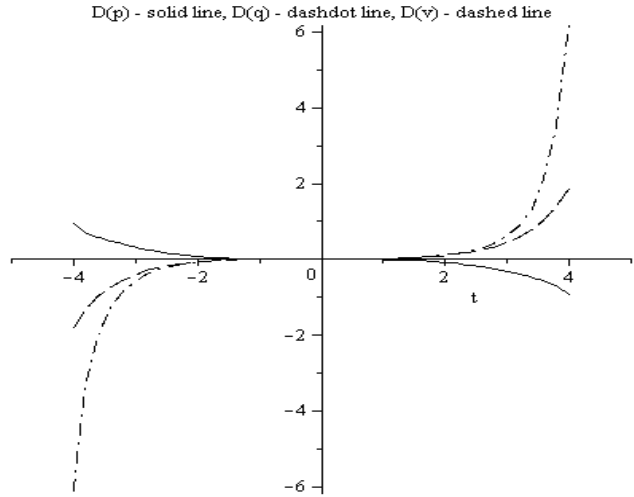

Fig. 41. The derivative of pressure of the positive component $\partial \tilde{p}_i / \partial \tilde{\xi}$ and negative component $\partial \tilde{p}_e / \partial \tilde{\xi}$, the derivative of the self-consistent potential $\partial \tilde{\varphi} / \partial \tilde{\xi}$ in quantum soliton.
The Figures 41 – 43 reflect the results of solution of Eqs. (7.1) – (7.7) for the case SYSTEM 3.

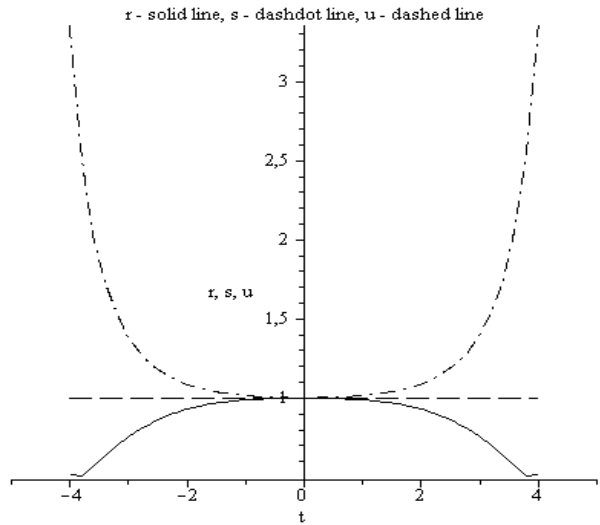 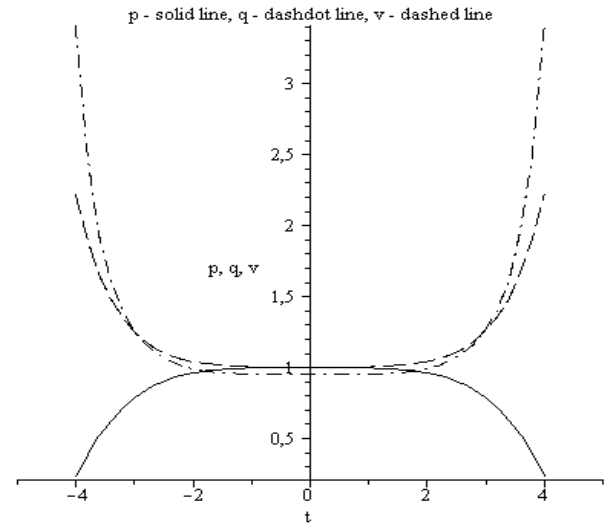

Fig. 42. r-density $\tilde{\rho}_i$, u-velocity $\tilde{u}$, s-density $\tilde{\rho}_e$ in quantum soliton.

Fig. 43. p - pressure $\tilde{p}_i$, q - pressure $\tilde{p}_e$, v - self consistent potential $\tilde{\varphi}$.

Fig. 42 displays the quantum object placed in bounded region of 1D space, all parts of this object are moving with the same velocity. Important to underline that no special boundary conditions were used for this and all following cases. Then this soliton is product of the self-organization of ionized matter. Fig. 41, 42 contain the answer for formulated above questions about stability of the object. Really the object is restricted by the negative shell. The derivative $\partial \tilde{\varphi} / \partial \tilde{\xi}$ is proportional to the self-consistent forces acting on the positive and negative parts of the soliton.

Consider for example the right side of soliton. The self consistent force of the electrical origin compresses the positive part of this soliton and provokes the movement of the negative part along the positive direction of the $\tilde{\xi}$ - axis ($t$ – axis in the accepted nomination). But the increasing of quantum pressure prevent to the destruction of soliton. Therefore the stability of the quantum object is result of the self-consistent influence of electric potential and quantum pressures. Interesting to notice that stability can be also achieved if soliton has *positive* shell and *negative* kernel but $\tilde{p}_i(0) < \tilde{p}_e(0)$.

Fig. 44, 45 display the typical quantum object placed in bounded region of 1D space, all parts of this object are moving with the same velocity. Figures 44, 45 reflect the following Cauchy conditions (written in Maple notations), SYSTEM 4:

```
v(0)=1,r(0)=1,s(0)=1/1838,u(0)=1,p(0)=1,q(0)=0.999, (L=1,T=1838)
D(v)(0)=0,D(r)(0)=0,D(s)(0)=0,D(u)(0)=0,D(p)(0)=0,D(q)(0)=0.
```

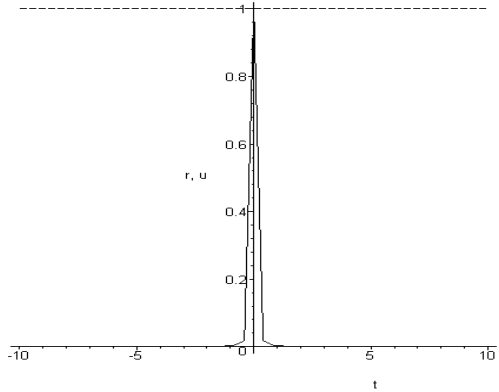 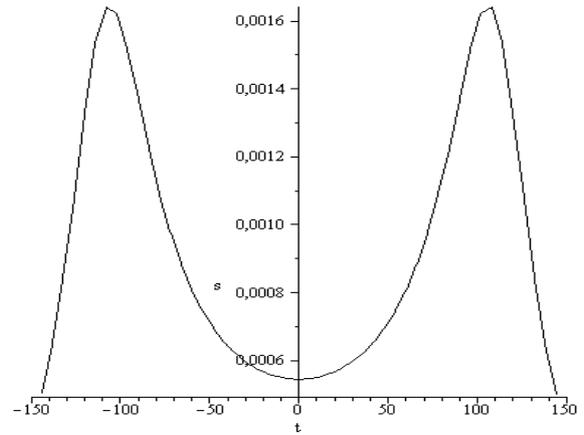

Fig. 44. r- density $\tilde{\rho}_i$, u- velocity $\tilde{u}$      Fig. 45. s- density $\tilde{\rho}_e$ in quantum soliton.

From calculations follow that increasing the difference $p_i(0) - p_e(0)$ lead to diminishing of the character domain occupied by soliton.

The classical construction with the positive kernel and negative shell is existing if $\tilde{p}_i(0) > \tilde{p}_e(0)$. In the opposite case $\tilde{p}_i(0) < \tilde{p}_e(0)$ mathematical modeling leads to the construction with the negative kernel and the positive shell for soliton. Moreover, calculations demonstrate the possibility of the soliton formations with very significant difference from the used scales. It means that lightning ball can be considered as "macroscopic atom-soliton". The previous calculations were performed for 1D Cartesian coordinate system.

Let us demonstrate the results of calculations (SYSTEM 4) in spherical coordinate system using the condition of spherical symmetry (Figs. 26 – 29, here $t$ is the dimensionless radial distance from the soliton's center). Obviously the character features of curves are not changed, but electron shell and nuclei are separated more significantly (compare the right side of Fig. 45 and Fig. 47).

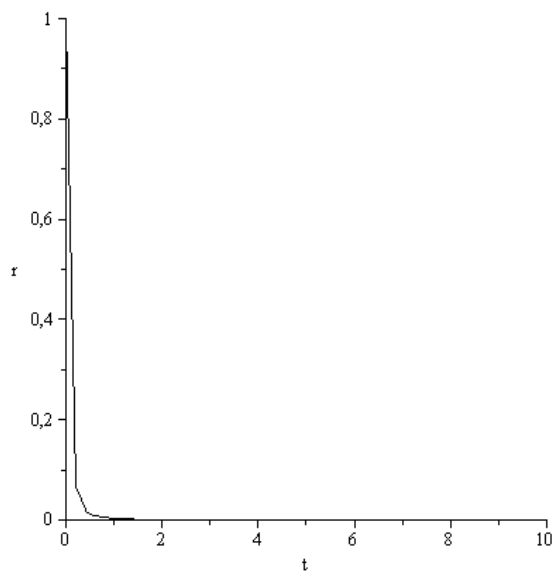 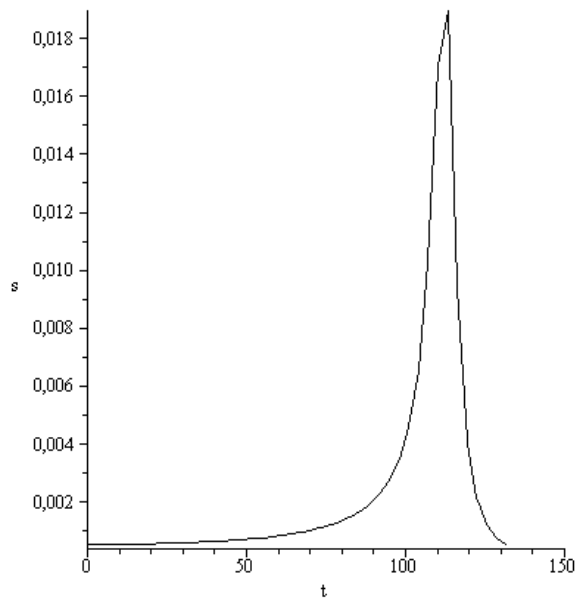

Fig. 46. r- density $\tilde{\rho}_i$ in quantum soliton.      Fig. 47. s- density $\tilde{\rho}_e$ in quantum soliton.

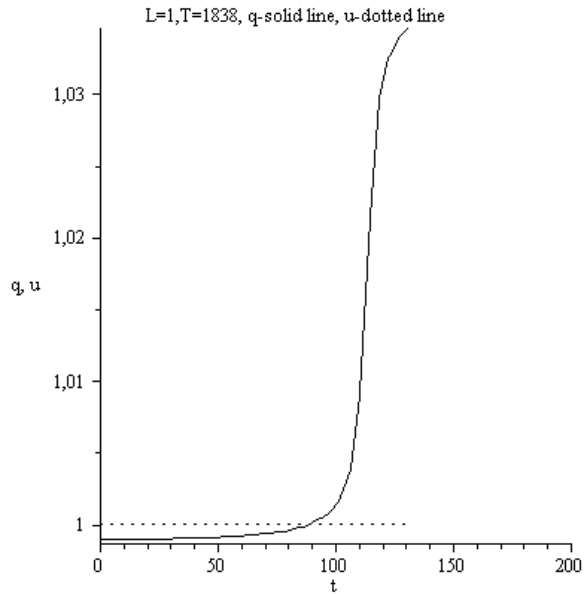 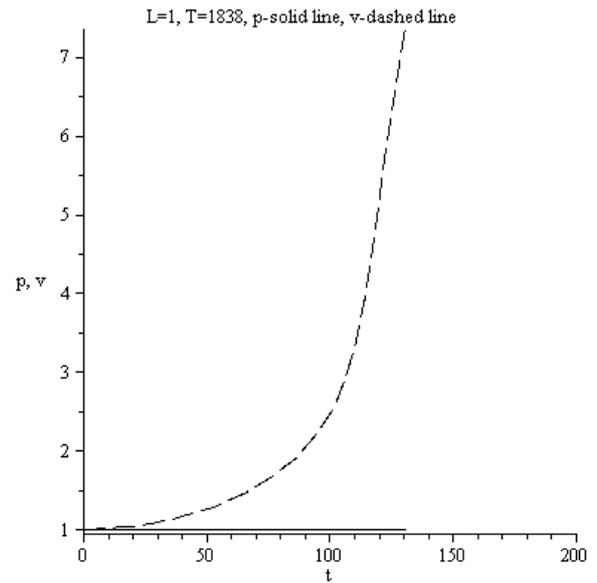

Fig. 48. $q$ - pressure $\tilde{p}_e$, u- velocity $\tilde{u}$ along the direction of motion of quantum soliton.

Fig. 49. p - pressure $\tilde{p}_i$, $v$ - self-consistent potential $\tilde{\varphi}$.

**Conclusion**

The problem of antigravitation, dark energy, dark matter reflects the crisis, crash of local physics on the whole and the local transport kinetic theory in particular. The origin of difficulties consists in total Oversimplification following from principles of local physics and reflects the general shortcomings of the local kinetic transport theory.

In other words the problem of Oversimplification is not the "trivial" simplification of the important problem. The situation is much more serious – total Oversimplification based on principles of local statistical physics, and obvious crisis, we see in astrophysics, simply reflects the general shortcomings of the local kinetic transport theory.

The unified generalized non-local quantum hydrodynamic theory is applied for mathematical modeling of objects in giant scale diapason from the Universe and galaxy scales to the atom structure. For the case of galaxies the theory leads to the flat rotation curves known from observations. The transformation of Kepler's regime into flat rotation curves for different solitons is shown. Peculiar features of the rotational speeds of galaxies and effects of the Hubble expansion need not in introduction of new essence like dark matter and dark energy.

In modeling of atom structure the solitons have the character of quantum objects (with the separated *negative or positive* kernel and *positive or negative* shells) if the initial rest pressures of non-local origin for the positive and negative components are not equal each other, and objects reach stability as result of equalizing of the corresponding pressure of the non-local origin and the self-consistent electric forces. These effects can be considered also as explanation of the existence of atom structures and lightning balls. The usual Schrödinger' quantum mechanics cannot be useful in this situation, because Schrödinger – Madelung quantum theory does not contain the energy equation on principal.

The aim of investigations undertaken by me here and before, consists in creation of the unified non-local theory of transport processes. It is not a new idea that physics is the unified construction but not the collection of the inconsistent facts. I hope this paper demonstrate the validity of this conception.

**Appendix. Non – local hydrodynamic system of equations describing the gravitational soliton motion along x-axis.**

Motion equation (z – projection):

$$\frac{\partial}{\partial \tilde{\xi}} (\tilde{\rho} \tilde{w} (\tilde{u} - 1)) + \frac{\partial}{\partial \tilde{y}} (\tilde{\rho} \tilde{v} \tilde{w}) + \frac{\partial}{\partial \tilde{z}} (\tilde{p} + \tilde{\rho} \tilde{w}^2) + \tilde{\rho} \frac{\partial \tilde{\Psi}}{\partial \tilde{z}} +$$

$$+ \frac{\partial}{\partial \tilde{\xi}} \left\{ \tilde{\tau} \left[ \begin{array}{l} -\frac{\partial}{\partial \tilde{\xi}} (\tilde{\rho} \tilde{w} (1 - \tilde{u})^2) + \frac{\partial}{\partial \tilde{y}} (\tilde{\rho} \tilde{v} \tilde{w} (1 - \tilde{u})) + \\ + \frac{\partial}{\partial \tilde{z}} (\tilde{\rho} \tilde{w}^2 (1 - \tilde{u})) + \frac{\partial \tilde{p}}{\partial \tilde{z}} + \tilde{\rho} \frac{\partial \tilde{\Psi}}{\partial \tilde{z}} (1 - \tilde{u}) - \frac{\partial}{\partial \tilde{\xi}} (\tilde{p} \tilde{w}) - \frac{\partial \tilde{\Psi}}{\partial \tilde{\xi}} \tilde{\rho} \tilde{w} \end{array} \right] \right\} -$$

$$- \tilde{\tau} \frac{\partial \tilde{\Psi}}{\partial \tilde{z}} \left( \frac{\partial}{\partial \tilde{\xi}} (\tilde{\rho} (\tilde{u} - 1)) + \frac{\partial}{\partial \tilde{y}} (\tilde{\rho} \tilde{v}) + \frac{\partial}{\partial \tilde{z}} (\tilde{\rho} \tilde{w}) \right) +$$

$$+ \frac{\partial}{\partial \tilde{y}} \left\{ \tilde{\tau} \frac{\partial}{\partial \tilde{\xi}} (\tilde{\rho} \tilde{v} \tilde{w} (1 - \tilde{u})) \right\} + \frac{\partial}{\partial \tilde{z}} \left\{ \tilde{\tau} \frac{\partial}{\partial \tilde{\xi}} (\tilde{\rho} \tilde{w}^2 (1 - \tilde{u})) \right\} +$$

$$+ \frac{\partial}{\partial \tilde{z}} \left\{ \tilde{\tau} \frac{\partial}{\partial \tilde{\xi}} (\tilde{p} (1 - \tilde{u})) \right\} -$$

$$- \frac{\partial}{\partial \tilde{y}} \left\{ \tilde{\tau} \left[ \frac{\partial}{\partial \tilde{y}} (\tilde{\rho} \tilde{v}^2 \tilde{w}) + \frac{\partial}{\partial \tilde{z}} (\tilde{\rho} \tilde{v} \tilde{w}^2) \right] \right\} - \frac{\partial}{\partial \tilde{z}} \left\{ \tilde{\tau} \left[ \frac{\partial}{\partial \tilde{y}} (\tilde{\rho} \tilde{v} \tilde{w}^2) + \frac{\partial}{\partial \tilde{z}} (\tilde{\rho} \tilde{w}^3) \right] \right\} -$$

$$- \frac{\partial}{\partial \tilde{z}} \left\{ \tilde{\tau} \left[ \frac{\partial}{\partial \tilde{\xi}} (\tilde{p} \tilde{u}) \right] \right\} - 2 \frac{\partial}{\partial \tilde{z}} \left\{ \tilde{\tau} \left[ \frac{\partial}{\partial \tilde{y}} (\tilde{p} \tilde{v}) \right] \right\} - 3 \frac{\partial}{\partial \tilde{z}} \left\{ \tilde{\tau} \left[ \frac{\partial}{\partial \tilde{z}} (\tilde{p} \hat{w}) \right] \right\} -$$

$$- \frac{\partial}{\partial \tilde{y}} \left\{ \tilde{\tau} \frac{\partial}{\partial \tilde{y}} (\tilde{p} \tilde{w}) \right\} -$$

$$- \frac{\partial}{\partial \tilde{y}} \left\{ \tilde{\tau} \left[ \frac{\partial \tilde{\Psi}}{\partial \tilde{y}} \tilde{\rho} \tilde{w} \right] \right\} - 2 \frac{\partial}{\partial \tilde{z}} \left\{ \tilde{\tau} \left[ \frac{\partial \tilde{\Psi}}{\partial \tilde{z}} \tilde{\rho} \tilde{w} \right] \right\} - \frac{\partial}{\partial \tilde{y}} \left\{ \tilde{\tau} \frac{\partial \tilde{\Psi}}{\partial \tilde{z}} \tilde{\rho} \tilde{v} \right\} = 0,$$

Motion equation (y – projection):

$$\frac{\partial}{\partial \tilde{\xi}}(\tilde{\rho}\tilde{v}(\tilde{u}-1)) + \frac{\partial}{\partial \tilde{y}}(\tilde{p}+\tilde{\rho}\tilde{v}^2) + \frac{\partial}{\partial \tilde{z}}(\tilde{\rho}\tilde{w}\tilde{v}) + \tilde{\rho}\frac{\partial \tilde{\Psi}}{\partial \tilde{y}} +$$

$$+\frac{\partial}{\partial \tilde{\xi}}\left\{\tilde{\tau}\left[\begin{array}{l}-\frac{\partial}{\partial \tilde{\xi}}(\tilde{\rho}\tilde{v}(1-\hat{u})^2) + \frac{\partial}{\partial \tilde{y}}(\tilde{\rho}\tilde{v}^2(1-\tilde{u})) + \frac{\partial}{\partial \tilde{z}}(\tilde{\rho}\tilde{w}\tilde{v}(1-\tilde{u})) + \\ +\frac{\partial \tilde{p}}{\partial \tilde{y}} + \tilde{\rho}\frac{\partial \hat{\Psi}}{\partial \tilde{y}}(1-\tilde{u}) - \frac{\partial}{\partial \tilde{\xi}}(\tilde{p}\tilde{v}) - \frac{\partial \tilde{\Psi}}{\partial \tilde{\xi}}\tilde{\rho}\tilde{v}\end{array}\right]\right\} -$$

$$-\tilde{\tau}\frac{\partial \tilde{\Psi}}{\partial \tilde{y}}\left(\frac{\partial}{\partial \tilde{\xi}}(\tilde{\rho}((\tilde{u}-1))) + \frac{\partial}{\partial \tilde{y}}(\tilde{\rho}\tilde{v}) + \frac{\partial}{\partial \tilde{z}}(\tilde{\rho}\tilde{w})\right) +$$

$$+\frac{\partial}{\partial \tilde{y}}\left\{\tilde{\tau}\frac{\partial}{\partial \tilde{\xi}}((\tilde{\rho}\tilde{v}^2+\tilde{p})(1-\tilde{u}))\right\} + \frac{\partial}{\partial \tilde{z}}\left\{\tilde{\tau}\frac{\partial}{\partial \tilde{\xi}}(\tilde{\rho}\tilde{w}\tilde{v}(1-\tilde{u}))\right\} -$$

$$-\frac{\partial}{\partial \tilde{y}}\left\{\tilde{\tau}\frac{\partial}{\partial \tilde{y}}(\tilde{\rho}\tilde{v}^3)\right\} - \frac{\partial}{\partial \tilde{z}}\left\{\tilde{\tau}\frac{\partial}{\partial \tilde{y}}(\tilde{\rho}\tilde{w}\tilde{v}^2)\right\} -$$

$$-\frac{\partial}{\partial \tilde{y}}\left\{\tilde{\tau}\frac{\partial}{\partial \tilde{z}}(\tilde{\rho}\tilde{v}^2\tilde{w})\right\} - \frac{\partial}{\partial \tilde{z}}\left\{\tilde{\tau}\frac{\partial}{\partial \tilde{z}}(\tilde{\rho}\tilde{w}^2\tilde{v})\right\} -$$

$$-\frac{\partial}{\partial \tilde{y}}\left\{\tilde{\tau}\frac{\partial}{\partial \tilde{\xi}}(\tilde{p}\tilde{u})\right\} - 3\frac{\partial}{\partial \tilde{y}}\left\{\tilde{\tau}\frac{\partial}{\partial \tilde{y}}(\tilde{p}\tilde{v})\right\} - 2\frac{\partial}{\partial \tilde{y}}\left\{\tilde{\tau}\frac{\partial}{\partial \tilde{z}}(\tilde{p}\tilde{w})\right\} -$$

$$-\frac{\partial}{\partial \tilde{z}}\left\{\tilde{\tau}\frac{\partial}{\partial \tilde{z}}(\tilde{p}\tilde{v})\right\} -$$

$$-2\frac{\partial}{\partial \tilde{y}}\left\{\tilde{\tau}\frac{\partial \tilde{\Psi}}{\partial \tilde{y}}\tilde{\rho}\tilde{v}\right\} - \frac{\partial}{\partial \tilde{z}}\left\{\tilde{\tau}\frac{\partial \tilde{\Psi}}{\partial \tilde{z}}\tilde{\rho}\tilde{v}\right\} - \frac{\partial}{\partial \tilde{z}}\left\{\tilde{\tau}\frac{\partial \tilde{\Psi}}{\partial \tilde{y}}\tilde{\rho}\tilde{w}\right\} = 0,$$

Continuity equation

$$\frac{\partial \tilde{\rho}}{\partial \tilde{\xi}} - \frac{\partial}{\partial \tilde{\xi}}(\tilde{\rho}\tilde{u}) +$$

$$+\frac{\partial}{\partial \tilde{\xi}}\left\{\tilde{\tau}\left[\frac{\partial}{\partial \tilde{\xi}}\left[\tilde{p}+\tilde{\rho}(\tilde{u}-1)^2\right] + \tilde{\rho}\frac{\partial \tilde{\Psi}}{\partial \tilde{\xi}} + \frac{\partial}{\partial \tilde{y}}[\tilde{\rho}\tilde{v}(\tilde{u}-1)] + \frac{\partial}{\partial \tilde{z}}[\tilde{\rho}\tilde{w}(\tilde{u}-1)]\right]\right\} -$$

$$-\frac{\partial}{\partial \tilde{y}}\left\{\tilde{\rho}\tilde{v} - \tilde{\tau}\left[\frac{\partial}{\partial \tilde{\xi}}[\tilde{\rho}\tilde{v}(\tilde{u}-1)] + \frac{\partial}{\partial \tilde{y}}(\tilde{p}+\tilde{\rho}\tilde{v}^2) + \frac{\partial}{\partial \tilde{z}}(\tilde{\rho}\tilde{w}\tilde{v}) + \tilde{\rho}\frac{\partial \tilde{\Psi}}{\partial \tilde{y}}\right]\right\} -$$

$$-\frac{\partial}{\partial \tilde{z}}\left\{\tilde{\rho}\tilde{w} - \tilde{\tau}\left[\frac{\partial}{\partial \tilde{\xi}}[\tilde{\rho}\tilde{w}(\tilde{u}-1)] + \frac{\partial}{\partial \tilde{z}}(\tilde{p}+\rho\tilde{w}^2) + \frac{\partial}{\partial \tilde{y}}(\tilde{\rho}\tilde{v}\tilde{w}) + \tilde{\rho}\frac{\partial \tilde{\Psi}}{\partial \tilde{z}}\right]\right\} = 0,$$

Poisson Equation:

$$\frac{\partial^2 \tilde{\Psi}}{\partial \tilde{\xi}^2} + \frac{\partial^2 \tilde{\Psi}}{\partial \tilde{y}^2} + \frac{\partial^2 \tilde{\Psi}}{\partial \tilde{z}^2} = 4\pi\tilde{\gamma}_N\left[\tilde{\rho} - \tilde{\tau}\left(-\frac{\partial \tilde{\rho}}{\partial \tilde{\xi}} + \frac{\partial}{\partial \tilde{\xi}}(\tilde{\rho}\tilde{u}) + \frac{\partial}{\partial \tilde{y}}(\tilde{\rho}\tilde{v}) + \frac{\partial}{\partial \tilde{z}}(\tilde{\rho}\tilde{w})\right)\right]$$

Motion equation (x – projection, x-axis is the direction of the soliton motion):

$$\frac{\partial}{\partial \tilde{\xi}}\left\{\tilde{\rho}\tilde{u}^2 + \tilde{p} - \tilde{\rho}\tilde{u}\right\} + \frac{\partial}{\partial \tilde{y}}(\tilde{\rho}\tilde{v}\tilde{u}) + \frac{\partial}{\partial \tilde{z}}(\tilde{\rho}\tilde{w}\tilde{u}) + \tilde{\rho}\frac{\partial \tilde{\Psi}}{\partial \tilde{\xi}} +$$

$$+ \frac{\partial}{\partial \tilde{\xi}}\left\{\tilde{\tau}\left[\begin{array}{l}\frac{\partial}{\partial \tilde{\xi}}\left(2\tilde{p}(1-\tilde{u}) - \tilde{\rho}\tilde{u}(1-\tilde{u})^2\right) + \\ + \frac{\partial}{\partial \tilde{y}}(\tilde{\rho}\tilde{v}\tilde{u}(1-\tilde{u})) + \frac{\partial}{\partial \tilde{z}}(\tilde{\rho}\tilde{w}\tilde{u}(1-\tilde{u})) + \tilde{\rho}\frac{\partial \tilde{\Psi}}{\partial \tilde{\xi}}(1-\tilde{u})\end{array}\right]\right\} -$$

$$- \tilde{\tau}\frac{\partial \tilde{\Psi}}{\partial \tilde{\xi}}\left(\frac{\partial}{\partial \tilde{\xi}}(\tilde{\rho}(\tilde{u}-1)) + \frac{\partial}{\partial \tilde{y}}(\tilde{\rho}\tilde{v}) + \frac{\partial}{\partial \tilde{z}}(\tilde{\rho}\tilde{w})\right) +$$

$$+ \frac{\partial}{\partial \tilde{y}}\left\{\tilde{\tau}\frac{\partial}{\partial \tilde{\xi}}(\tilde{\rho}\tilde{v}\tilde{u}(1-\tilde{u}))\right\} + \frac{\partial}{\partial \tilde{z}}\left\{\tilde{\tau}\frac{\partial}{\partial \tilde{\xi}}(\tilde{\rho}\tilde{w}\tilde{u}(1-\tilde{u}))\right\} -$$

$$- \frac{\partial}{\partial \tilde{y}}\left\{\tilde{\tau}\frac{\partial}{\partial \tilde{y}}(\tilde{\rho}\tilde{u}\tilde{v}^2)\right\} - \frac{\partial}{\partial \tilde{z}}\left\{\tilde{\tau}\frac{\partial}{\partial \tilde{y}}(\tilde{\rho}\tilde{u}\tilde{v}\tilde{w})\right\} -$$

$$- \frac{\partial}{\partial \tilde{y}}\left\{\tilde{\tau}\left[\frac{\partial}{\partial \tilde{z}}(\tilde{\rho}\tilde{u}\tilde{v}\tilde{w})\right]\right\} - \frac{\partial}{\partial \tilde{z}}\left\{\tilde{\tau}\left[\frac{\partial}{\partial \tilde{z}}(\tilde{\rho}\tilde{u}\tilde{w}^2)\right]\right\} -$$

$$- 2\frac{\partial}{\partial \tilde{\xi}}\left\{\tilde{\tau}\left[\frac{\partial}{\partial \tilde{y}}(\tilde{p}\tilde{v}) + \frac{\partial}{\partial \tilde{z}}(\tilde{p}\tilde{w})\right]\right\} -$$

$$- \frac{\partial}{\partial \tilde{\xi}}\left\{\tilde{\tau}\frac{\partial}{\partial \tilde{\xi}}(\tilde{p}\tilde{u})\right\} - \frac{\partial}{\partial \tilde{y}}\left\{\tilde{\tau}\frac{\partial}{\partial \tilde{y}}(\tilde{p}\tilde{u})\right\} - \frac{\partial}{\partial \tilde{z}}\left\{\tilde{\tau}\frac{\partial}{\partial \tilde{z}}(\tilde{p}\tilde{u})\right\} -$$

$$- \frac{\partial}{\partial \tilde{\xi}}\left\{\tilde{\tau}\left[\frac{\partial \tilde{\Psi}}{\partial \tilde{\xi}}\tilde{\rho}\tilde{u}\right]\right\} - \frac{\partial}{\partial \tilde{y}}\left\{\tilde{\tau}\left[\frac{\partial \tilde{\Psi}}{\partial \tilde{y}}\tilde{\rho}\tilde{u}\right]\right\} - \frac{\partial}{\partial \tilde{z}}\left\{\tilde{\tau}\left[\frac{\partial \tilde{\Psi}}{\partial \tilde{z}}\tilde{\rho}\tilde{u}\right]\right\} -$$

$$- \frac{\partial}{\partial \tilde{y}}\left\{\tilde{\tau}\left[\frac{\partial \tilde{\Psi}}{\partial \tilde{\xi}}\tilde{\rho}\tilde{v}\right]\right\} - \frac{\partial}{\partial \tilde{z}}\left\{\tilde{\tau}\left[\frac{\partial \tilde{\Psi}}{\partial \tilde{\xi}}\tilde{\rho}\tilde{w}\right]\right\} = 0,$$

Energy equation:

$$\frac{\partial}{\partial \tilde{\xi}}\left[\tilde{\rho}\tilde{v}_0^2(\tilde{u}-1)+5\tilde{p}\tilde{u}-3\tilde{p}\right]+2\tilde{\rho}\frac{\partial \tilde{\Psi}}{\partial \tilde{\xi}}\tilde{u}+$$

$$+\frac{\partial}{\partial \tilde{y}}\left[\tilde{\rho}\tilde{v}_0^2\tilde{v}+5\tilde{p}\tilde{v}\right]+2\tilde{\rho}\frac{\partial \tilde{\Psi}}{\partial \tilde{y}}\tilde{v}+\frac{\partial}{\partial \tilde{z}}\left[\tilde{\rho}\tilde{v}_0^2\tilde{w}+5\tilde{p}\tilde{w}\right]+2\tilde{\rho}\frac{\partial \tilde{\Psi}}{\partial \tilde{z}}\tilde{w}+$$

$$+\frac{\partial}{\partial \tilde{\xi}}\left\{\tilde{\tau}\left[\begin{array}{l}\frac{\partial}{\partial \tilde{\xi}}\left(-\tilde{\rho}\tilde{v}_0^2(1-\tilde{u})^2+7\tilde{p}\tilde{u}(1-\tilde{u})+3\tilde{p}(\tilde{u}-1)-\tilde{p}\tilde{v}_0^2-5\frac{\tilde{p}^2}{\tilde{\rho}}\right)+\\ +2\frac{\partial \tilde{\Psi}}{\partial \tilde{\xi}}\tilde{\rho}\tilde{u}(1-\tilde{u})-\tilde{\rho}\tilde{v}_0^2\frac{\partial \tilde{\Psi}}{\partial \tilde{\xi}}-5\tilde{p}\frac{\partial \tilde{\Psi}}{\partial \tilde{\xi}}\end{array}\right]\right\}+$$

$$+\frac{\partial}{\partial \tilde{\xi}}\left\{\tilde{\tau}\left[\frac{\partial}{\partial \tilde{y}}\left(\tilde{\rho}\tilde{v}_0^2\tilde{v}(1-\tilde{u})+5\tilde{p}\tilde{v}-7\tilde{p}v\tilde{u}\right)+2\frac{\partial \tilde{\Psi}}{\partial \tilde{y}}\tilde{\rho}\tilde{v}(1-\tilde{u})\right]\right\}+$$

$$+\frac{\partial}{\partial \tilde{\xi}}\left\{\tilde{\tau}\left[\frac{\partial}{\partial \tilde{z}}\left(\tilde{\rho}\tilde{v}_0^2\tilde{w}(1-\tilde{u})+5\tilde{p}\tilde{w}-7\tilde{p}\tilde{w}\tilde{u}\right)+2\frac{\partial \tilde{\Psi}}{\partial \tilde{z}}\tilde{\rho}\tilde{w}(1-\tilde{u})\right]\right\}+$$

$$+\frac{\partial}{\partial \tilde{y}}\left\{\tilde{\tau}\frac{\partial}{\partial \tilde{\xi}}\left(\tilde{\rho}\tilde{v}_0^2\tilde{v}(1-\tilde{u})+5\tilde{p}\tilde{v}-7\tilde{p}\tilde{u}\tilde{v}\right)\right\}+\frac{\partial}{\partial \tilde{z}}\left\{\tilde{\tau}\frac{\partial}{\partial \tilde{\xi}}\left(\tilde{\rho}\tilde{v}_0^2\tilde{w}(1-\tilde{u})+5\tilde{p}\tilde{w}-7\tilde{p}\tilde{u}\tilde{w}\right)\right\}-$$

$$-\frac{\partial}{\partial \tilde{y}}\left\{\tilde{\tau}\left[\frac{\partial}{\partial \tilde{y}}\left(\tilde{\rho}\tilde{v}_0^2\tilde{v}^2+7\tilde{p}\tilde{v}^2+\tilde{p}\tilde{v}_0^2+5\frac{\tilde{p}^2}{\tilde{\rho}}\right)+2\tilde{\rho}\frac{\partial \tilde{\Psi}}{\partial \tilde{y}}\tilde{v}^2+\tilde{\rho}\tilde{v}_0^2\frac{\partial \tilde{\Psi}}{\partial \tilde{y}}+5\tilde{p}\frac{\partial \tilde{\Psi}}{\partial \tilde{y}}\right]\right\}-$$

$$-\frac{\partial}{\partial \tilde{z}}\left\{\tilde{\tau}\left[\frac{\partial}{\partial \tilde{y}}\left(\tilde{\rho}\tilde{v}_0^2\tilde{v}\tilde{w}+7\tilde{p}\tilde{v}\tilde{w}\right)+2\tilde{\rho}\frac{\partial \tilde{\Psi}}{\partial \tilde{y}}\tilde{v}\tilde{w}+2\tilde{\rho}\frac{\partial \tilde{\Psi}}{\partial \tilde{\xi}}\tilde{u}\tilde{w}\right]\right\}-$$

$$-\frac{\partial}{\partial \tilde{y}}\left\{\tilde{\tau}\left[\frac{\partial}{\partial \tilde{z}}\left(\tilde{\rho}\tilde{v}_0^2\tilde{w}\tilde{v}+7\tilde{p}\tilde{w}\tilde{v}\right)+2\tilde{\rho}\frac{\partial \tilde{\Psi}}{\partial \tilde{z}}\tilde{w}\tilde{v}+2\tilde{\rho}\frac{\partial \tilde{\Psi}}{\partial \tilde{\xi}}\tilde{u}\tilde{v}\right]\right\}-$$

$$-\frac{\partial}{\partial \tilde{z}}\left\{\tilde{\tau}\left[\frac{\partial}{\partial \tilde{z}}\left(\tilde{\rho}\tilde{v}_0^2\tilde{w}^2+7\tilde{p}\tilde{w}^2+\tilde{p}\tilde{v}_0^2+5\frac{\tilde{p}^2}{\tilde{\rho}}\right)+2\tilde{\rho}\frac{\partial \tilde{\Psi}}{\partial \tilde{z}}\tilde{w}^2+\tilde{\rho}\tilde{v}_0^2\frac{\partial \tilde{\Psi}}{\partial \tilde{z}}+5\tilde{p}\frac{\partial \tilde{\Psi}}{\partial \tilde{z}}\right]\right\}+$$

$$+2\tilde{\tau}\frac{\partial \tilde{\Psi}}{\partial \tilde{\xi}}\left[\frac{\partial}{\partial \tilde{\xi}}(\tilde{\rho}\tilde{u}(1-\tilde{u}))-\tilde{\rho}\frac{\partial \tilde{\Psi}}{\partial \tilde{\xi}}\right]+2\tilde{\tau}\frac{\partial \tilde{\Psi}}{\partial \tilde{y}}\left[\frac{\partial}{\partial \tilde{\xi}}(\tilde{\rho}\tilde{v}(1-\tilde{u}))-\tilde{\rho}\frac{\partial \tilde{\Psi}}{\partial \tilde{y}}\right]+2\tilde{\tau}\frac{\partial \tilde{\Psi}}{\partial \tilde{z}}\left[\frac{\partial}{\partial \tilde{\xi}}(\tilde{\rho}\tilde{w}(1-\tilde{u}))-\tilde{\rho}\frac{\partial \tilde{\Psi}}{\partial \tilde{z}}\right]-$$

$$-2\tilde{\tau}\frac{\partial \tilde{\Psi}}{\partial \tilde{\xi}}\left[\frac{\partial}{\partial \tilde{\xi}}\tilde{p}\right]-2\tilde{\tau}\frac{\partial \tilde{\Psi}}{\partial \tilde{y}}\left[\frac{\partial}{\partial \tilde{y}}\tilde{p}\right]-2\tilde{\tau}\frac{\partial \tilde{\Psi}}{\partial \tilde{z}}\left[\frac{\partial}{\partial \tilde{z}}\tilde{p}\right]-$$

$$-2\tilde{\tau}\frac{\partial \tilde{\Psi}}{\partial \tilde{\xi}}\left[\frac{\partial}{\partial \tilde{y}}\tilde{\rho}\tilde{v}\tilde{u}\right]-2\tilde{\tau}\frac{\partial \tilde{\Psi}}{\partial \tilde{y}}\left[\frac{\partial}{\partial \tilde{y}}\tilde{\rho}\tilde{v}^2\right]-2\tilde{\tau}\frac{\partial \tilde{\Psi}}{\partial \tilde{z}}\left[\frac{\partial}{\partial \tilde{y}}\tilde{\rho}\tilde{v}\tilde{w}\right]-$$

$$-2\tilde{\tau}\frac{\partial \tilde{\Psi}}{\partial \tilde{\xi}}\left[\frac{\partial}{\partial \tilde{z}}\tilde{\rho}\tilde{w}\tilde{u}\right]-2\tilde{\tau}\frac{\partial \tilde{\Psi}}{\partial \tilde{y}}\left[\frac{\partial}{\partial \tilde{z}}\tilde{\rho}\tilde{w}\tilde{v}\right]-2\tilde{\tau}\frac{\partial \tilde{\Psi}}{\partial \tilde{z}}\left[\frac{\partial}{\partial \tilde{z}}\tilde{\rho}\tilde{w}^2\right]=0$$

Derivatives in the spherical coordinate system (usual notations):

$$\frac{\partial}{\partial x}=\sin\theta\cos\varphi\frac{\partial}{\partial r}+\frac{\cos\theta\cos\varphi}{r}\frac{\partial}{\partial \theta}-\frac{\sin\varphi}{r\sin\theta}\frac{\partial}{\partial \varphi},$$

$$\frac{\partial}{\partial y}=\sin\theta\sin\varphi\frac{\partial}{\partial r}+\frac{\cos\theta\sin\varphi}{r}\frac{\partial}{\partial \theta}+\frac{\cos\varphi}{r\sin\theta}\frac{\partial}{\partial \varphi},$$

$$\frac{\partial}{\partial z}=\cos\theta\frac{\partial}{\partial r}-\frac{\sin\theta}{r}\frac{\partial}{\partial \theta},$$

$$\frac{\partial^2}{\partial x^2} =$$
$$= \left[ \sin^2\theta \cos^2\varphi \frac{\partial^2}{\partial r^2} + \frac{2\sin\theta\cos\theta\cos^2\varphi}{r}\frac{\partial^2}{\partial r \partial \theta} - \frac{2\sin\theta\cos\theta\cos^2\varphi}{r^2}\frac{\partial}{\partial \theta} - \frac{2\cos\varphi\sin\varphi}{r}\frac{\partial^2}{\partial r \partial \varphi} \right] +$$
$$+ \left[ \frac{\cos^2\theta\cos^2\varphi}{r}\frac{\partial}{\partial r} + \frac{\cos^2\theta\cos^2\varphi}{r^2}\frac{\partial^2}{\partial \theta^2} - \frac{2\sin\varphi}{\sin\theta}\frac{\cos\theta\cos\varphi}{r^2}\frac{\partial^2}{\partial \theta \partial \varphi} + \frac{2\sin\varphi\cos\varphi}{r^2\sin^2\theta}\frac{\partial}{\partial \varphi} \right] +$$
$$+ \left[ \frac{\sin^2\varphi}{r}\frac{\partial}{\partial r} + \frac{\cos\theta}{r^2}\frac{\sin^2\varphi}{\sin\theta}\frac{\partial}{\partial \theta} + \frac{\sin^2\varphi}{r^2\sin^2\theta}\frac{\partial^2}{\partial \varphi^2} \right],$$

$$\frac{\partial^2}{\partial y^2} = \left[ \sin^2\theta \sin^2\varphi \frac{\partial^2}{\partial r^2} - \frac{2\sin\theta\cos\theta\sin^2\varphi}{r^2}\frac{\partial}{\partial \theta} + \frac{2\sin\varphi\cos\varphi}{r}\frac{\partial^2}{\partial r \partial \varphi} - \frac{2\sin\varphi\cos\varphi}{r^2\sin^2\theta}\frac{\partial}{\partial \varphi} \right] +$$
$$+ \left[ 2\sin\theta\frac{\cos\theta\sin^2\varphi}{r}\frac{\partial^2}{\partial r \partial \theta} + \frac{\cos^2\theta\sin^2\varphi}{r}\frac{\partial}{\partial r} + \frac{\cos^2\theta\sin^2\varphi}{r^2}\frac{\partial^2}{\partial \theta^2} + \frac{2\cos\varphi}{\sin\theta}\frac{\cos\theta\sin\varphi}{r^2}\frac{\partial^2}{\partial \theta \partial \varphi} \right] +$$
$$+ \left[ \frac{\cos^2\varphi}{r}\frac{\partial}{\partial r} + \frac{\cos\theta}{r^2}\frac{\cos^2\varphi}{\sin\theta}\frac{\partial}{\partial \theta} + \frac{\cos^2\varphi}{r^2\sin^2\theta}\frac{\partial^2}{\partial \varphi^2} \right],$$

$$\frac{\partial^2}{\partial z^2} = \cos^2\theta \frac{\partial^2}{\partial r^2} - \frac{2\cos\theta\sin\theta}{r}\frac{\partial^2}{\partial \theta \partial r} + \frac{\sin^2\theta}{r}\frac{\partial}{\partial r} + \frac{\sin^2\theta}{r^2}\frac{\partial^2}{\partial \theta^2} + \frac{2\sin\theta\cos\theta}{r^2}\frac{\partial}{\partial \theta}.$$